\documentclass[structabstract]{aa}  
\usepackage{graphicx}
\usepackage{txfonts}
\usepackage{natbib}
\usepackage{printlen}
\usepackage{color}
\newcommand{\comment}[1]{}
\newcommand{\vel}{\upsilon}
\newcommand{\ndI}{\hat I}
\begin{document}
   \title{Three-dimensional simulations of near-surface convection in main-sequence stars}
   \subtitle{II. Properties of granulation and spectral lines}

   \author{B. Beeck
          \inst{1,2}
          \and
          R.~H. Cameron\inst{1}
	  \and
	  A. Reiners\inst{2}
     	  \and
	  M. Sch\"ussler\inst{1}
          }

   \institute{Max-Planck-Institut f\"ur Sonnensystemforschung,
              Max-Planck-Stra\ss e 2, 37191 Katlenburg-Lindau, Germany\\
         \and
              Institut f\"ur Astrophysik, 
	      Universit\"at G\"ottingen, 
	      Friedrich-Hund-Platz 1,
	      37077 G\"ottingen, Germany\\
             }

   \date{Received: 22 February, 2013; accepted 16 August, 2013}
   \titlerunning{3D simulations of stellar surface layers}

\abstract
{The atmospheres of cool main-sequence stars are structured by convective flows from the convective envelope that penetrate the optically thin layers and lead to structuring of the stellar atmospheres analogous to solar granulation. The flows have considerable influence on the 3D structure of temperature and pressure and affect the profiles of spectral lines formed in the photosphere.}
{For the set of six 3D radiative (M)HD simulations of cool main-sequence stars described in the first paper of this series, we analyse the near-surface layers. We aim at describing the properties of granulation of different stars and at quantifying  the effects on spectral lines of the thermodynamic structure and flows of 3D convective atmospheres.}
{We detected and tracked granules in brightness images from the simulations to analyse their statistical properties, as well as their evolution and lifetime. We calculated spatially resolved spectral line profiles using the line synthesis code \texttt{SPINOR}. To enable a comparison to stellar observations, we implemented a numerical disc-integration, which includes (differential) rotation.}
{Although the stellar parameters change considerably along the model sequence, the properties of the granules are very similar. The impact of the 3D structure of the atmospheres on line profiles is measurable in disc-integrated spectra. Line asymmetries caused by convection are modulated by stellar rotation.}
{The 3D structure of cool stellar atmospheres as shaped by convective flows has to be taken into account when using photospheric lines to determine stellar parameters.}

   \keywords{convection -- hydrodynamics -- stars:late-type -- stars:low-mass}

   \maketitle
%
\section{Introduction}
Cool main-sequence stars have outer convection zones or are fully convective. The convective flows, which carry the bulk of the energy flux from the interiors of the stars through their convection zones to their surfaces, shoot over into the stably stratified, optically thin photospheric layers and have a strong impact on the structure of the atmospheres. The convective motions in the near-surface layers create the granulation pattern. Hot upflow regions (granules) with a horizontal size close to the local density scale height are surrounded by a network of intergranular lanes of cool, dense plasma transporting the material back down in strong, turbulent downdrafts \citep[for a review, see][]{Nordlund09}.\par
While solar granulation can be spatially resolved with optical instruments, the direct observation of granulation in other main-sequence stars is impossible. To infer properties of stellar granulation, one usually analyses spectral line profiles: the velocity field and its correlation with temperature (and pressure) entail line broadening, asymmetries, and shifts \citep[see, e.\,g.,][]{ND90b, Gray09}. However, as abundances, (differential) rotation, and magnetic fields also affect the spectral line profiles, such an analysis is often ambiguous, and the problem of determining all stellar parameters from a spectrum is highly degenerate. A good description of the 3D structure of stellar granulation is therefore essential for a precise analysis of stellar spectra. Following the pioneering work of \citet{ND90a}, \citet{ArtGad94}, \citet{Vort6}, and others, we simulated and analysed the near-surface layers of cool stars, using parameters close to those of real main-sequence objects.\par
In the first paper of this series \citep[][Paper~I, hereafter]{Paper1}, we analysed the depth dependence of the mean structure of the simulated main-sequence stars with spectral types between F3 and M2. Here, we concentrate on the properties of the stellar photospheres, especially of the granulation (Sect.~\ref{sec:gran}), the limb darkening (Sect.\ref{sec:ld}), and the effect of convection on photospheric spectral lines (Sect. \ref{sec:lines}). In a third paper, we plan to investigate the effect of a magnetic field on the atmospheric structure and on the resulting spectral lines.
%
\section{Simulations}\label{sec:sim}
The simulations presented in this paper were performed with the 3D radiative MHD code \texttt{MURaM} \citep{Vogd,MURaM1}. The dimensions of the computational domain were chosen according to the expected granule size and pressure scale heights. To facilitate an easy comparison, a very similar simulation setup was chosen for the six simulations analysed here. For more details on the simulation setup, see Paper~I.\par
In the \texttt{MURaM} code, the radiative transfer is calculated by applying the opacity binning method. The simulations analysed here used four opacity bins calculated on the basis of opacity distribution functions provided by the \texttt{ATLAS9} code \citep{ATLAS9}. The regrouping of the opacities was done according to the $\tau$-sorting method (bin limits at $\log \tau_{\mathrm{ref}}=0,2,4$). The reference atmosphere for the $\tau$-sorting was obtained by running the simulation applying opacities binned for similar stellar parameters (see Paper~I). For more details of the treatment of radiation in the \texttt{MURaM} code, see \citet{Vog04}.\par

%

%
\section{Granulation}\label{sec:gran}
\begin{figure*}
\centering
\begin{tabular}{cccc}
 & \includegraphics[width=6cm]{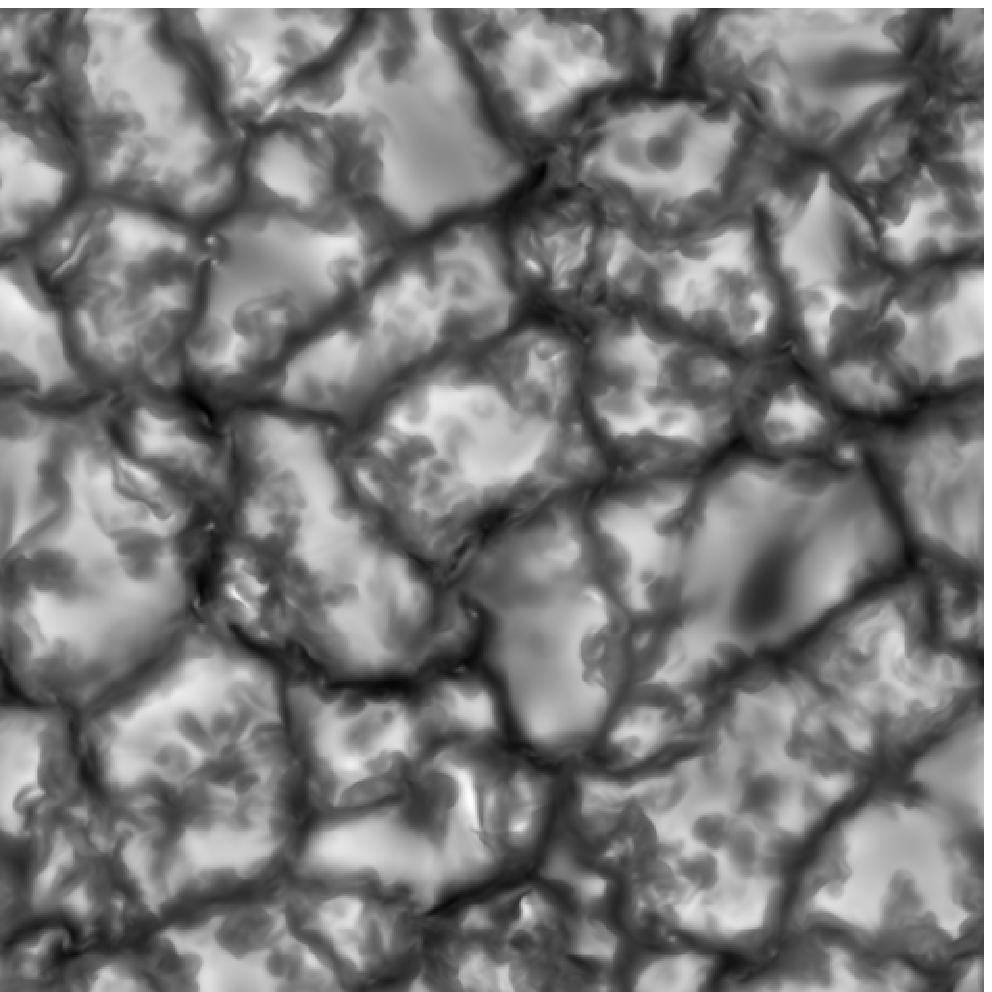} & \parbox[b][6cm][c]{1cm}{\Large$\stackrel{(S1)}{\longrightarrow}$} & \includegraphics[width=6cm]{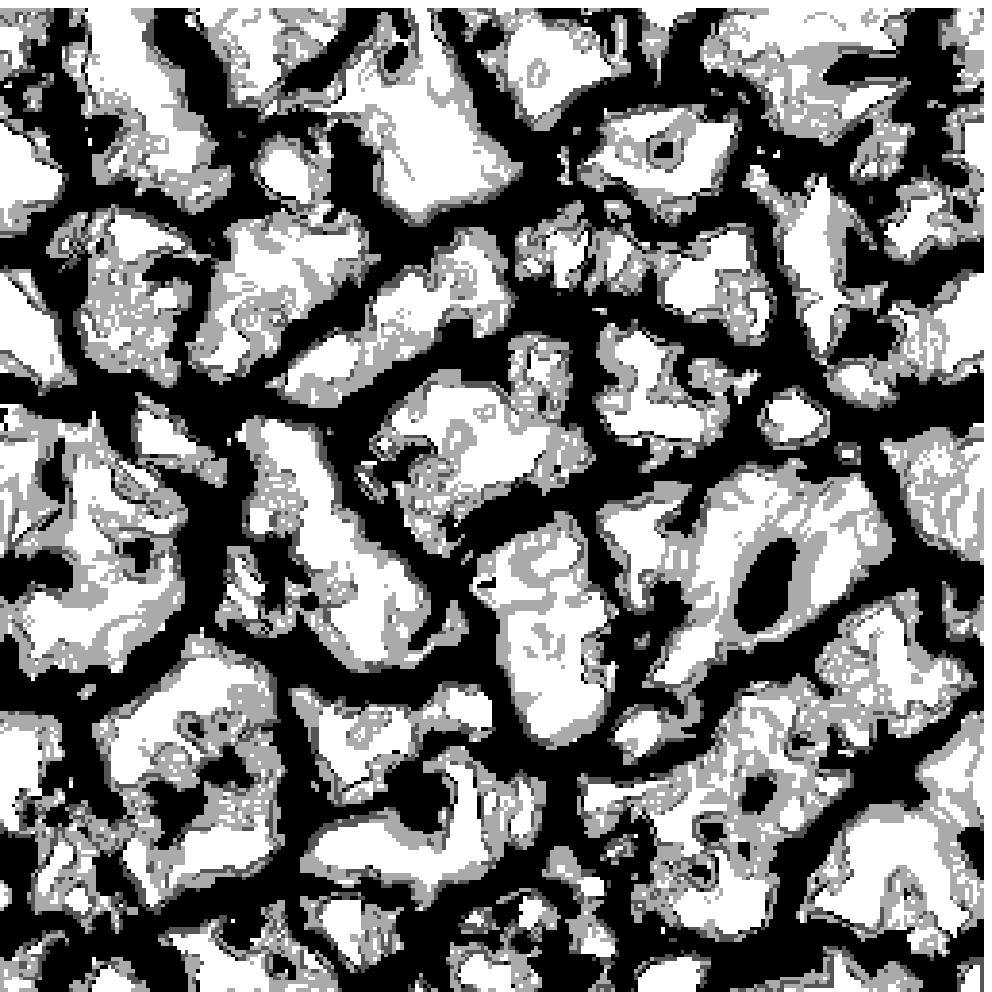}\\ \parbox[b][6cm][c]{1cm}{\Large$\stackrel{(S2)}{\longrightarrow}$} & \includegraphics[width=6cm]{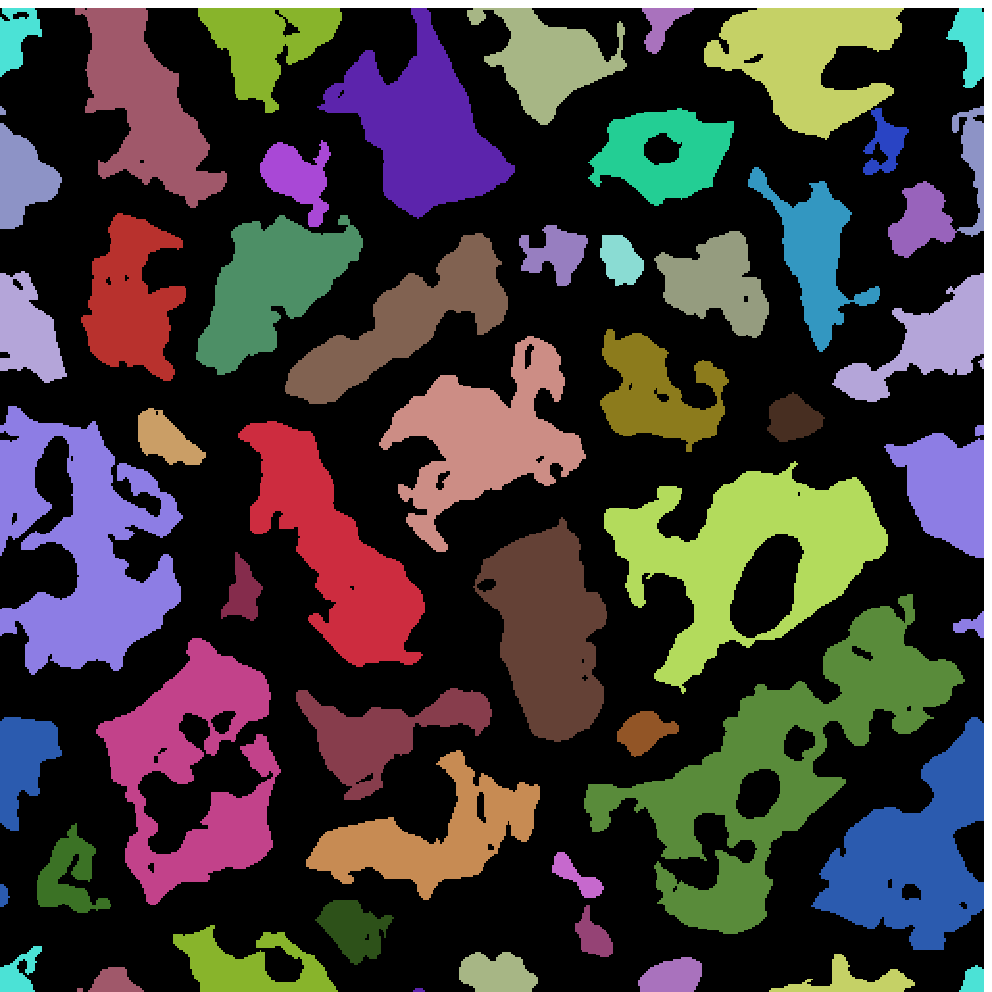} &
\parbox[b][6cm][c]{1cm}{\Large$\stackrel{(S3)}{\longrightarrow}$} & \includegraphics[width=6cm]{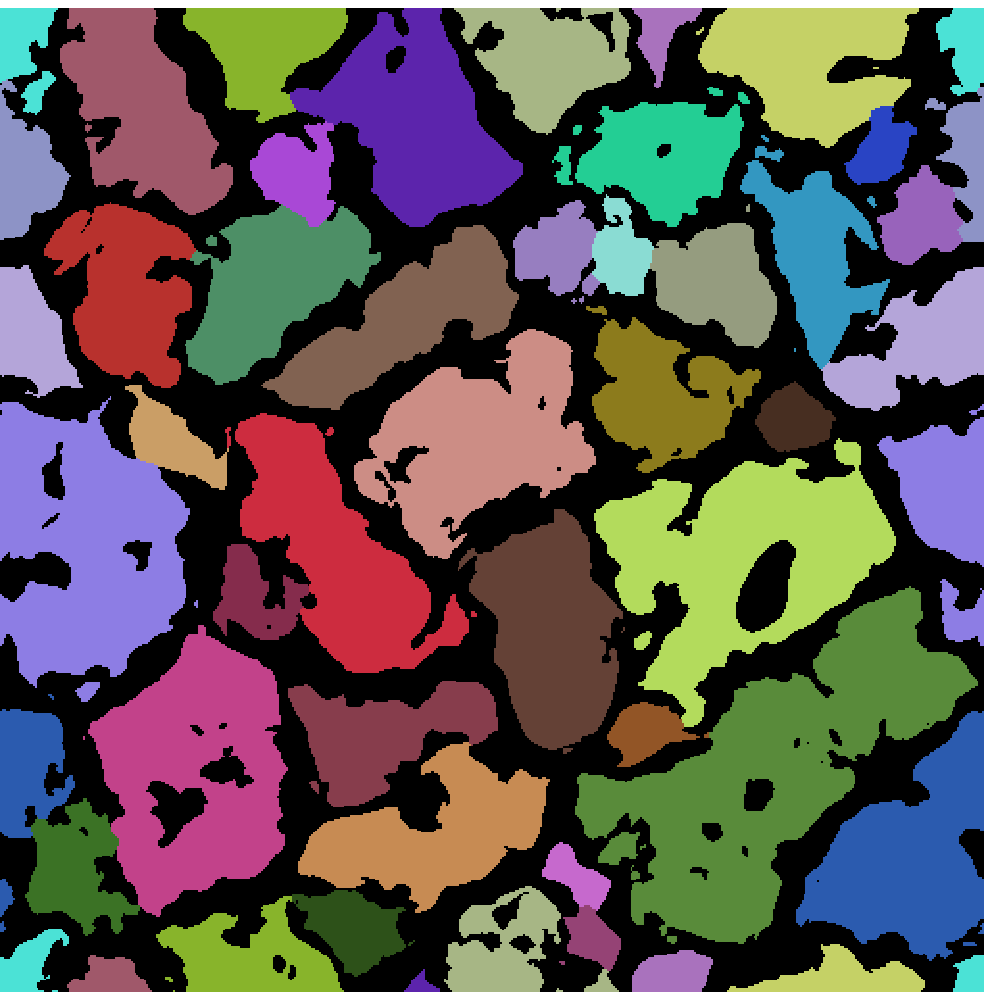} \\ \parbox[b][6cm][c]{1cm}{\Large$\stackrel{(T1)}{\longrightarrow}$} & \includegraphics[width=6cm]{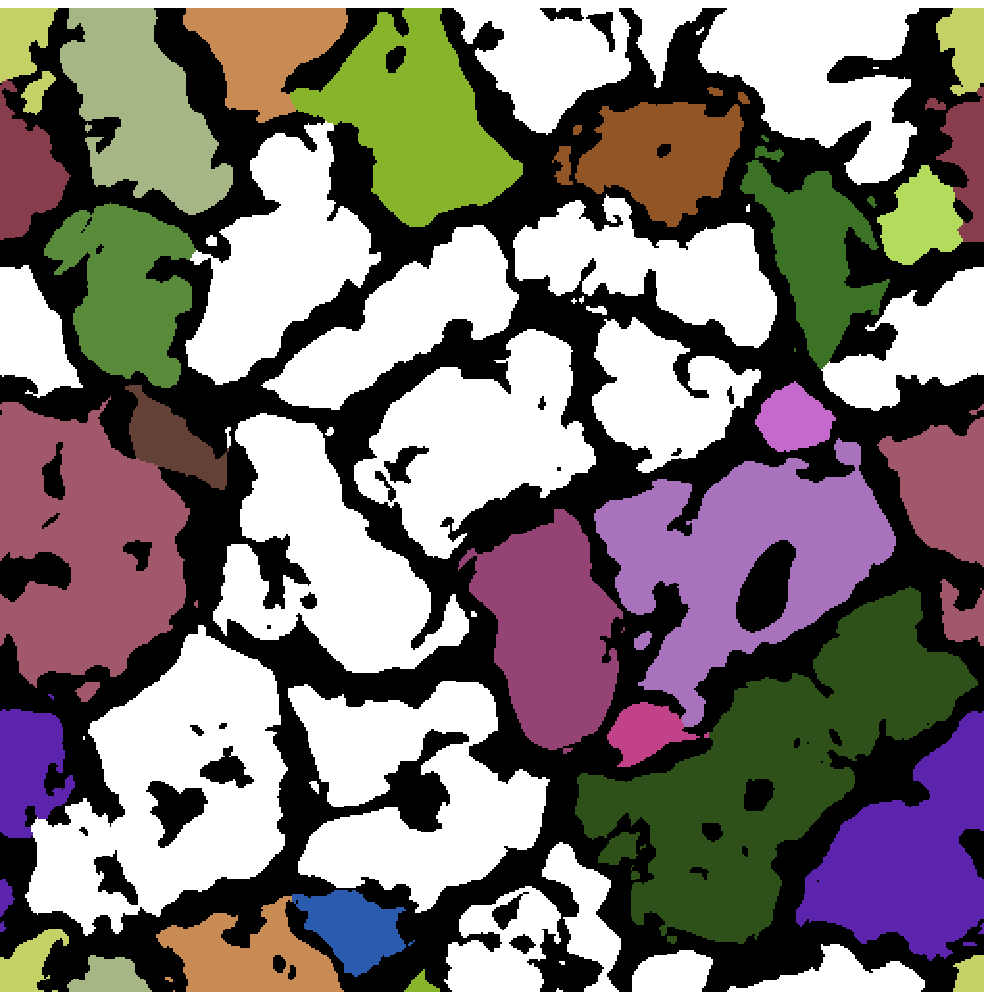} & \parbox[b][6cm][c]{1cm}{\Large$\stackrel{(T2-5)}{\longrightarrow}$} & \includegraphics[width=6cm]{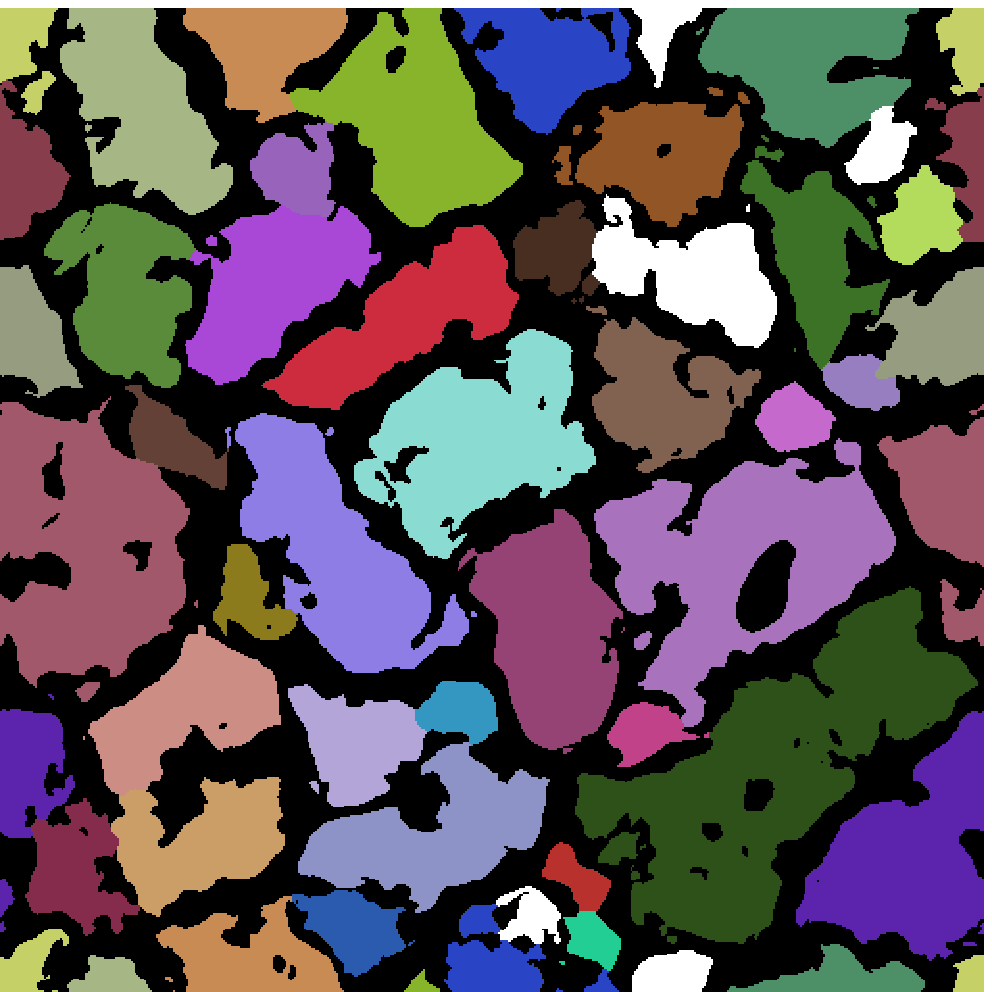}\\
\end{tabular}
\caption{Illustration of the segmentation and tracking algorithm (snapshot of the G2V simulation). {\it Top panels:} raw intensity image ({\it left}); map of subimage categories (the number $a$ is shown in grey scale; {\it right}). {\it Middle panels}: granule segmentation before the multiple-level tracking (MLT; coloured areas = granules; black = intergranular lanes; colours were chosen randomly and for clarity; {\it left}); segmented image after MLT ({\it right}). {\it Bottom panels:} trackable (coloured) and non-trackable (white) granules after the first segmentation/tracking run ({\it left}) and after the fifth run ({\it right}).}\label{fig:algo}
\end{figure*}
\subsection{Granule segmentation}\label{sec:met:gran}
As a basis for the analysis of the granulation patterns, we used a time series of 500 synthetic (bolometric) intensity images for each of the six simulation runs (for details on the simulations considered, see Paper I). The images are separated by $20\delta t$ where $\delta t$ is the simulation time step (depending on the star, $2.9\,\mathrm{s}\le 20\delta t\le 4.3\,\mathrm{s}$), which is well below the typical lifetime of the stellar granules (of the order of several minutes).\par 
We implemented a granule segmentation and tracking algorithm, which was designed to follow granules in intensity images. In order to scale to the different intensity values of the stars the algorithm uses the normalised (bolometric) intensity fluctuation,
\begin{equation}\label{eqn:def:I}
\ndI = (I-\langle I\rangle)/\sigma_I\,,\quad\mathrm{with}\,\,\, \sigma_I=\sqrt{\langle I^2\rangle - \langle I \rangle ^2}\,,
\end{equation}
where $\langle\dots\rangle$ denotes the temporal and spatial mean. This normalisation simplifies the comparison between stars with different surface temperatures and temperature contrasts.\par
In the first step, $S1$, of the algorithm, the image is divided into subimages of $2\times 2$ pixels. For each subimage $i$, the mean, $\langle \ndI \rangle_{i}$, and the standard deviation, $\sigma_i(\ndI)$, of the normalised intensity contrast are calculated. The algorithm then utilises the fact that granules are extended bright features, while intergranular lanes are dark and rather sharp (i.\,e. the intensity gradient is large). Accordingly, depending on its values of $\langle \ndI\rangle_{i}$ and $\sigma_{i}(\ndI)$, each subimage is put into four categories, characterised by a number $a$ between 0 and 1. A high value of $a$ means that the probability that the subimage is part of a granule is high. The categories are defined as:
\begin{equation}
a= \left\{ \begin{array}{r@{\,\,\,\,}l} 0, & \mathrm{if}\,\, \langle\ndI\rangle_i < I_{\mathrm{inter}}\,\,\,\mathrm{or}\,\,\, \sigma_i(\ndI) > \sigma_h\\
                 0.3, & \mathrm{if}\,\, I_{\mathrm{inter}} \le \langle\ndI\rangle_{i} \le I_{\mathrm{gran}}\,\,\,\mathrm{and}\,\,\, \sigma_h>\sigma_{i}(\ndI) \ge \sigma_m\\
                 0.7, & \mathrm{if}\,\, I_{\mathrm{inter}} \le \langle\ndI\rangle_{i} \le I_{\mathrm{gran}}\,\,\,\mathrm{and}\,\,\, \sigma_m>\sigma_{i}(\ndI) \ge \sigma_l\\
                 1, &  \mathrm{if}\,\, \langle\ndI\rangle_{i} > I_{\mathrm{gran}}\,\,\,\mathrm{or}\,\,\, \sigma_i(\ndI) <\sigma_l\end{array}\right.\,.
\end{equation}
The threshold values $I_{\mathrm{gran}}$, $I_{\mathrm{inter}}$, $\sigma_l$, $\sigma_m$, and $\sigma_h$ enter the segmentation as parameters.\par
The ``map'' resulting from this categorising (cf. upper right panel of Fig.~\ref{fig:algo}) has half the resolution of the original image. In the next step, $S2$, this map is smoothed by a Gaussian ($\sigma = 2\,\mathrm{pixels}$), interpolated back onto the original grid size, and clipped at a threshold value of typically 0.5. The continuous regions obtained in this step already roughly correspond to granules in the original images but are usually somewhat smaller. In the last step, $S3$, using a multiple-level tracking (MLT) method \citep[cf.][]{BW01}, all pixels with  $\ndI > I_{\mathrm{inter}}$ which are not yet part of a detected granule are either assigned to one of the granules already detected or considered as small-scale intergranular structures, which were excluded with a minimum-size criterion. The first four images of the series shown in Figure~\ref{fig:algo} give an illustration of the segmentation algorithm.\par
For a statistical analysis of granule properties (such as size, mean intensity, etc.) in snapshots this algorithm provides a suitable basis. However, it is not sufficient for tracking granules in time since there are often cases where the distinction between a single granule with substructure and a group of two or more granules is not clear. The interpretation of such cases by the algorithm could fluctuate from snapshot to snapshot, which would give rise to numerous granules which are identified only in one snapshot. To overcome this problem and enable tracking granules, a modification of the scheme is necessary. First, the series of images were subdivided into packages of 10 sequential images. After running the segmentation algorithm described above for all 10 images, a simple comparison of overlapping granule area in subsequent images yields a first tracking step ($T1$). Structures which exist through the whole subseries of images are considered as unambiguous cases of trackable (``stable'') granules. Next, the segmentation is run again only considering the parts of the 10 images where no trackable granules have been detected in the previous step. A slight variation of the segmentation parameters enables the algorithm to find further granules in this run. After about five iterations 80 -- 90\% of the granules can be tracked through the series of 10 subsequent images. The remaining granules are probably in a phase of evolution where their definition is particularly difficult (splitting or merging). The last two images of the series shown in Figure~\ref{fig:algo} show the trackable granules after the first segmentation/tracking run and after the last (fifth) run, respectively. Most of the granules are detected by the algorithm. There are, however, a few untrackable granules: the largest white region appearing in the lower right panel of Figure~\ref{fig:algo}, for instance, seems to belong to a larger granule-like, bright area. A closer inspection of the original image and the intermediate steps of the algorithm suggests, however, that this larger area is composed of three parts. While the middle and the right part do not appear clearly divided from each other by a dark lane, there is a lane-like structure between the left-hand side part of the area and the rest. The algorithm will therefore find this left-hand side part as one granule living through the short subseries of 10 images for most parameter sets whereas the rest of this granule-like area will sometimes appear as two, sometimes as one detected granule and remain untrackable.\par

Finally, the packages are merged and granules which exist in subsequent packages are identified (search for overlapping area, similar in size). The described procedure efficiently finds long-lived granules, but it also limits the lifetime resolution, since, by definition in this algorithm, the granules can only have lifetimes which are integer multiples of 10 subsequent images. This corresponds to a minimum lifetime of roughly 29 to 43\,s, depending on the simulated star.\par
In order to use the same data set and algorithm for the statistical analysis (Sect.~\ref{sec:granprop}) of snapshots as for the evolution and lifetime analysis of the granules (Sect.~\ref{sec:timescale}), we limited ourselves to a rather short sequence of 500 images spanning roughly half an hour of stellar time. As a good compromise between statistical independence and statistical significance, we considered every $15^{\mathrm{th}}$ image for the statistical part of the analysis, which corresponds to a time step of the order of one minute of stellar time. Since this is already smaller than the typical evolution time scale of the granules ($\sim$ several minutes), a finer sampling would not improve the statistical significance. For the statistical analysis in Sect.~\ref{sec:granprop}, all granules (trackable and not trackable) were considered, whereas for the determination of the lifetime and time averaged properties, the untrackable granules were excluded from the sample.\par
 
\subsection{General appearance of the granulation pattern}
\begin{figure*}
\centering
\includegraphics[width=5.7cm]{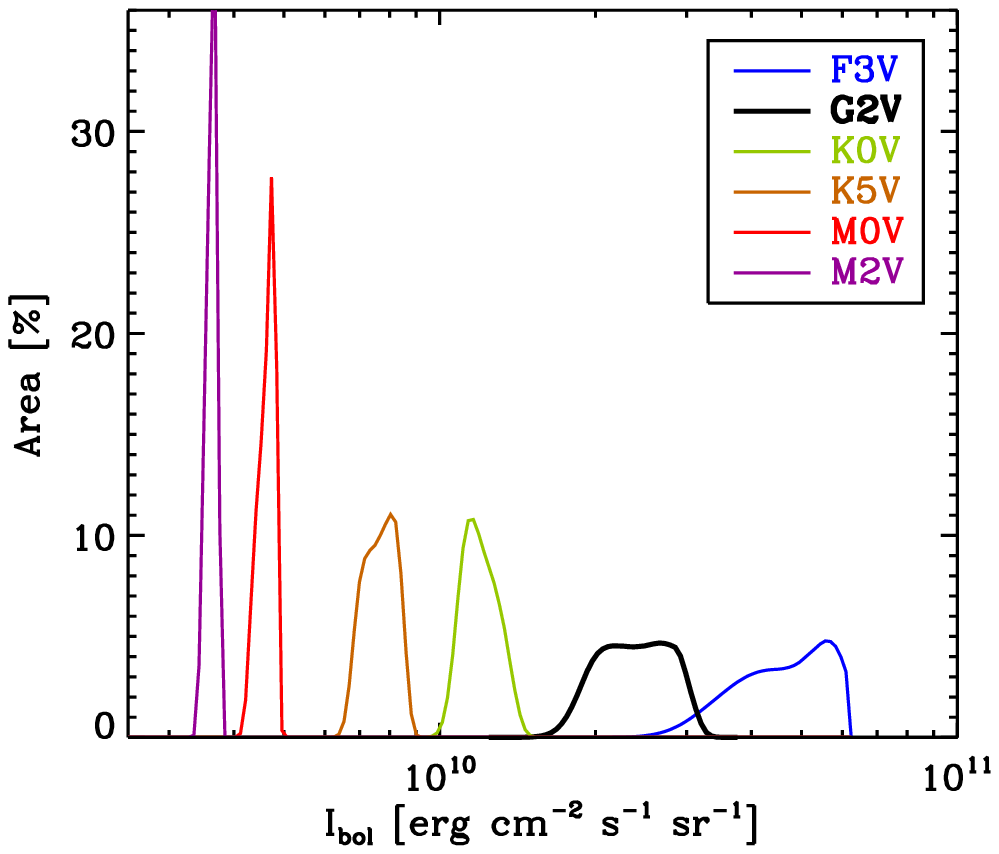}~%
\includegraphics[width=5.7cm]{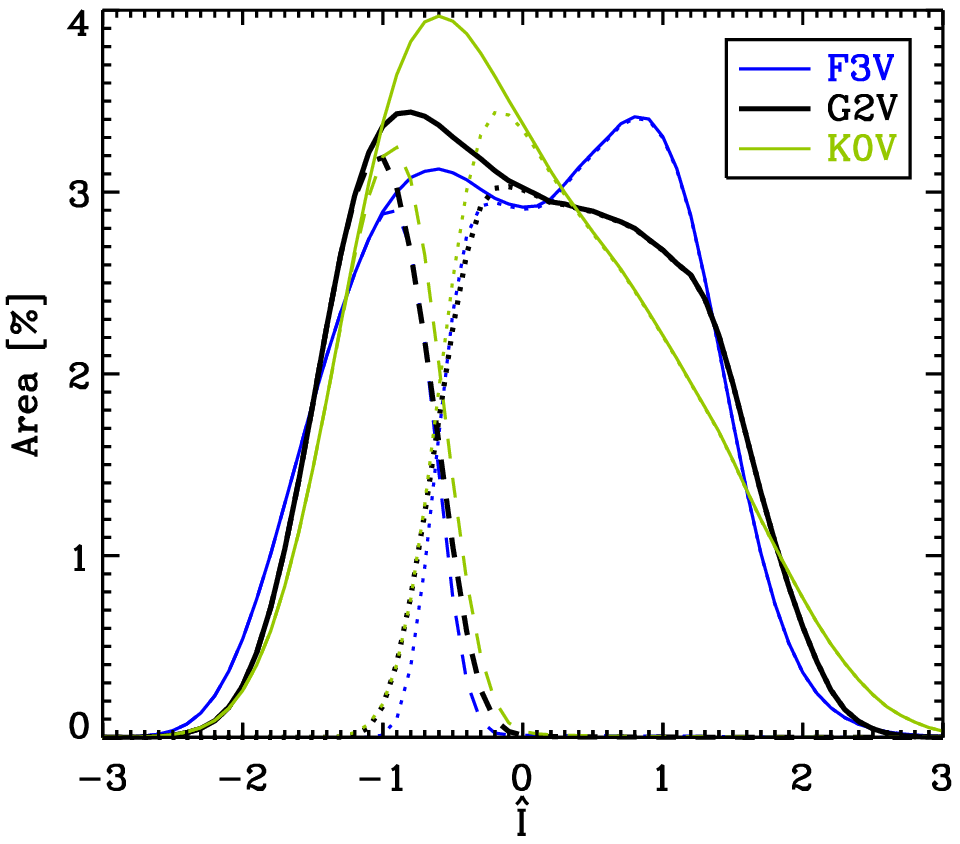}~%
\includegraphics[width=5.7cm]{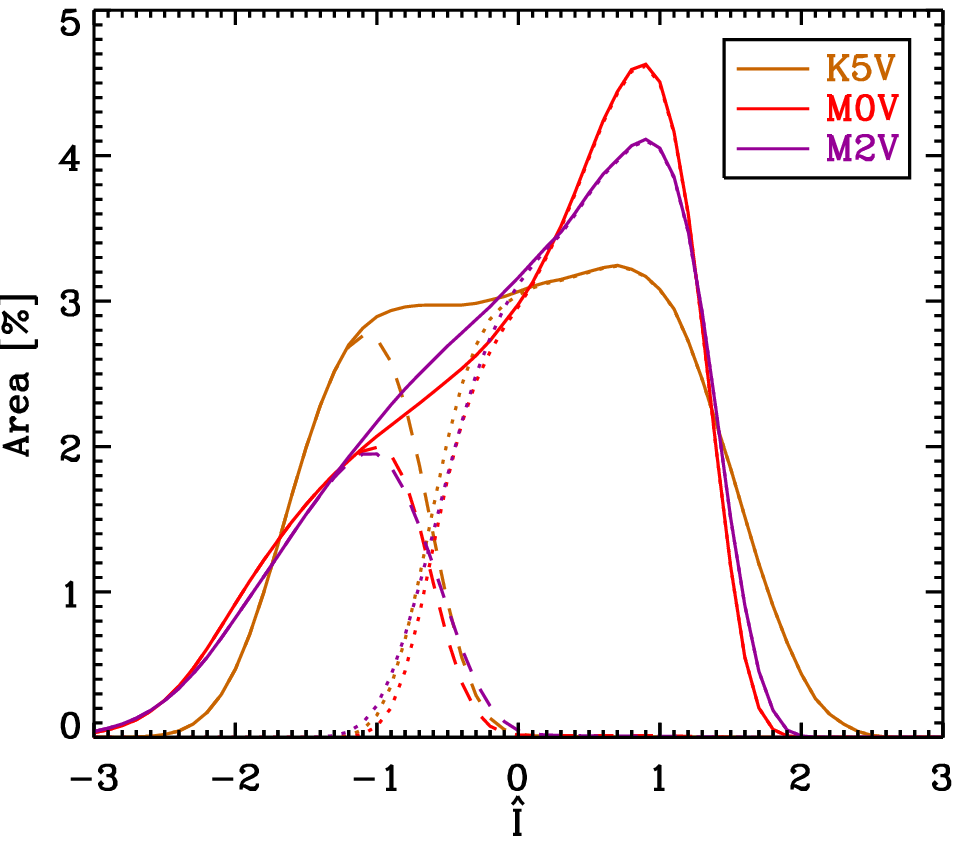}\\
\caption{Intensity histograms. {\it Left panel:} histograms of the bolometric intensity. The bins are logarithmically equidistant with a bin size of 0.01 dex. {\it Middle and right panels:} histograms of the normalised intensity fluctuation $\ndI$ (see Eq.~(\ref{eqn:def:I})). The bins are linearly equidistant with a bin size of 0.1. Solid curves show the histograms of the entire area, dotted and dashed curves in the middle and right panel show the histograms of granules and intergranular lanes, respectively, as detected by the algorithm described in Sect.~\ref{sec:met:gran}.}\label{fig:int_hist}
\end{figure*}
All six simulations considered here show spatial patterns in the vertically emerging (bolometric) intensity reminiscent of solar granulation (see Figure~2 of Paper~I). The visual appearance indicates a qualitative change of the convection within the model sequence around spectral type K, which will be analysed further in this section.\par
The left panel of Figure~\ref{fig:int_hist} shows histograms of the bolometric intensity, illustrating the difference of about a factor of ten between the average intensity of the hottest and the coolest model, as well as the decreasing intensity contrast (width of the distribution). In what follows, we use the normalised intensity fluctuation $\ndI$ as defined in Eq.~(\ref{eqn:def:I}). For this quantity, the differences in mean intensity and contrast are removed by the normalisation.\par
The middle and right panel of Fig.~\ref{fig:int_hist} show histograms of the normalised intensity fluctuations, $\ndI$, (solid curves) calculated from a large number of statistically independent snapshots. In all cases, the distributions show signs of bimodality, i.\,e. there is a bright and a dark component. Without taking into account possible asymmetries of the two individual components, the balance between them varies along the model sequence: the brighter component gets relatively weaker from F3V to K0V and then stronger again from K0V to M2V.\par
The dashed and dotted curves in Fig.~\ref{fig:int_hist} show histograms of separate distributions of $\ndI$ for intergranular lanes and granules, respectively, as defined by our segmentation algorithm. The fractional area covered by granules is almost the same for the
different simulations, implying an inverse correlation between the
amplitudes and widths of the distriutions shown in Fig.~\ref{fig:int_hist}.
The intensity distribution of the granular component is broader and has a lower amplitude in the three hotter models than in the cooler ones. It becomes more asymmetric from F3V to K0V with its peak shifting towards lower intensities. From K0V to M2V, the granular component becomes more symmetrical again and somewhat narrower. In contrast, the intergranular component is relatively narrow and has a high amplitude in the hottest models and becomes broader in the K- and M-star simulations. Consequently, along the model sequence, the granular component becomes weaker in the combined distributions from F3V to K0V and than stronger from K0V to M2V. In the M-star simulations, the intergranular component merges with the flank of bright component, which results in an apparently unimodal distribution with a peak at the typical granule intensity.
\begin{figure*}
\begin{tabular}{cc}
{\bf F3V} & {\bf K5V}\\
\includegraphics[width=5.5cm]{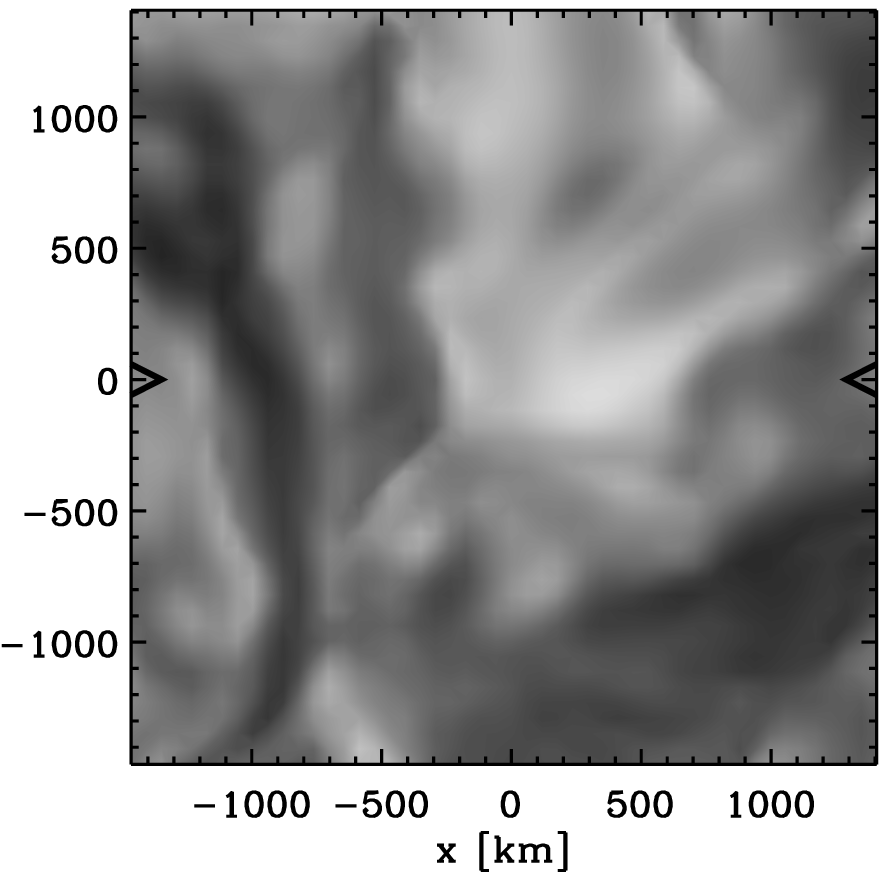}&%
\includegraphics[width=5.5cm]{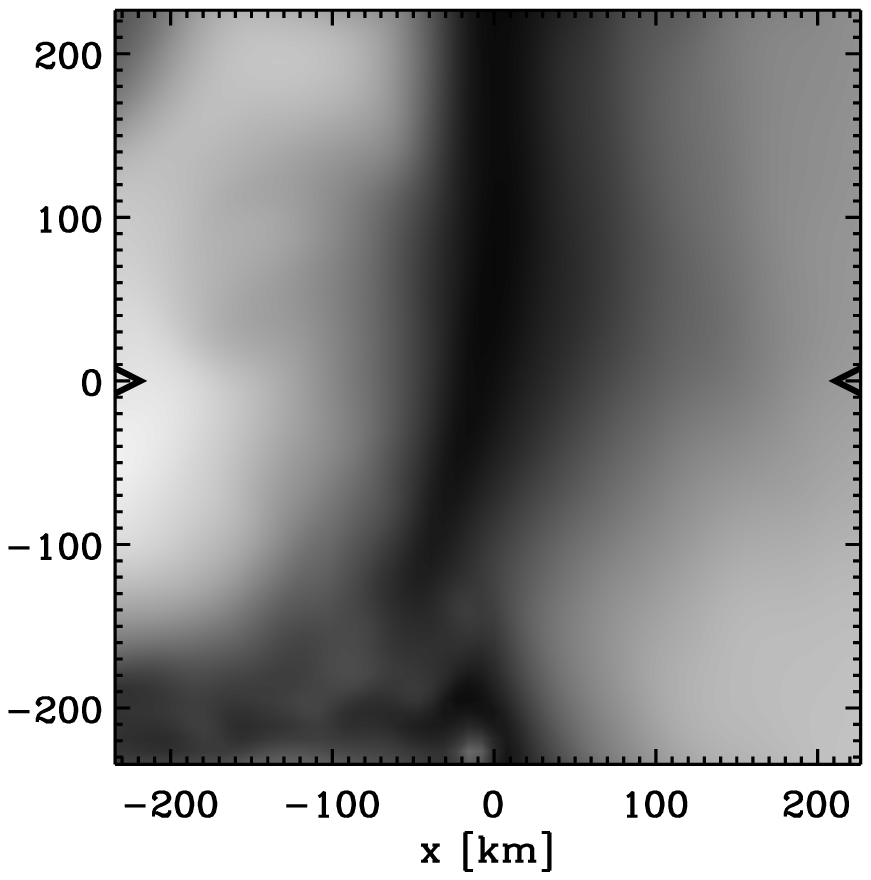}\\
\includegraphics[width=8.5cm]{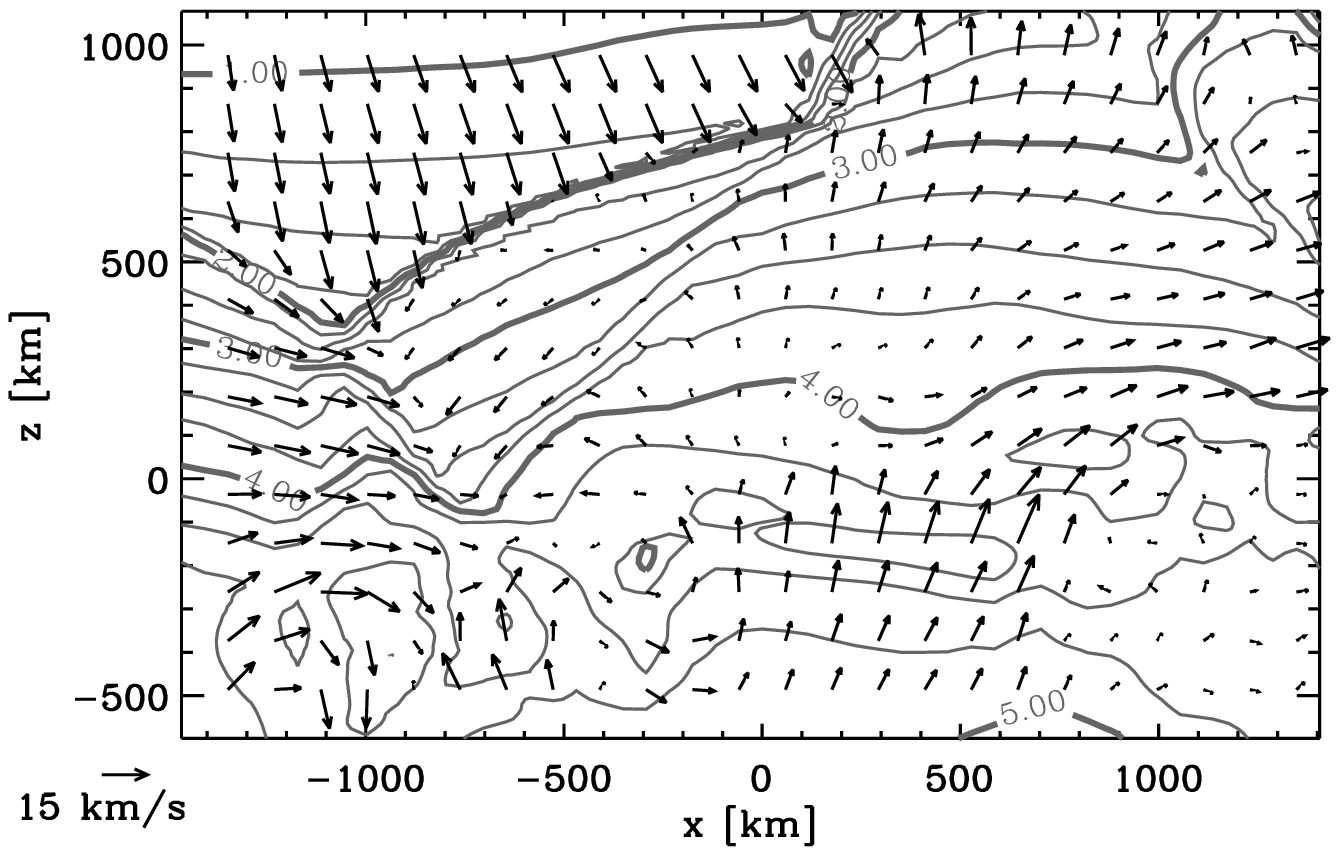}&%
\includegraphics[width=8.5cm]{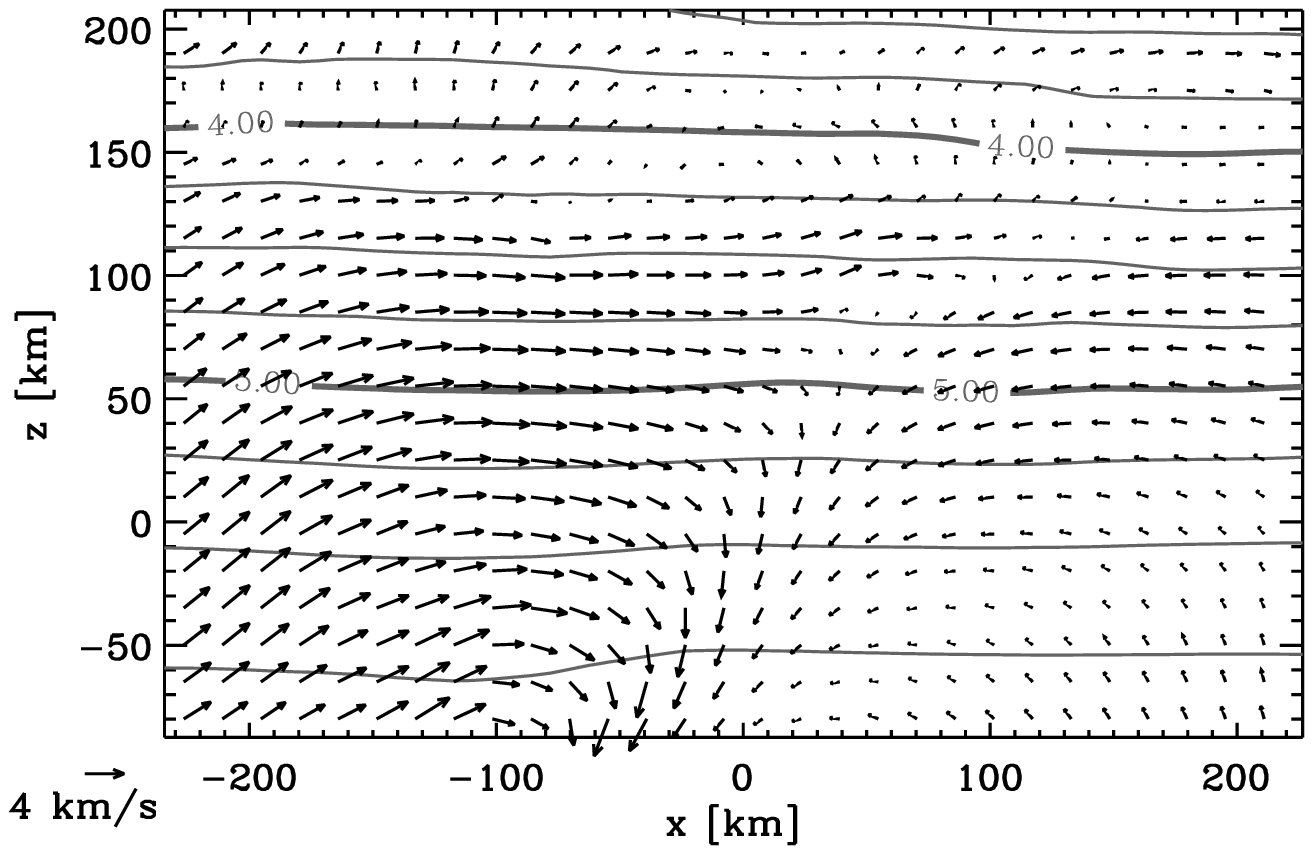}\\
\includegraphics[width=8.5cm]{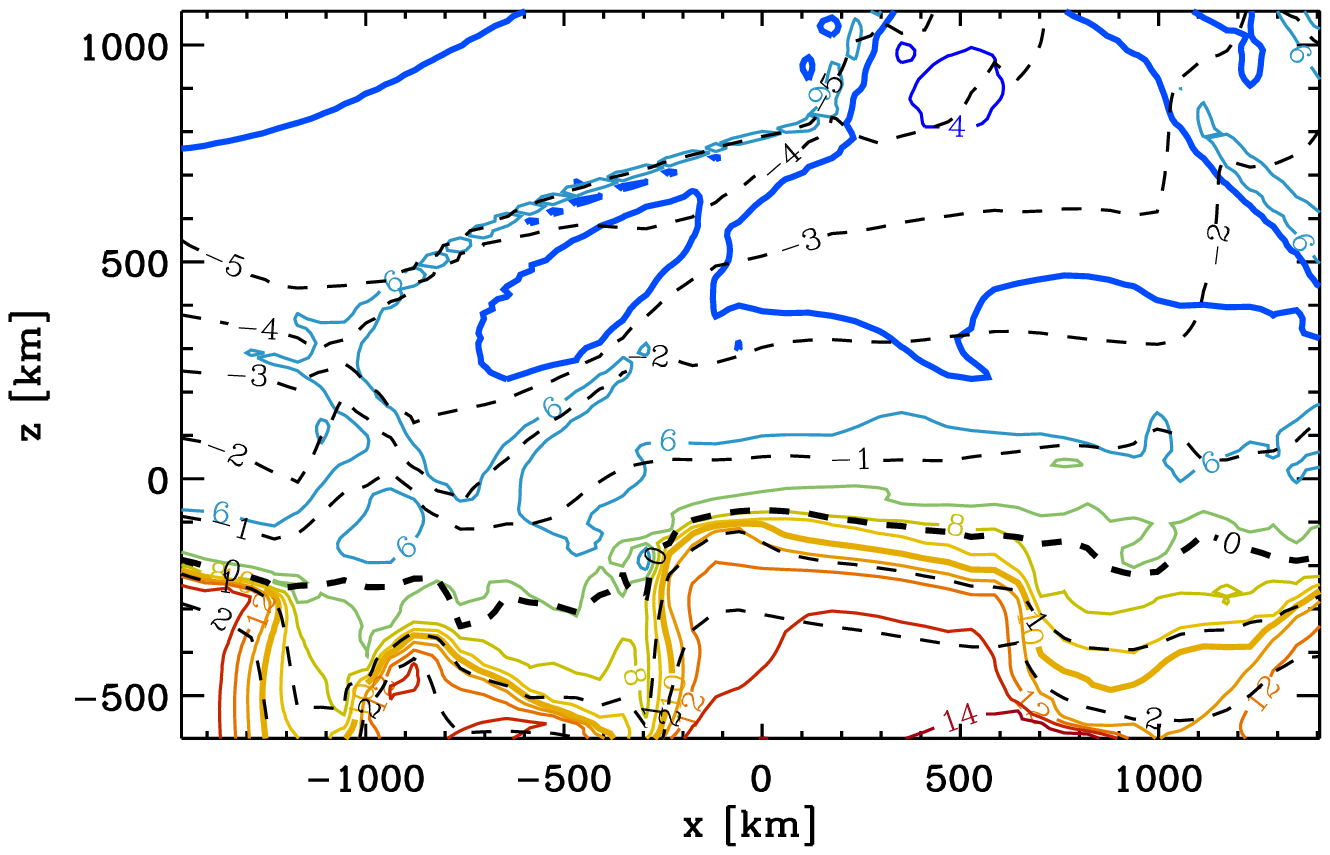}&%
\includegraphics[width=8.5cm]{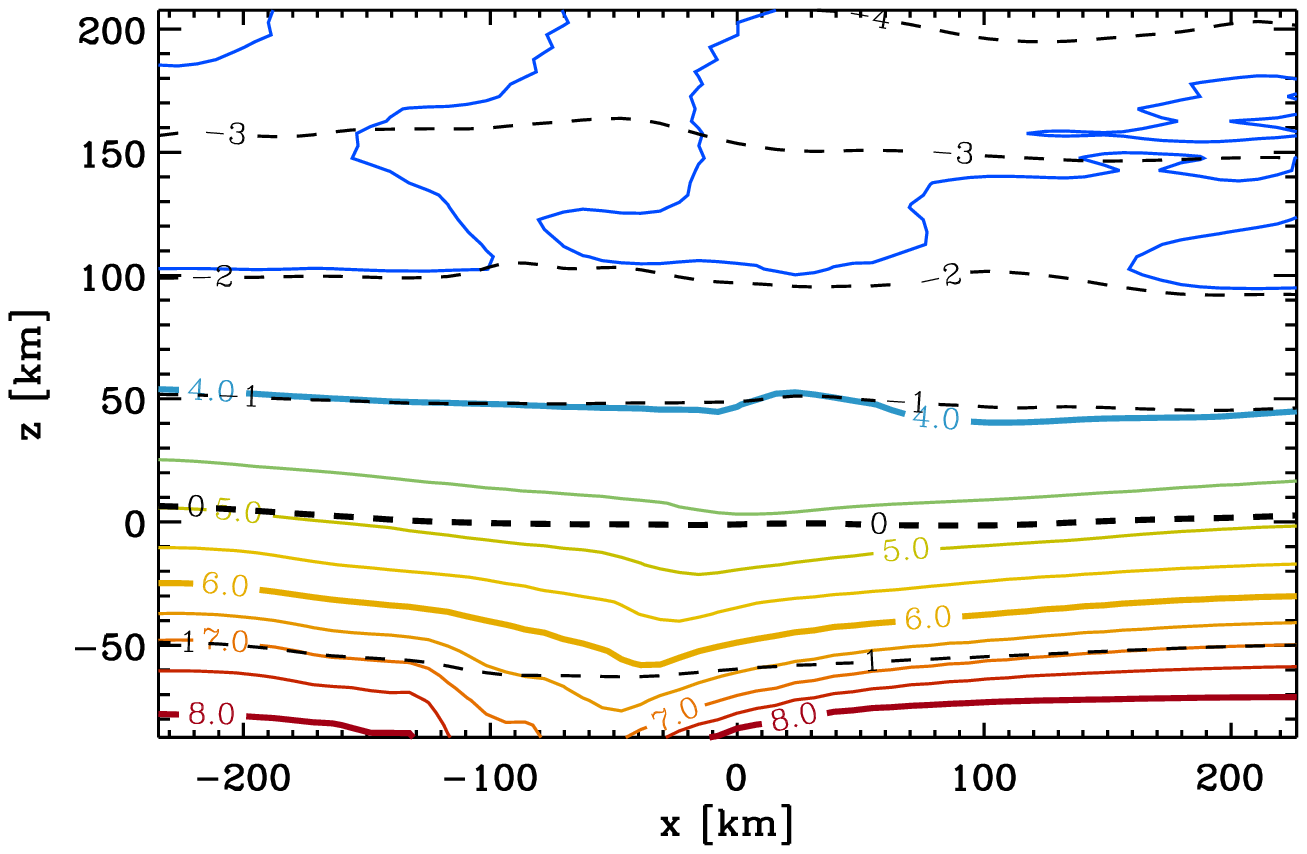}\\
\end{tabular}
\caption{Comparison between F3V and K5V granulation. {\it Top panels:} Images of the bolometric intensity in small areas of simulation snapshots corresponding to about the size of one granule in the respective simulations. The images are 50 and 58 grid points across, for the F3V and K5V simulations, respectively. The black triangles mark the position $y=0$ of the vertical cuts shown in the lower panels. {\it Middle panels:} Vertical cuts through the near-surface layers; arrows indicate the velocity field, grey contours show the isobars (labels give the logarithm of the pressure in $\mathrm{dyn}\,\mathrm{cm}^{-2}$). {\it Bottom panels:} Same cuts as middle panels showing the temperature field (coloured solid contours, labels in $1000\,\mathrm{K}$) and levels of constant Rosseland optical depth $\tau_{\mathrm{R}}$ (dashed contours, labels give logarithm of $\tau_{\mathrm{R}}$).}\label{fig:vgl_conv}
\end{figure*}
The decreasing intensity contrast $\sigma_I/\langle I \rangle$ with decreasing effective temperature is a consequence of the lower energy flux ($\propto T_{\mathrm{eff}}^4$) and higher density of the emitting layers. Higher density results in a higher heat capacity per volume which means that the enthalpy flux can be sustained by a smaller temperature contrast between up- and downflows (see Paper I). The smooth appearance and lack of bright substructure in K- and M-star granules can be explained by the veiling of the granulation by an optically thick layer \citep[cf.][]{ND90a}: for decreasing effective temperature, the transition from convective to radiative heat transport occurs over a greater range of pressure scale heights and sets in at greater optical depth. This leads to a thin subsurface layer in which diffusive radiative energy transport plays an important role. In most cases, the top part of this layer (the region around $\tau_{\mathrm{R}}\approx 1$) is stable against convection. The vertical motions which are present in this layer mainly represent overshoot from the unstable layers below so that the correlation between upflow velocity and temperature is less pronounced than in the convection zone. The granules of cooler stars are veiled by this optically thick convectively stable layer, which brakes convective flows and smears out the inhomogeneities in the temperature of the upflows by horizontal radiative diffusion.\par
Compared to these ``veiled'' granules, granules of G- and F-type stars are ``naked'' in the sense that the bulk of the energy is carried by convection up to the height where the atmosphere becomes optically thin. The transition to radiative energy transport occurs very rapidly owing to the strong temperature dependence of the opacity in the temperature range in which the cooling sets in. The corresponding layer is an order of magnitude smaller than the typical granule diameter. The cool downflows, however, remain optically thin into deeper layers and over a larger depth range. This leads to a strong corrugation of the optical surface. Together with the high (sometimes super-sonic) flow velocities, which entail strong deviations from hydrostatic equilibrium, this makes the 3D structure of F- and G-type stars more complex than that of cooler stars.\par
Figure~\ref{fig:vgl_conv} illustrates the differences between the ``naked'' granulation of the F3V star with much brightness substructure, strong deviations from hydrostatic equilibrium, a complex velocity field, and a corrugated optical surface, on the one hand, and the ``veiled'' granulation of the K5V star with a lack of brightness substructure, small deviations from hydrostatic equilibrium, a less complex velocity field, and a rather flat optical surface, on the other hand (see also Fig.~A.1 of Paper I).
\subsection{Vortex motions}\label{sec:vort}
\begin{figure*}
\centering
\includegraphics[width=5.8cm]{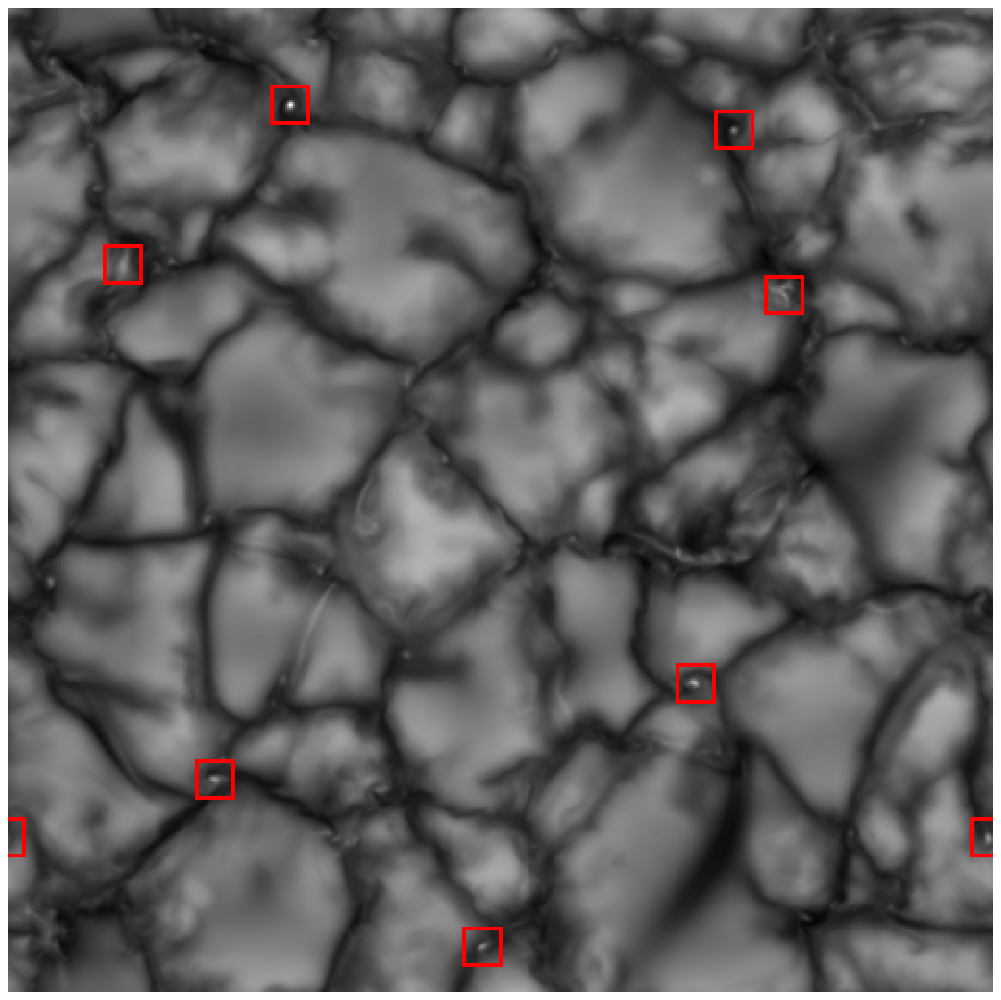}~%
\includegraphics[width=5.8cm]{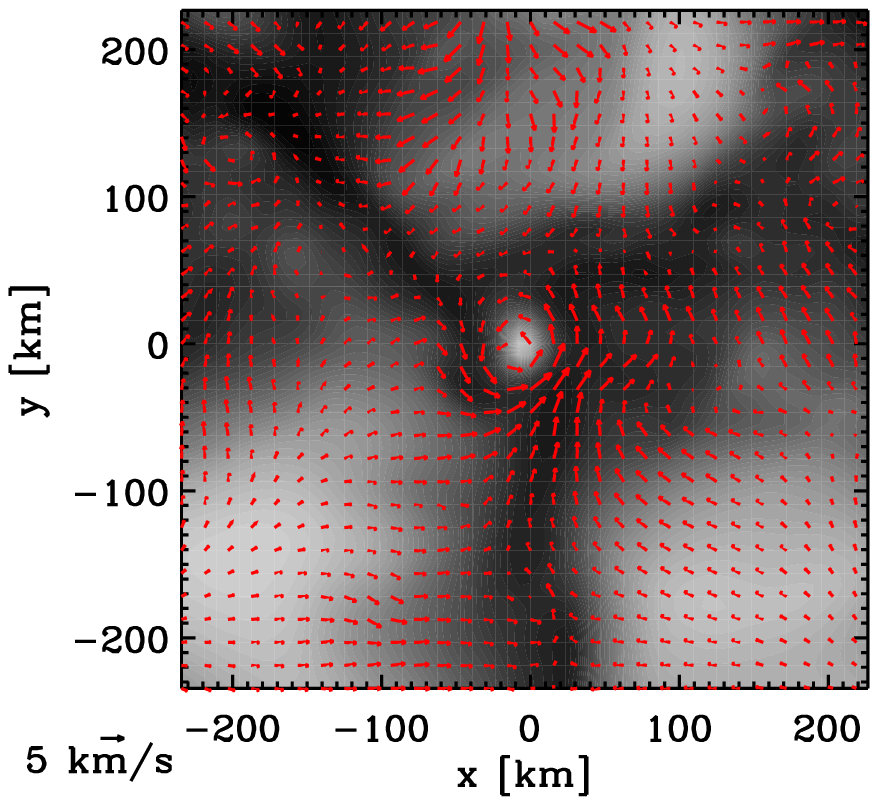}~%
\includegraphics[width=5.8cm]{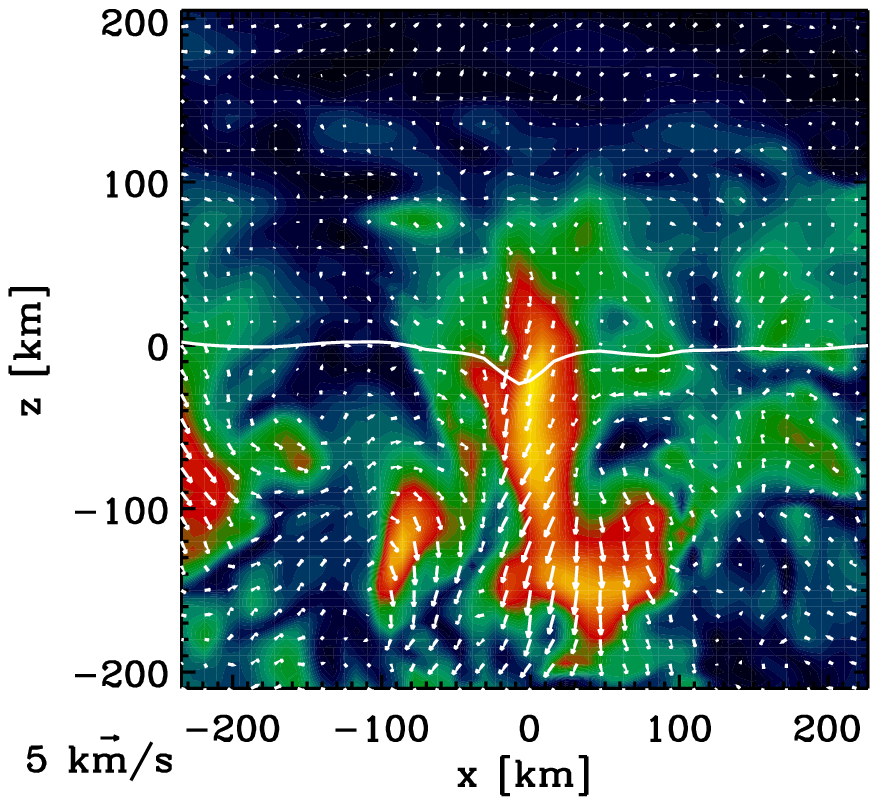}~%
\caption{Vertical vortices. {\it Left panel:} intensity map with bright points ($\ndI > 1.0$) in a snapshot of the K5V simulation, all of which are associated with vertical vortices. {\it Middle panel:} zoom into the intensity map of a vertical vortex; the horizontal flow at $z=0$ is shown as red arrows. {\it Right panel:} vertical cut through the vortex shown in the middle panel; the colour map shows the modulus of the vorticity $|\vec\omega|=|\nabla \times \vec \vel|$ (red/yellow indicating regions of high vorticity), the white arrows show the projection of the velocity onto the plane of the cut, the solid white line indicates the optical surface ($\tau_{\mathrm{R}}=1$).}\label{fig:vortex}
\end{figure*}
Horizontal and vertical vortex flows of (sub-)granular scale have been detected in high-resolution solar observations \citep[e.\,g.][]{Vort1, Vort2, Vort4, Vort3, Vort5} and also in solar simulations \citep[e.\,g.][]{Vort6, Moll11}. \citet{Lu06} reported swirling downflows also in simulations of cool main-sequence stars of spectral type M.\par
We find vertical and horizontal vortex motions in all our simulations. Horizontal vortices are usually located along the edges of granules, where they are driven by the shear between the horizontal outflow from the granule centres and the more randomly oriented flows in the optically thin layers above. Vertical vortices occur in strong downdrafts as consequence of angular momentum conservation in the highly stratified medium. Strong vertical vortices are partially evacuated owing to the centrifugal force, which leads to a local depression of the optical surface. Analogous to small-scale bright magnetic features (magnetic bright points), for which the depressions of the optical surface are caused by the magnetic pressure, the vortex flows sometimes can appear as features of enhanced intensity due to side-wall heating of the interior of the depression. The vortices in solar \texttt{MURaM} simulations are described in detail by \citet{Moll11, Moll12}.\par
The formation of bright vortex features requires a sufficiently high spatial resolution of the simulation. In our models, the horizontal resolution scales with the granule size and the vertical resolution with the local pressure scale height (see Paper~I) in order to be dynamically comparable to each other. We observe bright vortices to be most common in the K-star simulations (especially K0V). This is probably an opacity effect. Generally, the opacity, $\kappa(p,T)$, depends on pressure and temperature. However, at pressures around $10^{-4}$ to $10^{-5}\,\mathrm{dyn\,cm^{-2}}$ there is a rather temperature-insensitive regime of $\kappa$ for $4000\,\mathrm{K}\lesssim T \lesssim 5000\,\mathrm{K}$. The absorption coefficient, $\kappa\varrho$, thus mainly depends on density. Photospheric transitions in this temperature range are not accompanied by the usual strong gradient of $\kappa\varrho$. This is the case for the intergranular lanes of the K stars. Therefore, the location of the photosphere, in these cases, is more sensitive to density than to temperature. The partial evacuation caused by vertical vortices in K stars locally moves the photosphere inward to higher temperatures, entailing a brightening of the cores of these vortices. In the other stars, $\kappa\varrho$ is more sensitive to temperature and the depth of the local photosphere in most vortices only changes little and the brightening is much less pronounced. \par
Figure~\ref{fig:vortex} illustrates the vertical vortices in the K5V simulation. In the left panel, a snapshot with eight sharp intensity maxima above $\ndI >1.0$ is shown. All these features are associated with vortices, but there are many weaker intensity structures in this image which are also linked to vortex motion. The middle and right panels of Figure~\ref{fig:vortex} illustrate the velocity field and depression of the optical surface for a single vertical vortex. \par
\subsection{Granule properties}\label{sec:granprop}
\begin{table*}
\centering
\caption{Detected granules.}\label{tab:gran1}
\begin{tabular}{lrrrrrr}
\hline\hline
Simulation & F3V & G2V & K0V & K5V & M0V & M2V \\\hline
\# of snapshots & 33 & 33 & 33 & 33 & 33  & 33 \\
time span [min] & 28.2 & 24.8 & 34.9 & 28.4 & 28.4 & 24.8 \\ 
\# of granule snapshots\,$^{\mathrm{a}}$ & 1788 & 1075 & 1684 & 1251 & 1598 & 1328 \\
granule filling factor [\%] & $68.9 \pm 0.4$ & $68.1 \pm 0.8$ & $67.9 \pm 0.8$ & $69.0 \pm 0.5$ & $72.3 \pm 0.8$ &  $72.6 \pm 0.9$ \\\hline
\# of granules\,$^{\mathrm{b}}$ & 522 & 255 & 562 & 354 & 573 & 518\\
\# of granules ($t_{\mathrm{life}}>3\,\min$) & 69 & 48 & 101 & 66 & 66 & 34\\\hline
\end{tabular}
\begin{list}{}{}
\item[$^{\mathrm{a}}$] ``granule snapshot'' refers to a granule as it appears in a snapshot (no tracking/evolution)
\item[$^{\mathrm{b}}$] ``granule'' refers to granules tracked through a part of the considered image series 
\end{list}
\end{table*}
\begin{figure}
\centering
\includegraphics[width=8.5cm]{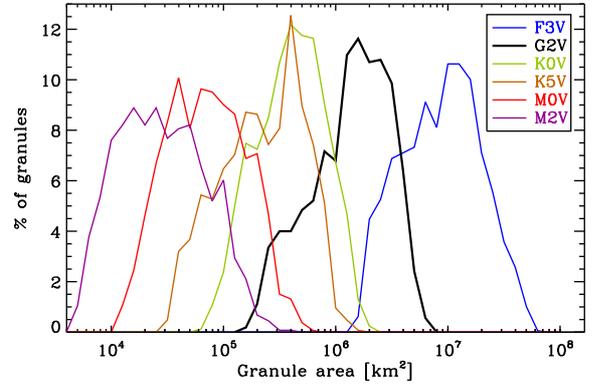}
\caption{Histograms of granule area. The bins are logarithmically equidistant with a bin size of 0.1 dex.}\label{fig:size_hist}
\end{figure}
\begin{figure}
\centering
\includegraphics[width=8.5cm]{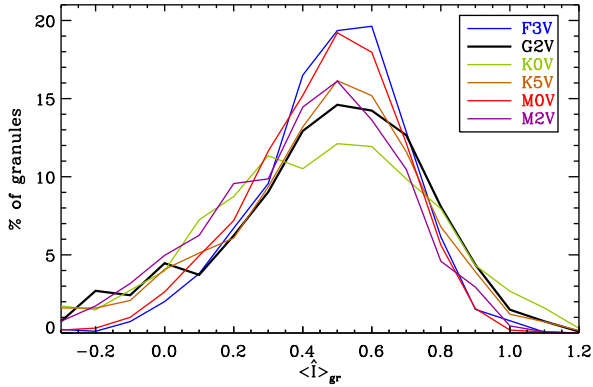}
\caption{Histograms of granule brightness $\langle \ndI \rangle_{\mathrm{gr}}$, i.\,e. the normalised intensity fluctuation $\ndI$ averaged over single (snapshots of) granules (definition of $\ndI$, see Eq.~(\ref{eqn:def:I})). The bins are linearly equidistant with a bin size of 0.1.}\label{fig:int_hist_gr}
\end{figure}
\begin{figure}
\centering
\includegraphics[width=8.5cm]{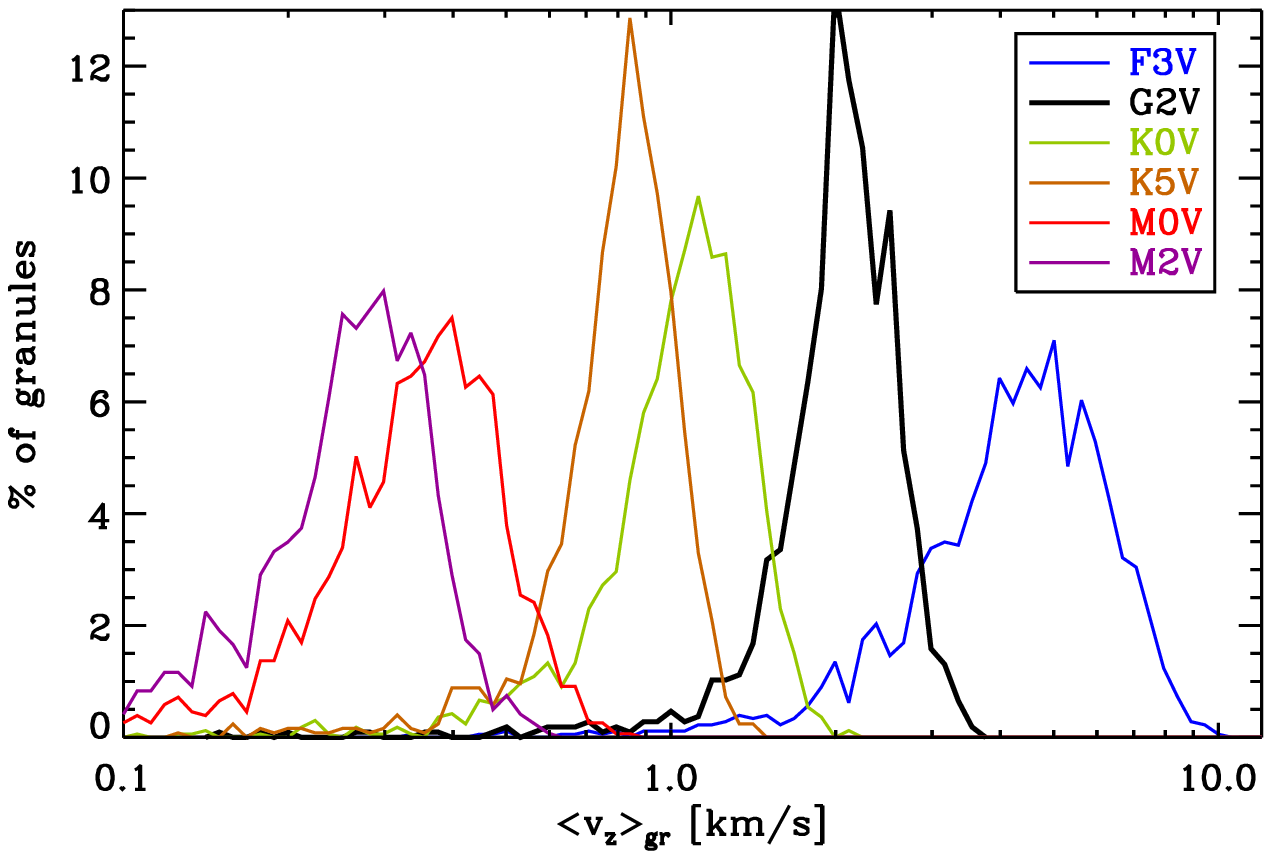}
\caption{Histograms of $\langle \vel_z\rangle_{\mathrm{gr}}$, the vertical velocity averaged over the area of a (snapshot of a) granule. The bins are logarithmically equidistant with a bin size of 0.025 dex.}\label{fig:mnvz_hist}
\end{figure}
In this section, we analyse the statistical properties of granules as they appear in individual snapshots of the simulation runs, i.\,e. we do not consider the evolution or time-averaged properties of individual granules but their properties at a given point in time. From 33 snapshots for each simulation about 1000 to 2000 granules were detected (values see table~\ref{tab:gran1}). The time interval between two subsequent images considered was $\Delta t= 300\,\delta t$ where $\delta t$ is the simulation time step ($\Delta t$ is thus on the order of 1 minute of stellar time). Since many granules live much longer than $\Delta t$, individual granules are likely to appear several times in this sample at different stages of their evolution.\par
\begin{figure*}
\centering
\begin{tabular}{cc}
\includegraphics[width=8.65cm]{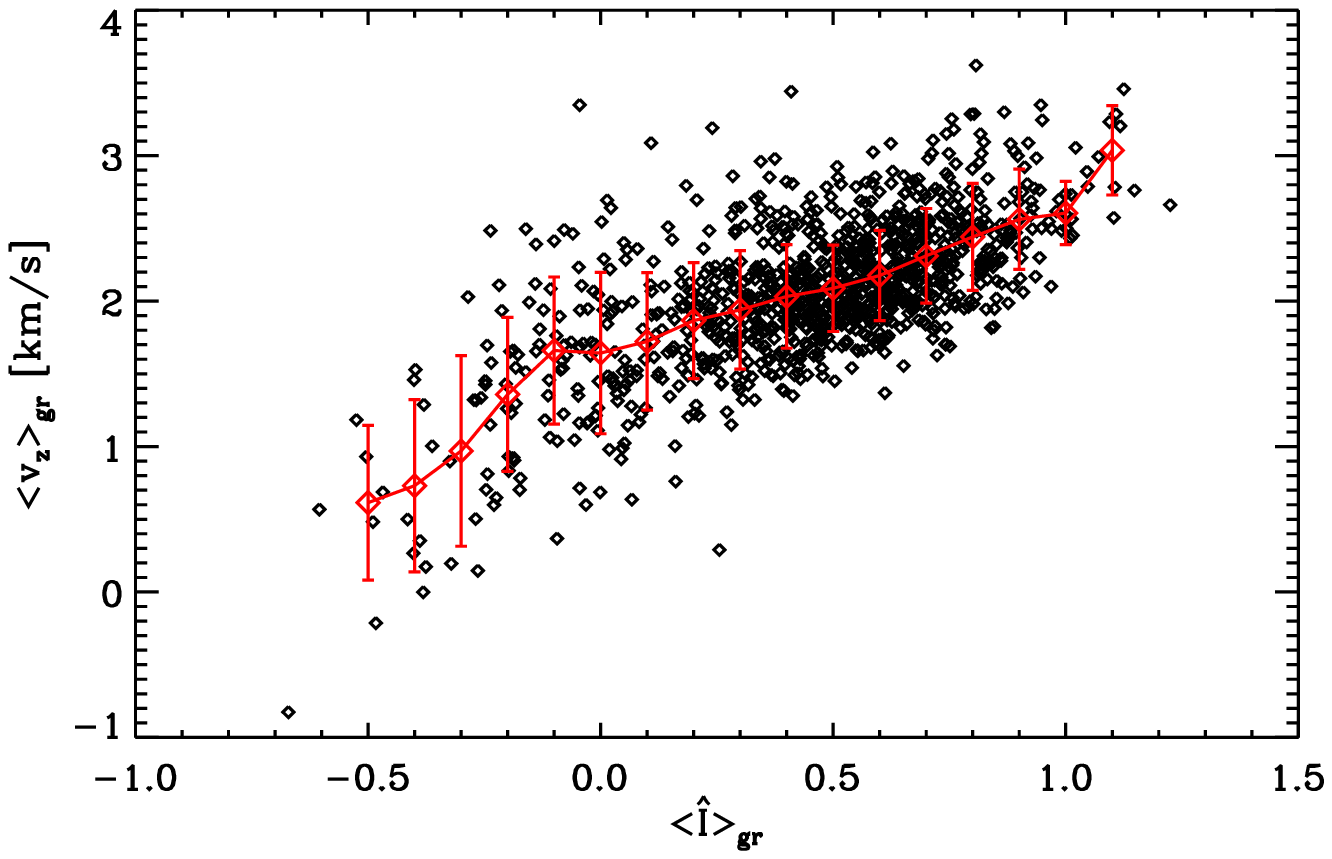} & \includegraphics[width=8.65cm]{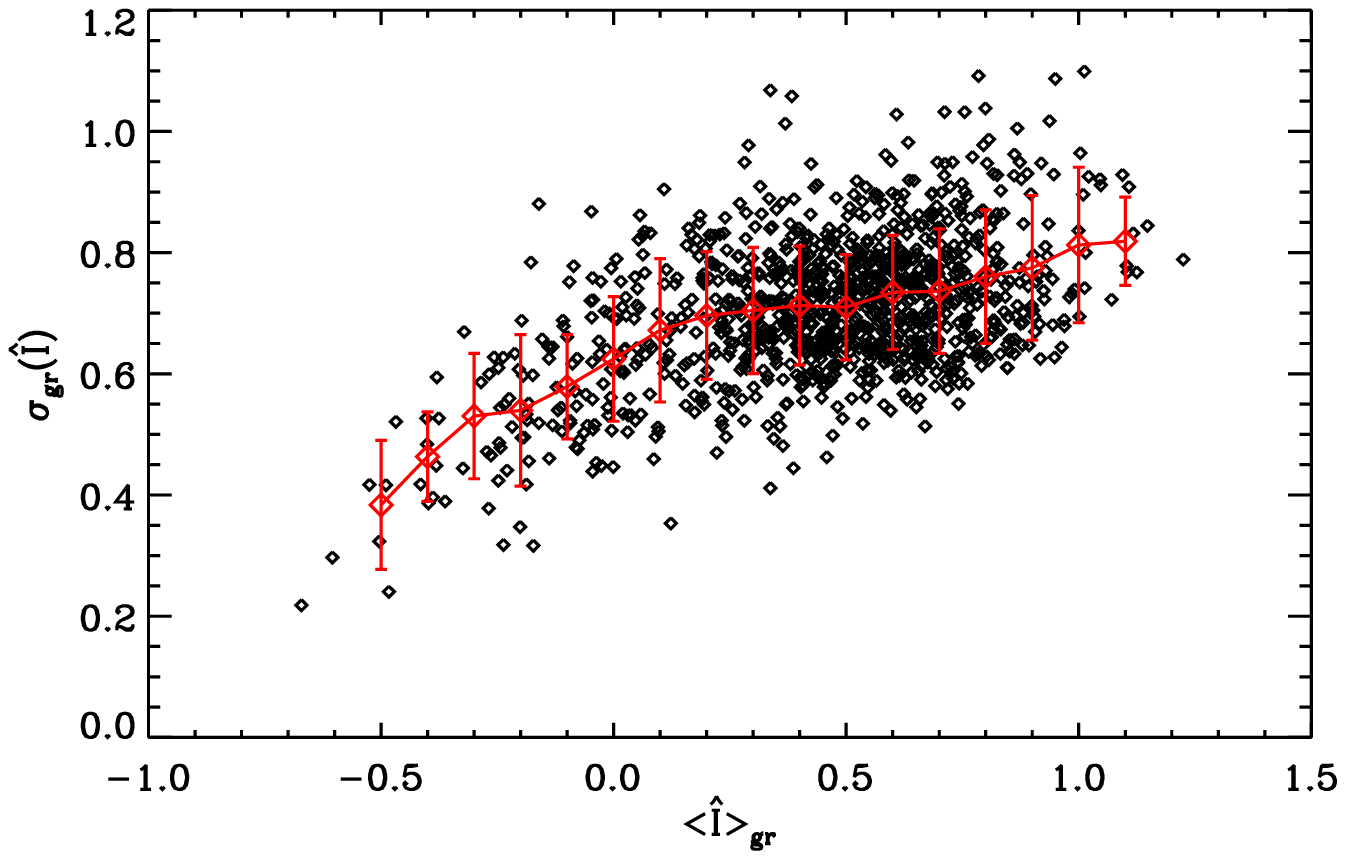} \\
\includegraphics[width=8.65cm]{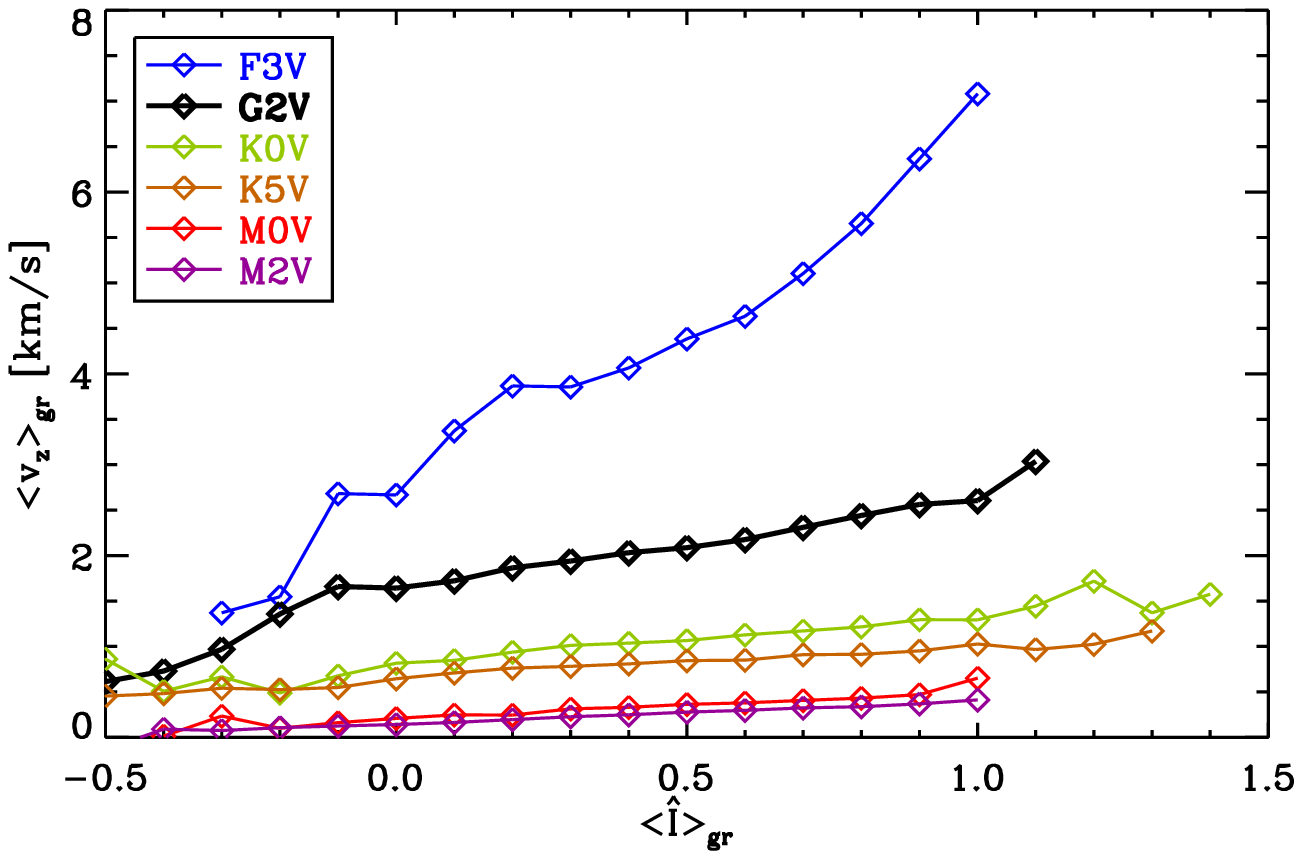} & \includegraphics[width=8.65cm]{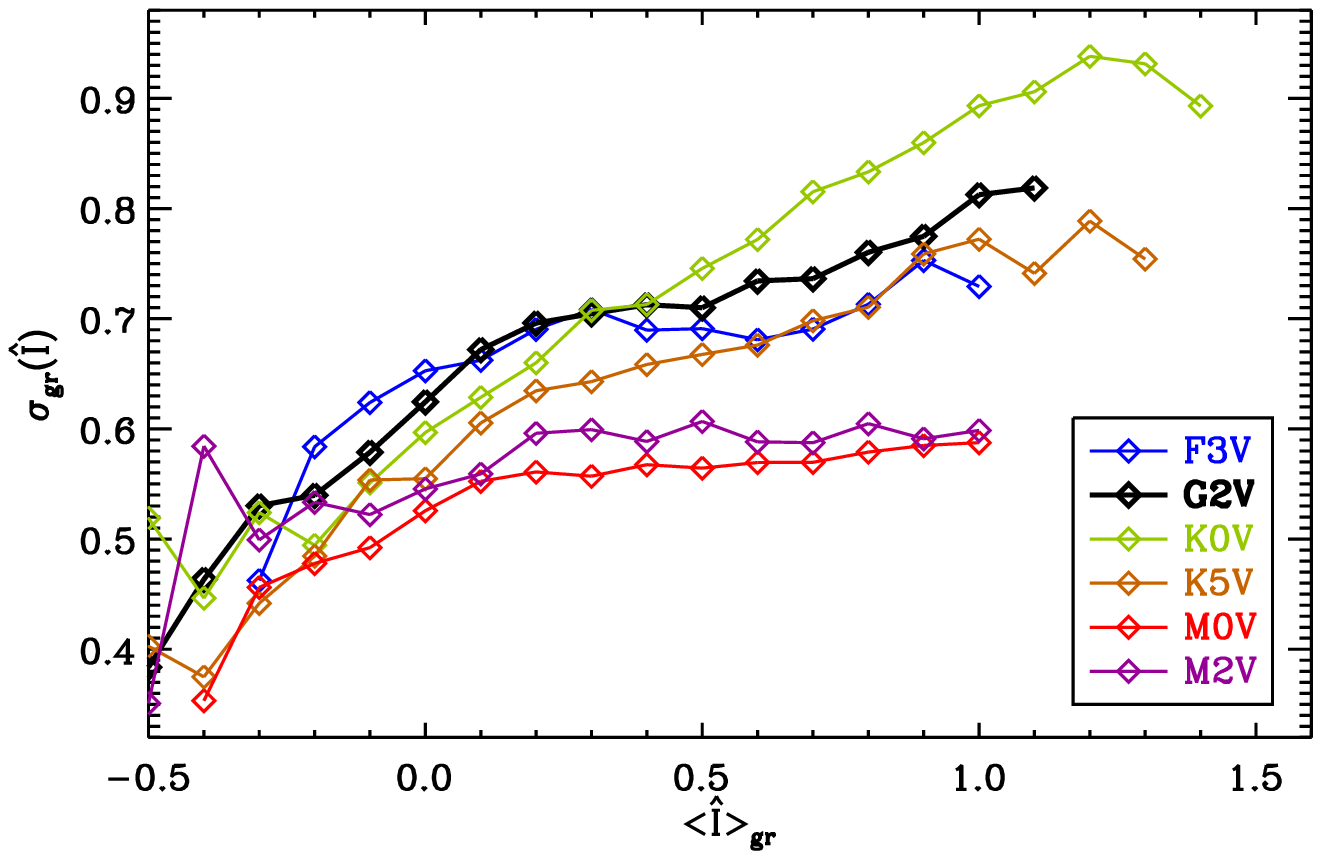} \\
\end{tabular}
\caption{Correlations between granule brightness and other granule properties. {\it Left panels}: Correlation between granule brightness, $\langle \ndI \rangle_{\mathrm{gr}}$, and average vertical velocity. {\it Right panels:} Correlation between $\langle \ndI \rangle_{\mathrm{gr}}$ and standard deviation of normalised intensity $\sigma_{\mathrm{gr}}(\ndI)$ as measure for the substructure within the granules ({\it right panels}). The {\it upper panels} show scatter plots for the solar run (G2V) where each black diamond represents a single granule. The red diamonds are binned averages (bin size 0.1 in $\ndI$) the error bars show the $1$-$\sigma$ scatter. In the {\it lower panels} the binned data are displayed for each of the six simulations.}\label{fig:brightness_cor}
\end{figure*}
For the F-, G-, and K-star simulations, the filling factor of granules as detected by our segmentation algorithm is between 67.5 and 69\% (including non-trackable granules). In the M stars, the granules have a slightly higher filling factor of about 72\%, which confirms the impression that the intergranular lanes of the M stars are narrower than the ones of hotter stars. These values are very close to the filling factor of 64 and 69\% for the upflow area about one pressure scale height below the surface in all simulations (see Fig.~7 of Paper~I). The asymmetry between up- and downflow area is mirrored in the asymmetry between granular and intergranular area, which we find are fairly robust and independent of the parameters used in our segmentation method.\par
\begin{figure*}
\begin{tabular}{cc}
\includegraphics[width=8.65cm]{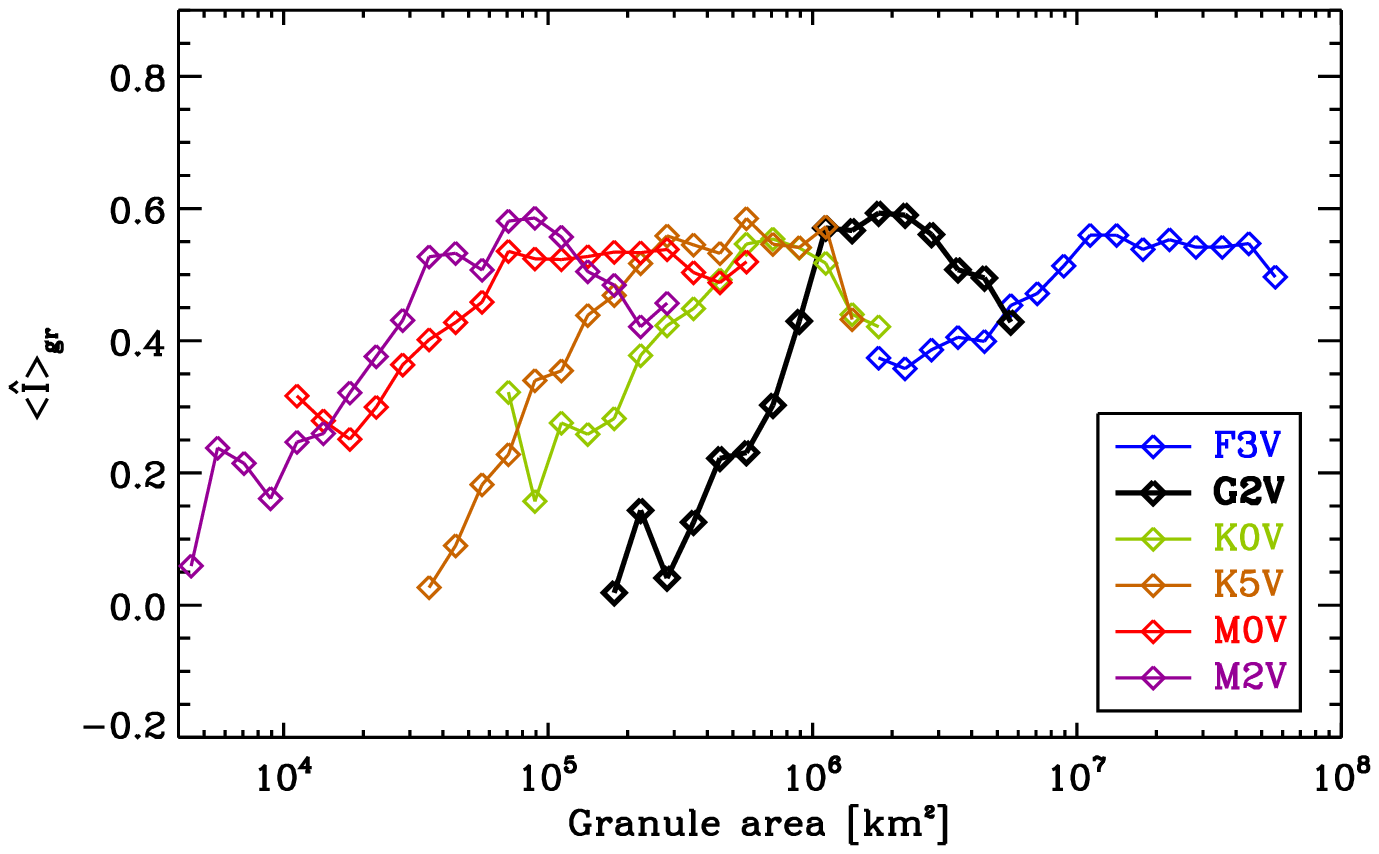} & \includegraphics[width=8.65cm]{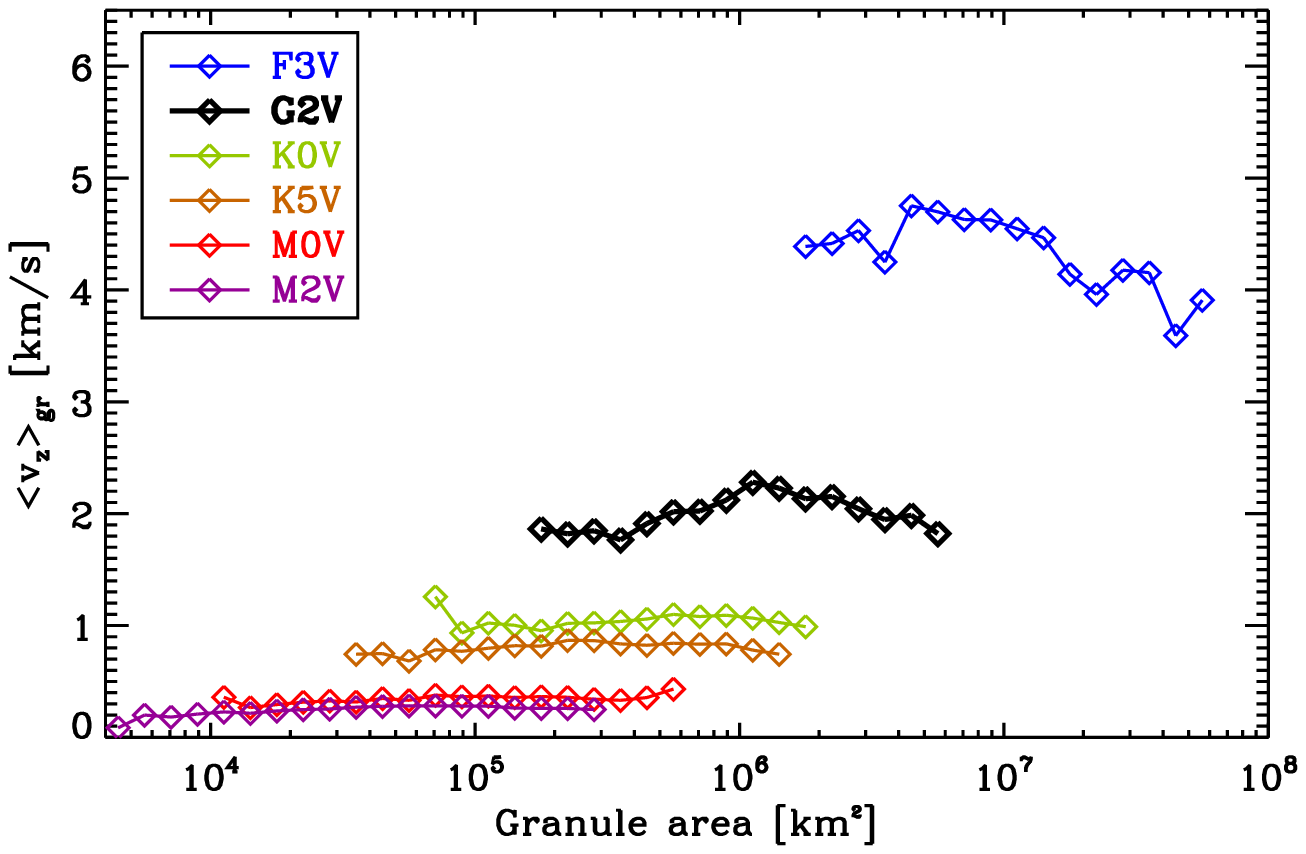} \\
\end{tabular}
\caption{Correlations of granule area with granule brightness ({\it left}) and with vertical velocity ({\it right}) in binned scatter plots (cf. Fig.\,\ref{fig:brightness_cor}).}\label{fig:area_cor}
\end{figure*}
Figure~\ref{fig:size_hist} shows histograms of the granule area. The size of the granules decreases for cooler, more compact stars (see discussion in Paper I). The spread of the size distributions of almost two orders of magnitude in area (one order of magnitude in granule diameter) is similar for all stars. For the Sun (G2V simulation), we find granule areas between 0.2 and 6 $\mathrm{Mm}^2$, corresponding to diameters (assuming circular granule shape) of 0.5 to 2.5 Mm (0.6 to 3.5''). This is roughly consistent with the upper part of the size distributions in observations \citep[e.\,g.][]{Hirz99a}. However, observationally obtained size distributions of solar granules show an increasing number of granules at decreasing size down to diameters of less than 0.5'', while our numerically obtained distribution peaks at roughly 2''. At least part of this discrepancy can be due to a different definition of a granule or to artefacts of the image reconstruction for the observations. A detailed comparison would require a degrading of our synthetic intensity maps and the segmentation of these ``synthetic observations'' and real observations of the same quality with the same algorithm. This is beyond the scope of this paper.\par

Figure~\ref{fig:int_hist_gr} shows histograms of the granule brightness. The quantity $\langle \ndI \rangle_{\mathrm{gr}}$ is the normalised intensity fluctuation $\ndI$ as defined in Eq.~(\ref{eqn:def:I}), averaged over the area of a granule. The histograms are similar for all spectral types, in spite of the different overall intensity distribution (see Fig.~\ref{fig:int_hist}). This can partly be attributed to the fact that the intensity fluctuation $\ndI$ is normalised by its standard deviation but also indicates the physical similarity of the convective upflow regions.\par

Figure~\ref{fig:mnvz_hist} shows histograms of $\langle \vel_z \rangle_{\mathrm{gr}}$, which is the upflow velocity $\vel_z(\tau_{\mathrm{R}}=1)$ at the corrugated optical surface spatially averaged over the area of a granule. A very small fraction ($\lesssim 1\%$) of the detected granules have a negative mean vertical velocity; these can either be wrong detections or granules in a very late evolutionary state. In our logarithmic representation of the histograms only granules with an average upflow speed of more than $0.1\,\mathrm{km\,s^{-1}}$ are shown (more than 95\% of the detected granules). The distributions show a marked peak at a velocity that can be regarded as the ``typical convective velocity'', $\vel_{\mathrm{conv}}$. This peak shifts from about $4\,\mathrm{km\,s^{-1}}$ for the F3V simulation to about $0.3\,\mathrm{km\,s^{-1}}$ for M2V, and is roughly proportional to the rms value of $\vel_z$ at the optical surface (see Paper I).\\
We find a correlation of the brightness of a granule with its upflow velocity as well as with the amount of substructure in the granule. As a quantitative measure for the latter, we take the standard deviation $\sigma_{\mathrm{gr}}(\ndI)$ of the normalised intensity fluctuation $\ndI$ within the granule area. Figure~\ref{fig:brightness_cor} shows the correlation between granule brightness, vertical velocity, and $\sigma_{\mathrm{gr}}(\ndI)$ as scatter plots for the solar run (G2V) and as binned scatter plots for all simulations. Brighter granules tend to have stronger mean upflow speeds and more substructure. The brightness of a granule is proportional to the convective energy flux directly below the optical surface, which in turn is roughly proportional to the vertical velocity. A correlation between mean velocity and brightness is therefore expected. The brighter granules also tend to have more brightness variation than the dimmer ones in all simulations. This is caused by a combination of many effects, such as inhomogeneities in the upflows, a stronger corrugation of the optical surface due to the higher average velocities in the brighter granules, more pronounced shock waves (in the F-star simulation), and a less effective horizontal radiative diffusion owing to the shorter time span in which the convective elements rise through the near-surface layers where radiation becomes important.\par

Figure~\ref{fig:area_cor} shows the correlations of brightness and granule-averaged upflow velocity with the area of the granules. We find a correlation between area and mean brightness for the smaller granules, whereas for the larger ones the brightness saturates and correlation is lost. This is consisitent with solar observations \citep[see, e.\,g,.,][]{Hirz97}. At least for large granules, there is negative correlation between size and mean vertical velocity (cf. Figure~3 in Paper~I). This might be due to buoyancy breaking: granules typically grow as more material wells up from below. Once their size exceeds a critical value of a few density scale heights the granule vanishes or splits \citep{Nordlund09}.
\begin{figure}
\centering
\includegraphics[width=8.5cm]{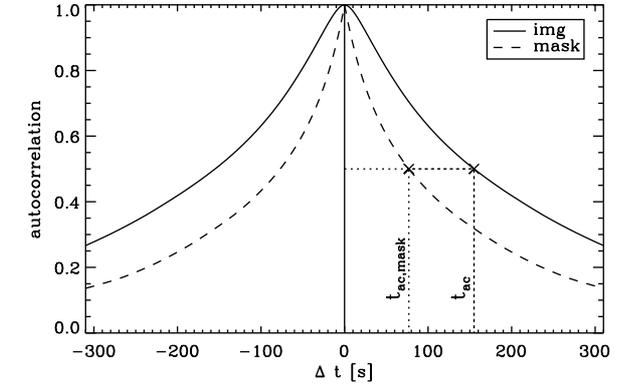} 
\caption{Autocorrelation function of the bolometric intensity images (img) and binary segmented masks (mask) as function of the time lag $\Delta t$ for the solar simulation (G2V). The autocorrelation times $t_{\mathrm{ac}}$ and $t_{\mathrm{ac},\mathrm{mask}}$ (half width at half maximum) are marked.}\label{fig:corrtime}
\end{figure}
\subsection{Granule lifetime}\label{sec:timescale}
In this section, we aim to determine the time scale on which the granules evolve and the typical lifetime of granules. From solar observations, granule lifetimes of the order of several minutes have been reported \citep[see, e.\,g.,][]{Title89, Hirz99a}.\par 
\begin{figure*}
\centering
\begin{tabular}{ccc}
\includegraphics[width=5.8cm]{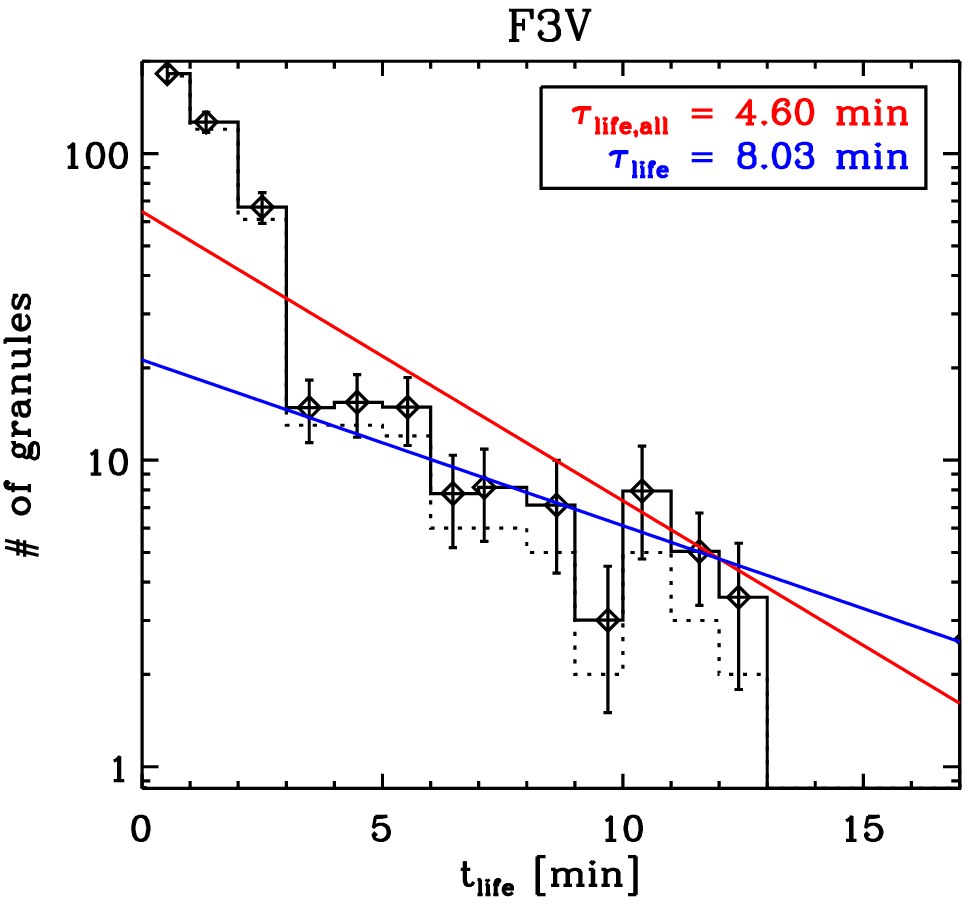} & %
\includegraphics[width=5.8cm]{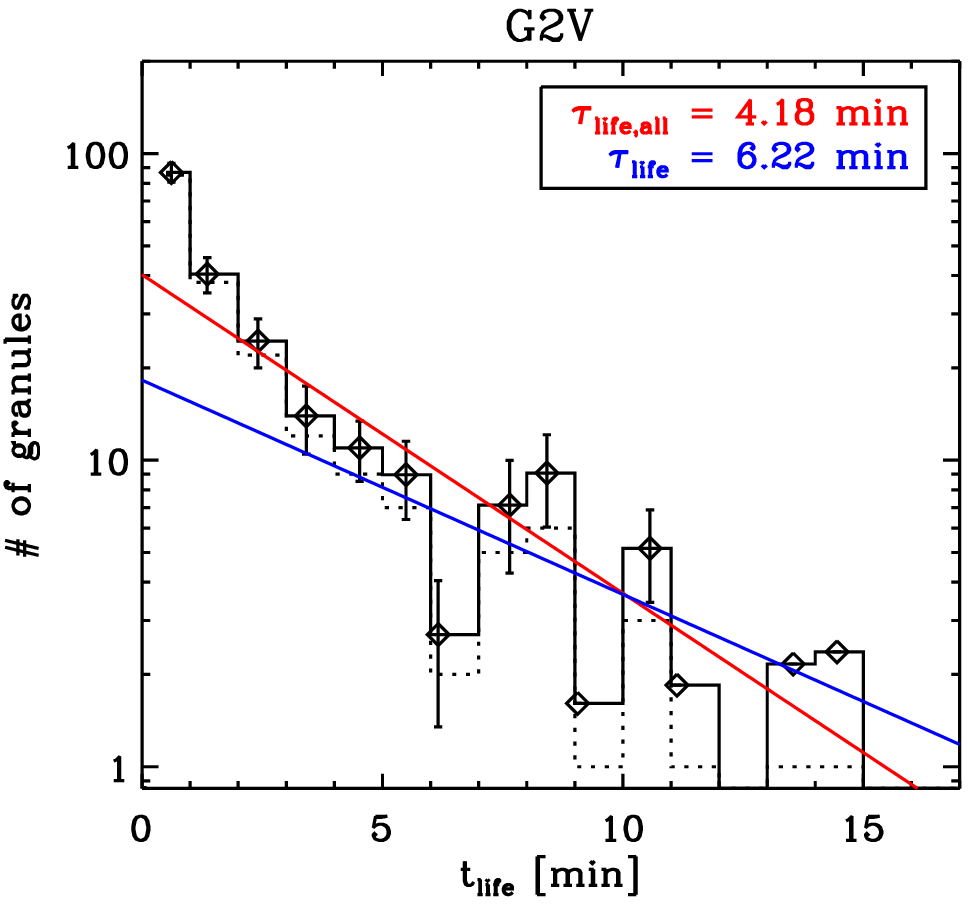} & %
\includegraphics[width=5.8cm]{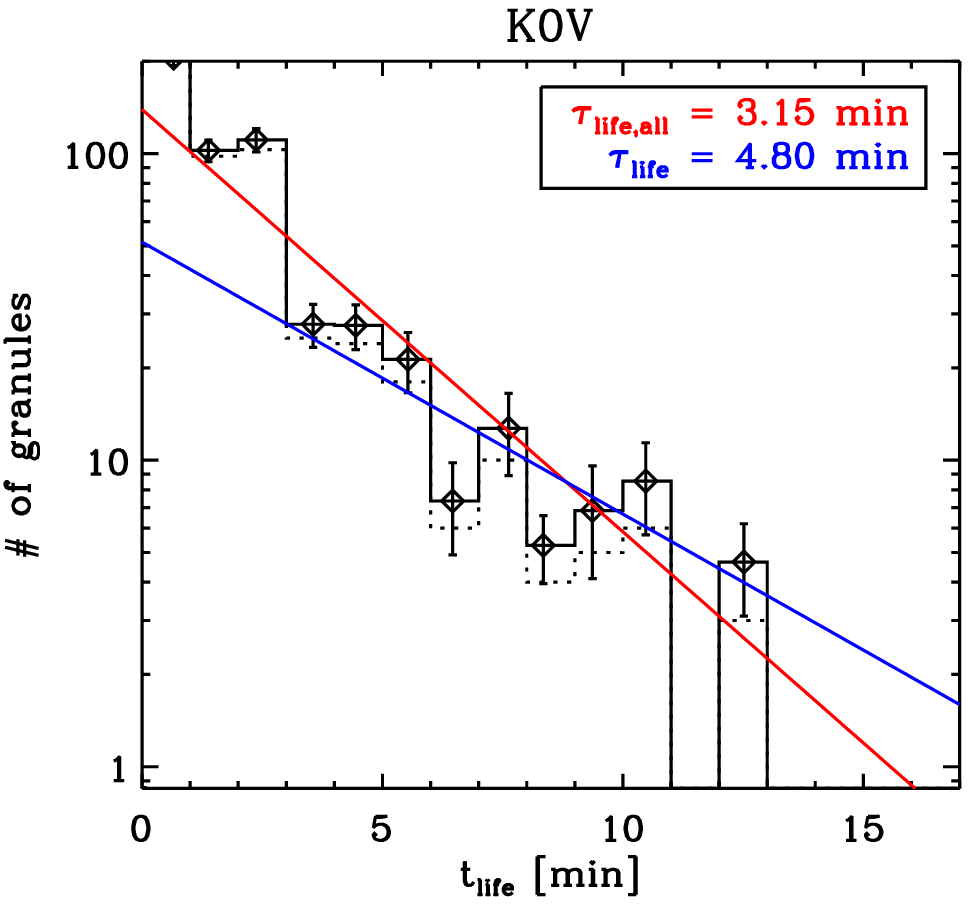}\\[2mm]
\includegraphics[width=5.8cm]{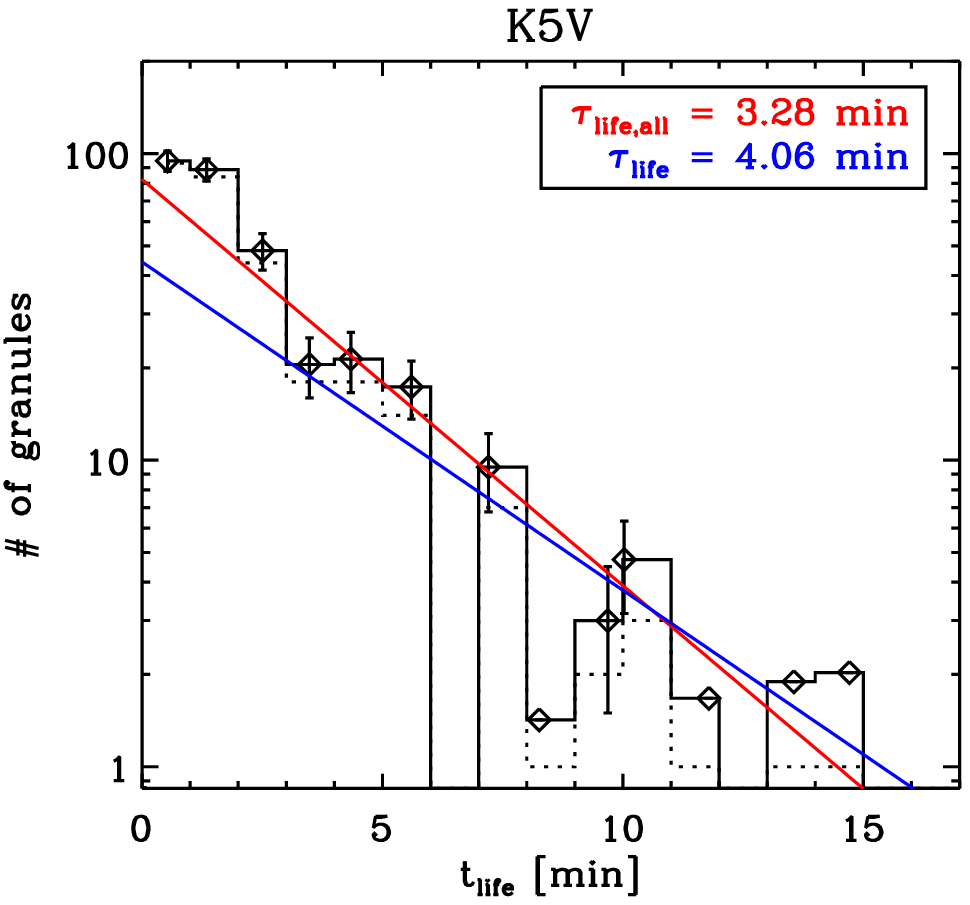} & %
\includegraphics[width=5.8cm]{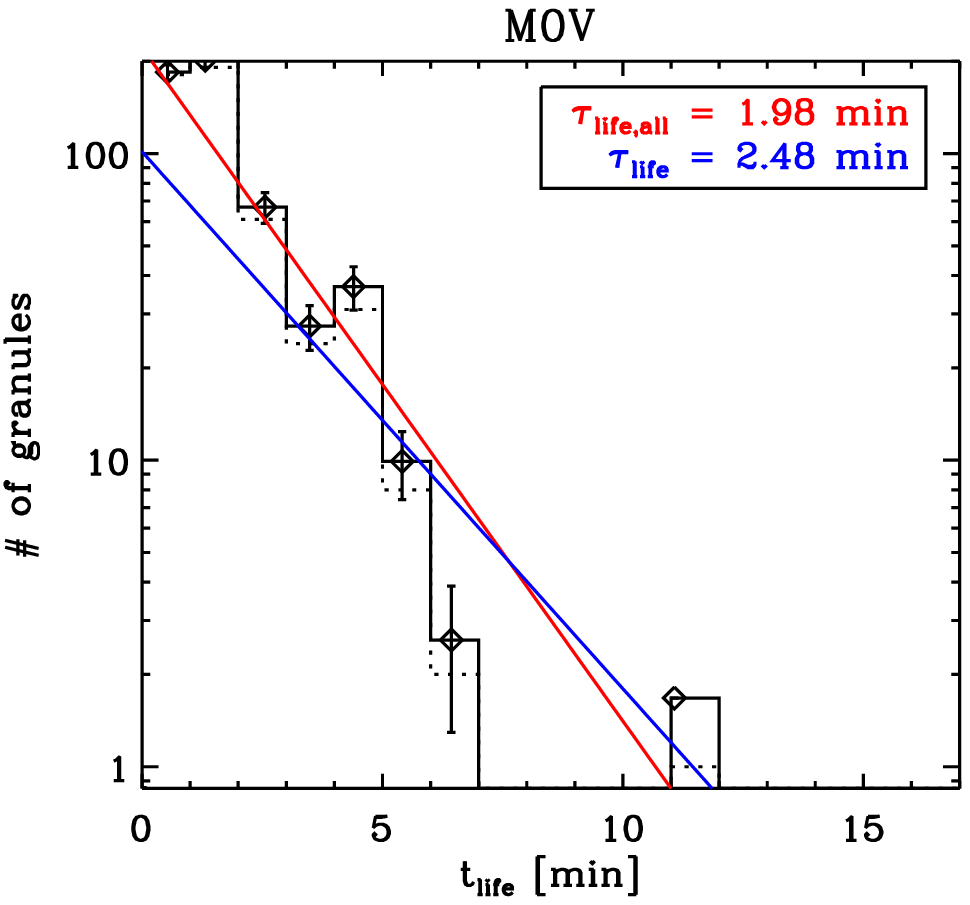} & %
\includegraphics[width=5.8cm]{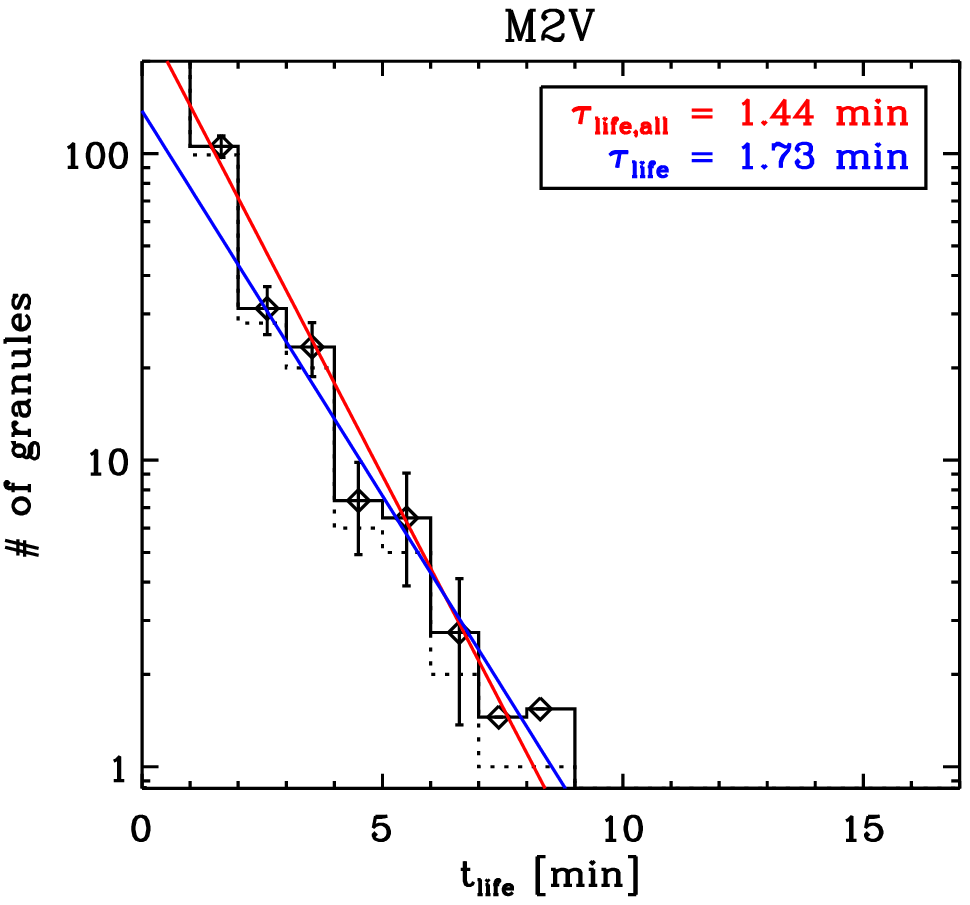}\\
\end{tabular}
\caption{Lifetime histograms. The dotted black lines show the actual number of detected granules, $N_{i}^{\ast}$, while the solid black lines show the corrected granule number, $N_i$, of Eq.~(\ref{eqn:ltcorrection}). The solid coloured lines are exponential least-square fits to all histogram bins (red), and to all but the first three bins (blue), respectively.}\label{fig:gr_lifetime}
\end{figure*}
One possible approach to define an evolution time scale of the granulation pattern is autocorrelation \citep{Title89}. We calculated the autocorrelation function for a time series of intensity images from our simulations as well as for a sequence of binary masks of the segmented images (granules\,=\,1, intergranular lanes\,=\,0). We define the autocorrelation times $t_{\mathrm{ac}}$ and $t_{\mathrm{ac},\mathrm{mask}}$, respectively, as the time lag for which the autocorrelation has dropped by half. Figure~\ref{fig:corrtime} shows the autocorrelation function of the solar simulation (G2V). The advantage of using the autocorrelation time is that it can be computed without previous image segmentation and does not depend on arbitrary definitions of granules. It does, however, not necessarily reflect the time scale on which {\it individual} granules evolve: a pattern of randomly drifting, but otherwise unchanging granules would show a finite autocorrelation time although the individual granules show no evolution at all. The autocorrelation time therefore gives a pattern evolution time, which, in most cases, will be much shorter than the mean evolution time or the lifetime of individual granules. Table~\ref{tab:gran2} lists the autocorrelation times $t_{\mathrm{ac}}$ and $t_{\mathrm{ac},\mathrm{mask}}$ for all simulations.\par
The lifetime of individual granules can be determined by tracking in time. For each simulation, about 250 to 600 granules were tracked with our algorithm (see Sect.~\ref{sec:met:gran} and Table~\ref{tab:gran1}). Our definition of the beginning and end of a ``life'' of a granule roughly follows \citet{Hirz99a}: a granule starts its life either by fragmentation, merger, or apperance and dies by fragmentation, merger, or fading away. If a merger (or splitting) occurs which involves two very unequally large granules (with a critical area ratio of 15) only the smaller one ends (begins) its life, while the larger one survives. If both granules involved in a merger (splitting) are similar in size, both granules ``die'' (are ``born''). Figure~\ref{fig:gr_lifetime} shows histograms of the lifetime $t_{\mathrm{life}}$ of the $\sim$ 250 to 600 granules tracked with these criteria in each of the six time series. We only consider granules, whose lives lie entirely within the series. This implies that, for a time series starting at $t=0$ and ending at $t=T$, a granule with a lifetime of $t_i$ has to be born between $t=0$ and $t=T-t_i$ in order to be considered. The resulting histogram (shown as dotted line) therefore depends on the length $T$ of the time series. In order to remove this dependence, we corrected the histogram bins centred arround $t_i$ according to
\begin{equation}\label{eqn:ltcorrection}
N_{i}= N_{i}^{\ast}\cdot\frac{T}{T-t_i}\,\,,
\end{equation}  
where $N_{i}^{\ast}$ is the number of detected granules within the lifetime bin $i$ (centred around $t_i$) and $N_{i}$ is the corrected number. These corrected histograms are shown as solid lines in Figure~\ref{fig:gr_lifetime}.\par
\begin{table}
\caption{Autocorrelation times and granule lifetimes in minutes}\label{tab:gran2}
\centering
\begin{tabular}{lrrrr}
\hline\hline
Simulation & $t_{\mathrm{ac}}$ & $t_{\mathrm{ac},\mathrm{mask}}$ & $\tau_{\mathrm{life}}$ & $\tau_{\mathrm{life},\mathrm{all}}$\\\hline
F3V & 1.55 & 0.87 & 8.03 & 4.60 \\
G2V & 2.58 & 1.28 & 6.22 & 4.18 \\
K0V & 1.90 & 0.80 & 4.80 & 3.15 \\
K5V & 1.84 & 0.96 & 4.06 & 3.28 \\
M0V & 1.32 & 0.64 & 2.48 & 1.98 \\
M2V & 1.13 & 0.54 & 1.73 & 1.44 \\\hline
\end{tabular}
\end{table}
We find many small granules with a very short lifetime of less than 3 minutes, some of which are probably granule substructures or wrong detections. If one excludes these granules, the resulting histograms can be fitted quite well with an exponential function:
\begin{equation}\label{eqn:explt}
N(t_{\mathrm{life}})=N_0\,\exp(-t_{\mathrm{life}}/\tau_{\mathrm{life}})\,\,.
\end{equation}
Note that the short-lived granules are much more numerous (see table~\ref{tab:gran1}). For the sample from the solar simulation, we obtain $\tau_{\mathrm{life}}=6.22\,\min$ considering the 48 granules with $t_{\mathrm{life}}\ge 3\,\min$ and $\tau_{\mathrm{life},\mathrm{all}}= 4.18\,\min$ considering all 255 granules. This is consistent with the results by \citet{Title89} and \citet{Hirz99a}, who found exponential laws for the lifetime of observed solar granules with a time constant $\tau_{\mathrm{life}}$ between 2.53 and 6.25\,min (depending on the method of granule detection). \citet{DM04} analysed three data sets of solar granulation and obtained values between 2.6 and $4.1\,\min$ for the long-lived granules. However, he argues that the distributions are better described by a stretched exponential or that the granule population should be divided into two parts with lifetimes of $t_{\mathrm{life}}< 2.5\,\min$ and $t_{\mathrm{life}}>2.5\,\min$, respectively. This is also consistent with our histograms.
\begin{figure*}
\begin{tabular}{cc}
\includegraphics[width=8.5cm]{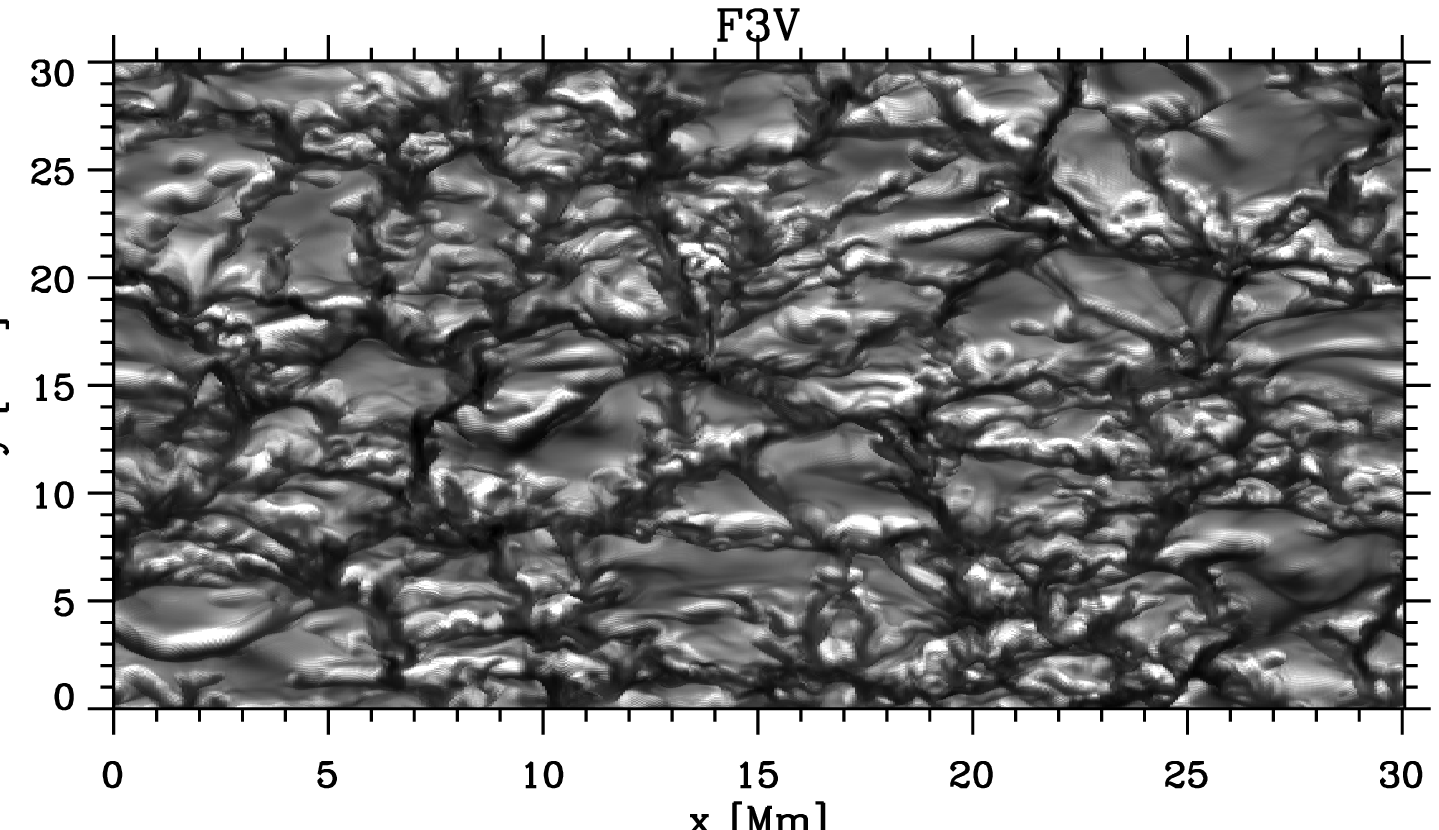} & \includegraphics[width=8.5cm]{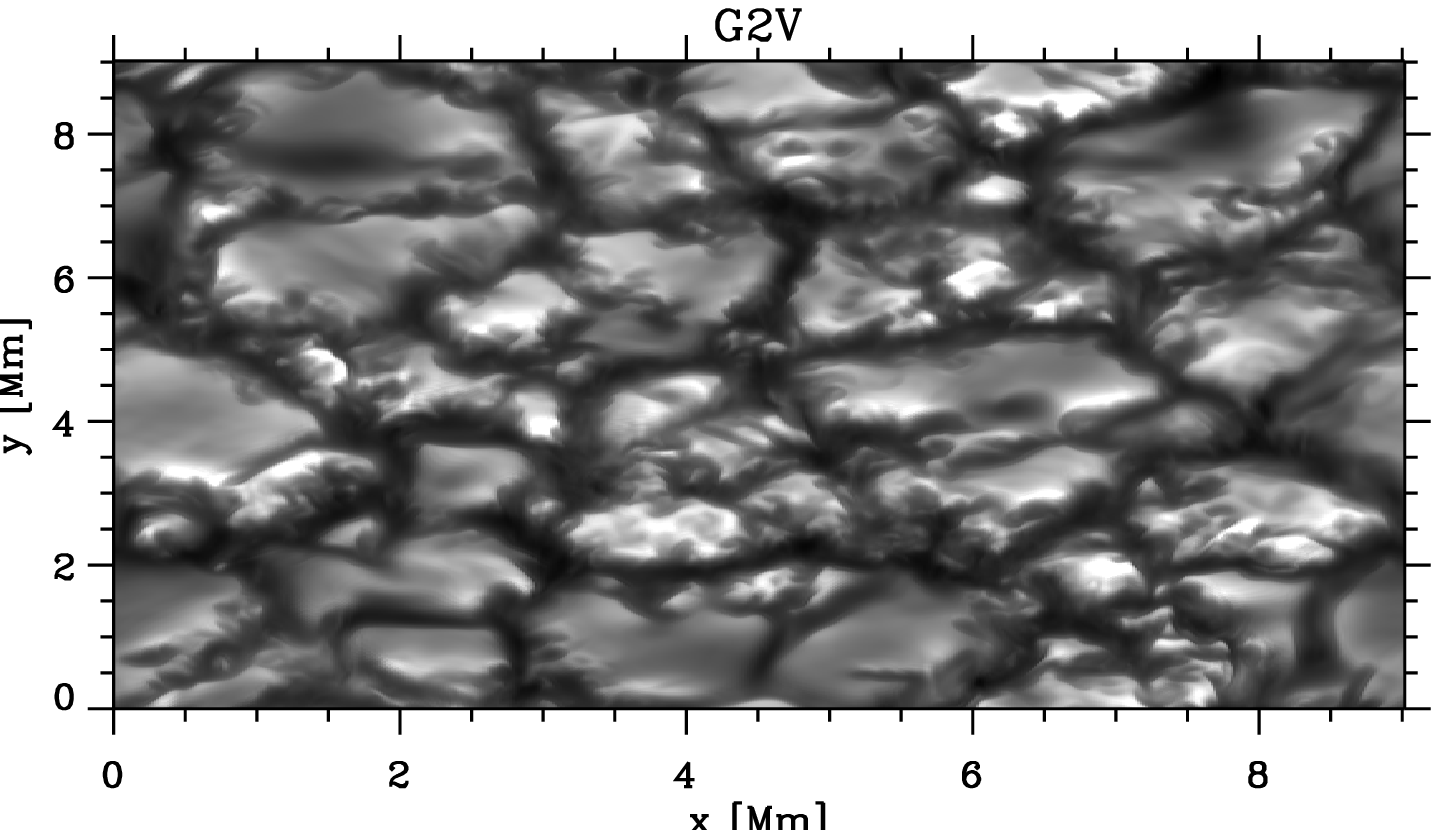}\\[3mm]
\includegraphics[width=8.5cm]{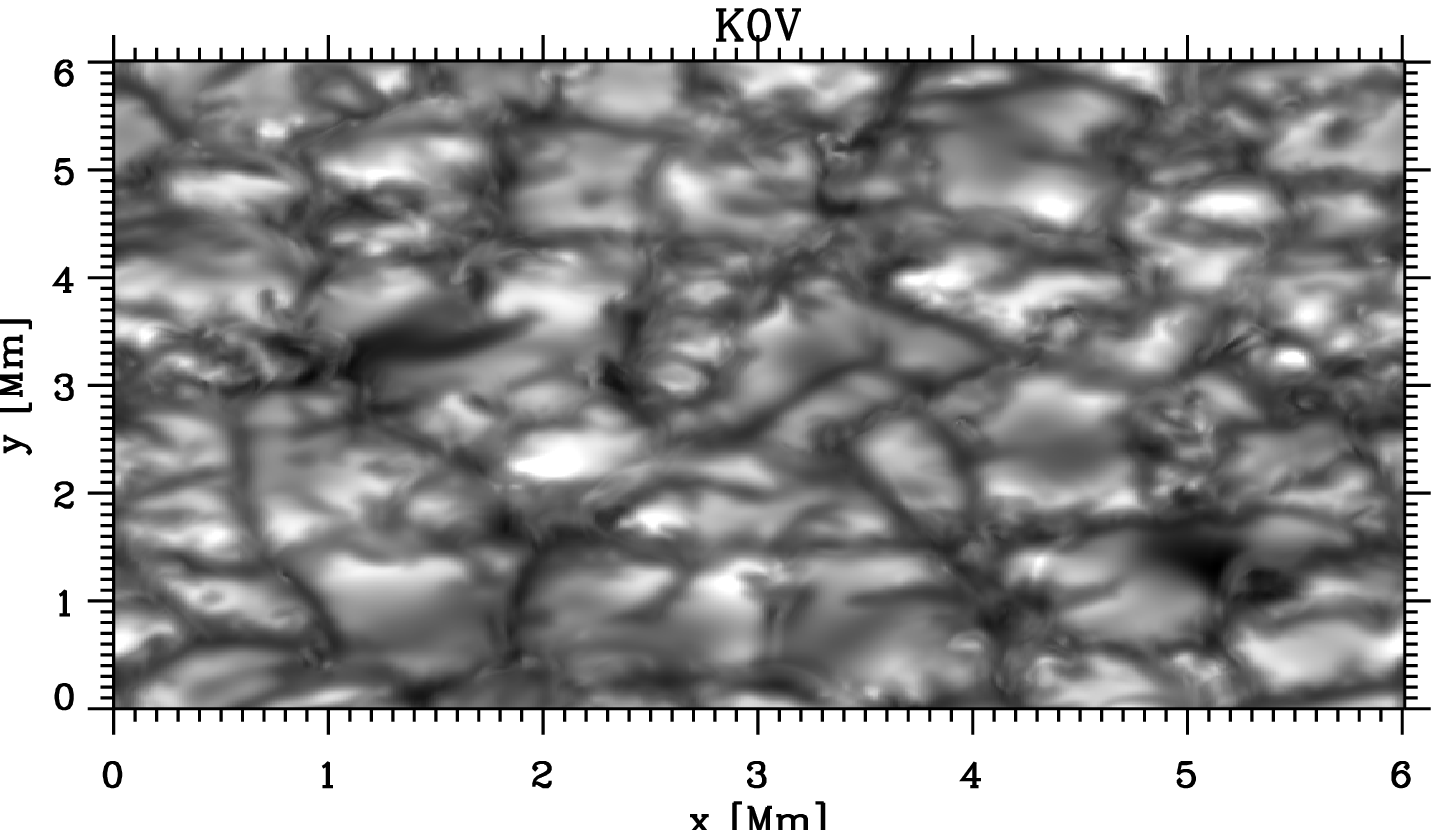} & \includegraphics[width=8.5cm]{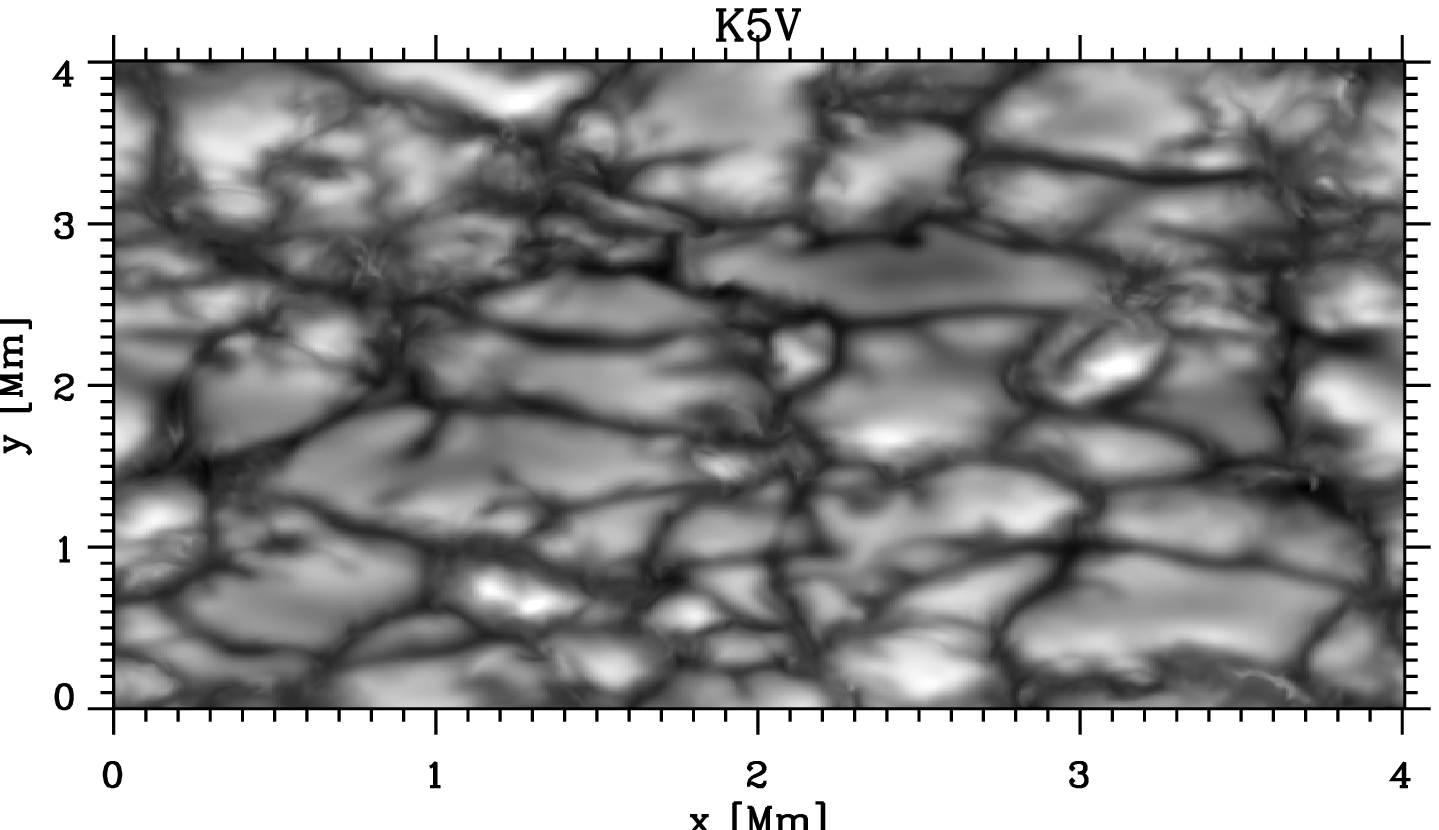}\\[3mm] 
 \includegraphics[width=8.5cm]{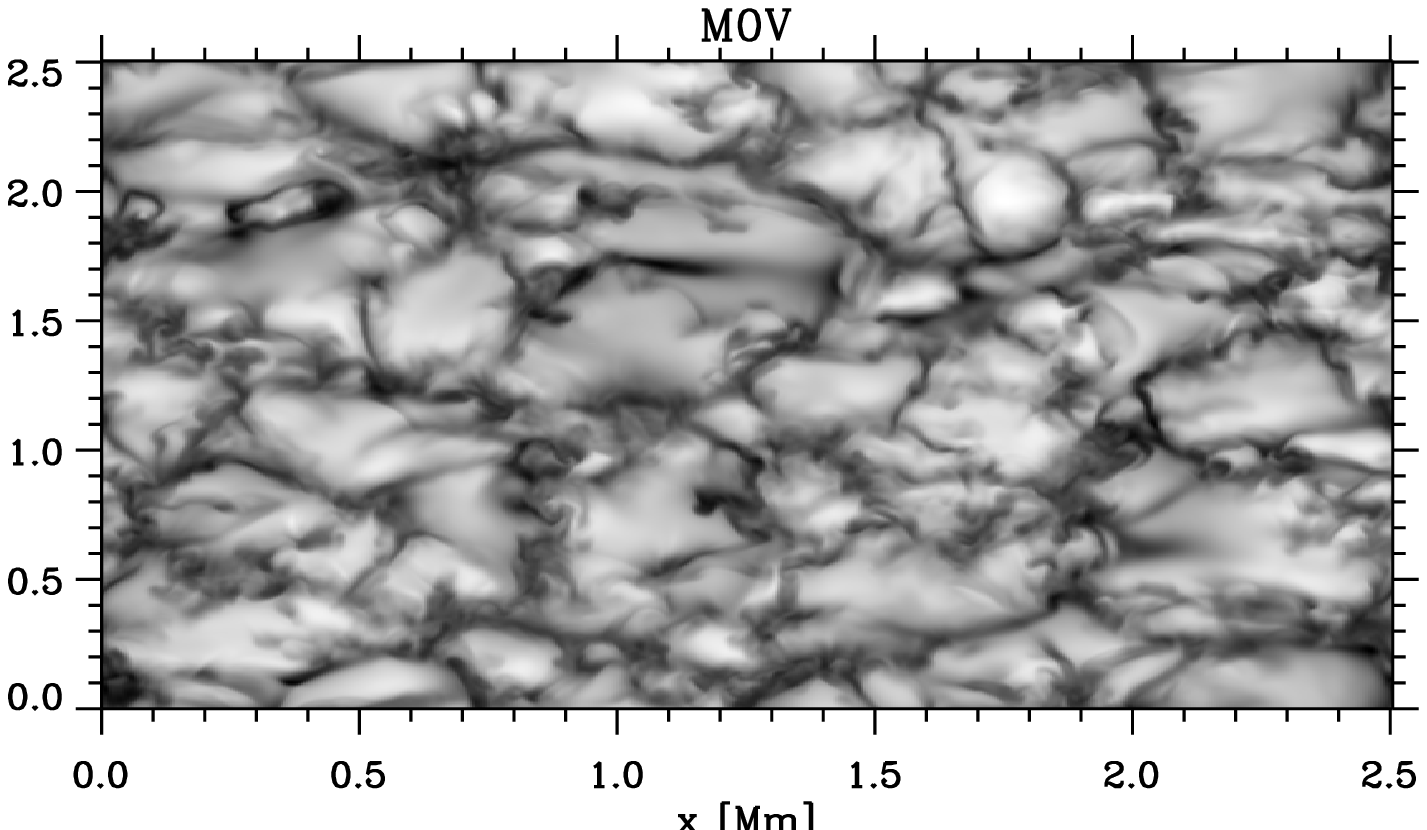} & \includegraphics[width=8.5cm]{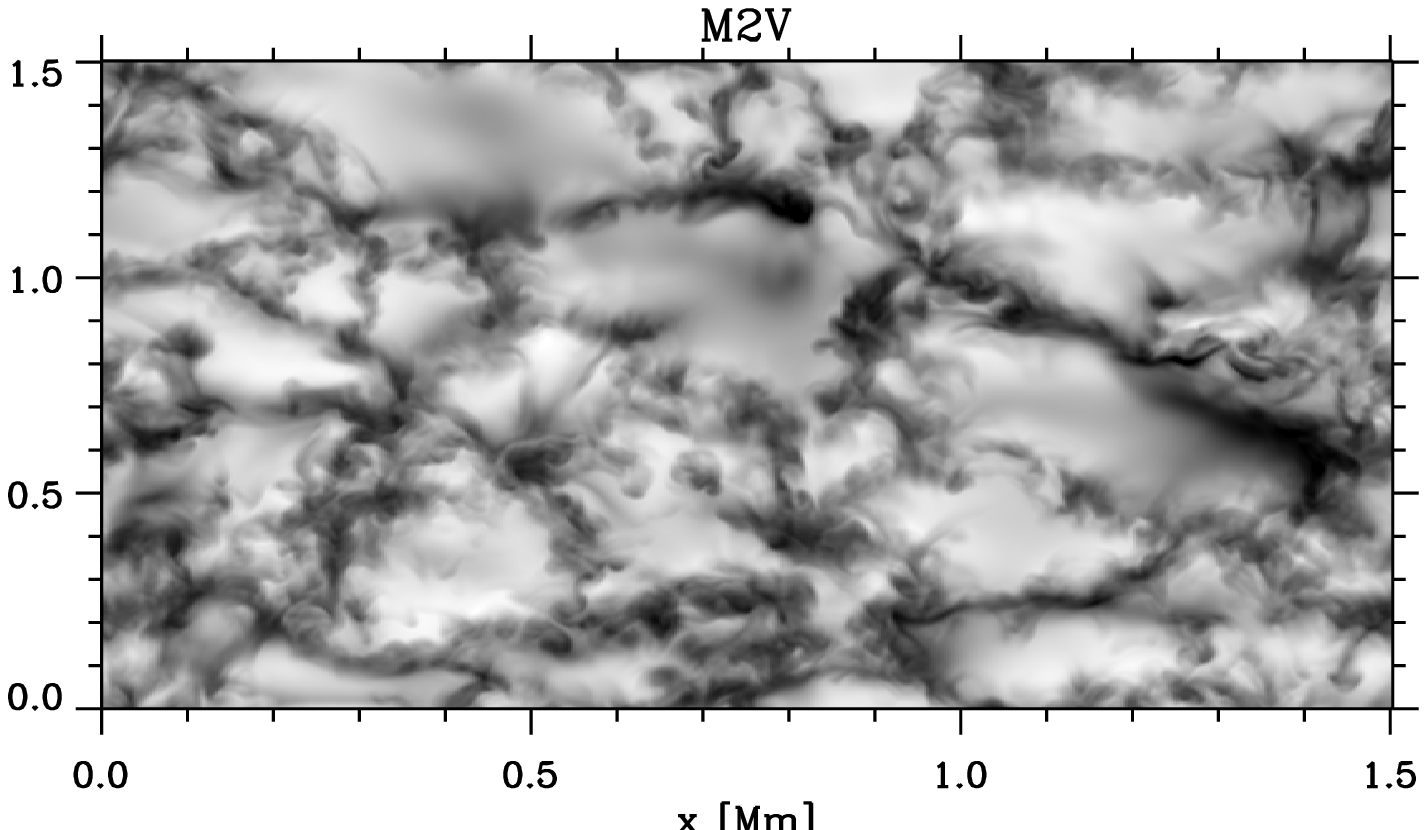}\\[3mm]
\end{tabular}
\caption{Maps of the intensity at $\mu=0.5$ for a wavelength band in the blue (400\,--\,410\,nm). The grey scale is saturated at $\pm 3$ standard deviations.}\label{fig:inclined}
\end{figure*}
In Table~\ref{tab:gran2}, we give the four different time scales described in the previous paragraphs. There is an obvious trend towards shorter lifetimes for cooler stars. While the longer-lived granules of the F3V-star simulation live on average for 8 minutes, the M2V-star granules have a mean lifetime of less than 2 minutes. The autocorrelation time shows the same trend, except for F3V. This is probably due to the granule substructure and shock waves exhibited by this simulation, which both evolve on shorter time scales than the granules.\par
Apart from the fact that many of the short-lived granules ($t_{\mathrm{life}}< 3\,\min$) are very small, we do not find significant correlations between the granule lifetime and the temporal averages of the granule properties (area, intensity, mean vertical velocity). This, in the solar case, is in contrast to the results of \citet{DM04}, who found weak correlations between granule lifetime and mean area, mean intensity and maximum intensity of individual granules. However, our statistics (especially for granules with $t_{\mathrm{life}}> 5\min$) are relatively poor.\par
%
\section{Limb darkening}\label{sec:ld}
In this short section, we analyse the centre-to-limb variation of the intensity (limb darkening) and of the rms intensity contrast of the simulated stars. Analysis of the centre-to-limb variation of spectral lines follows in Sect.~\ref{sec:clv}.\par
\begin{figure}
\centering
\includegraphics[width=4.4cm]{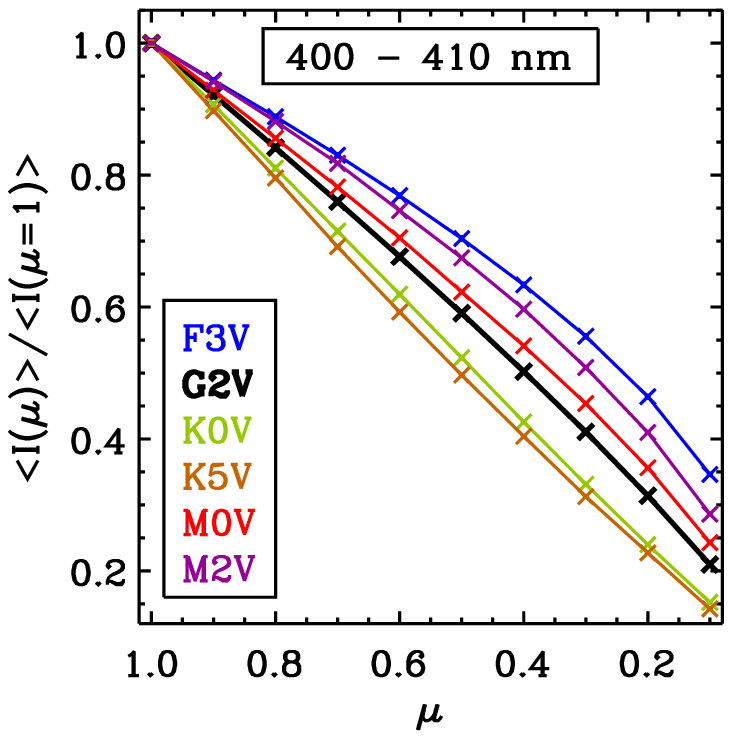}%
\includegraphics[width=4.4cm]{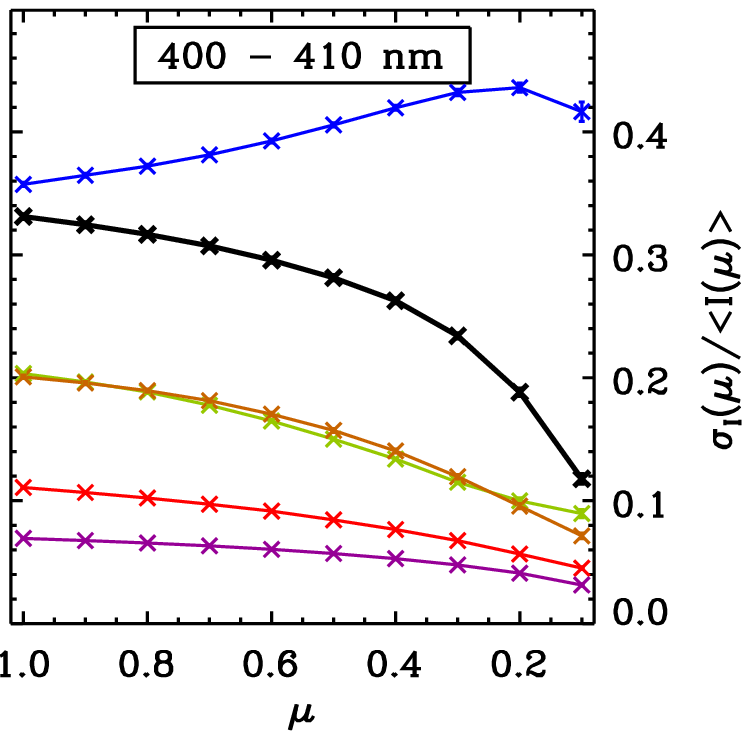}\\
\includegraphics[width=4.4cm]{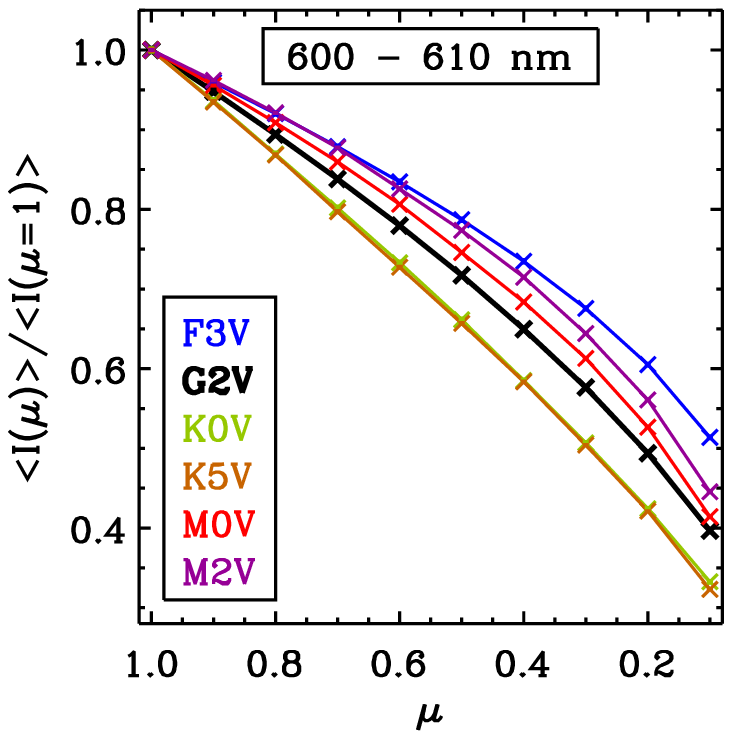}%
\includegraphics[width=4.4cm]{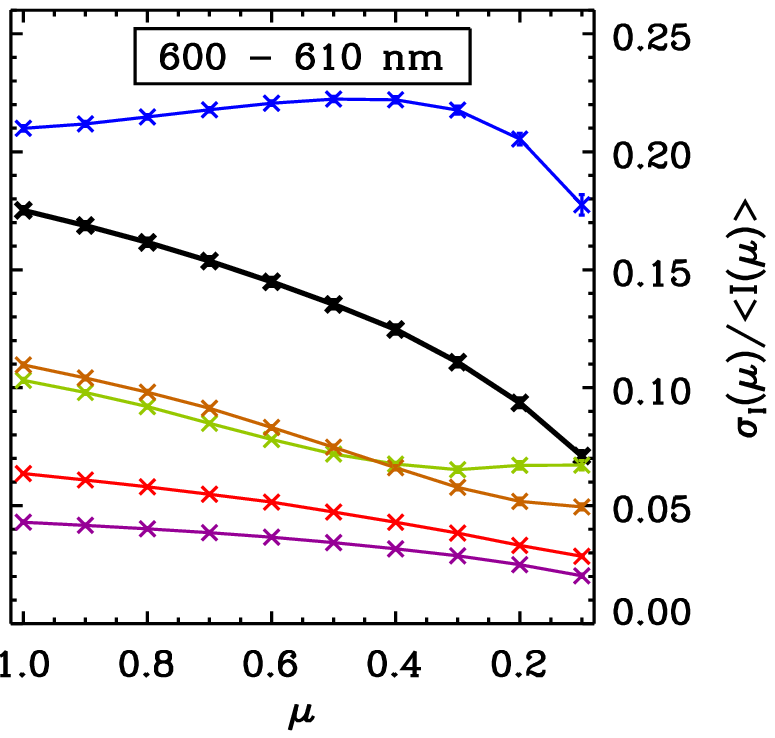}\\
\includegraphics[width=4.4cm]{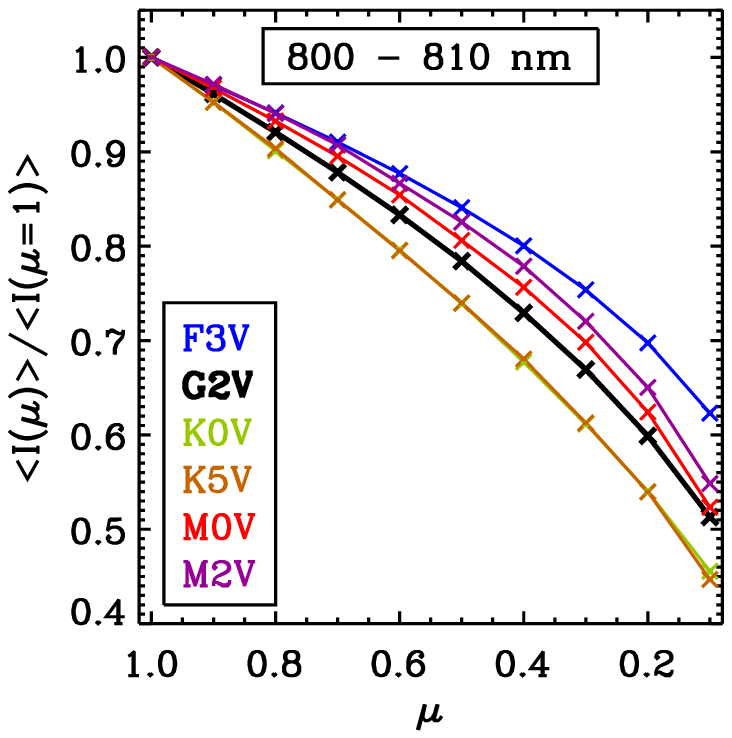}%
\includegraphics[width=4.4cm]{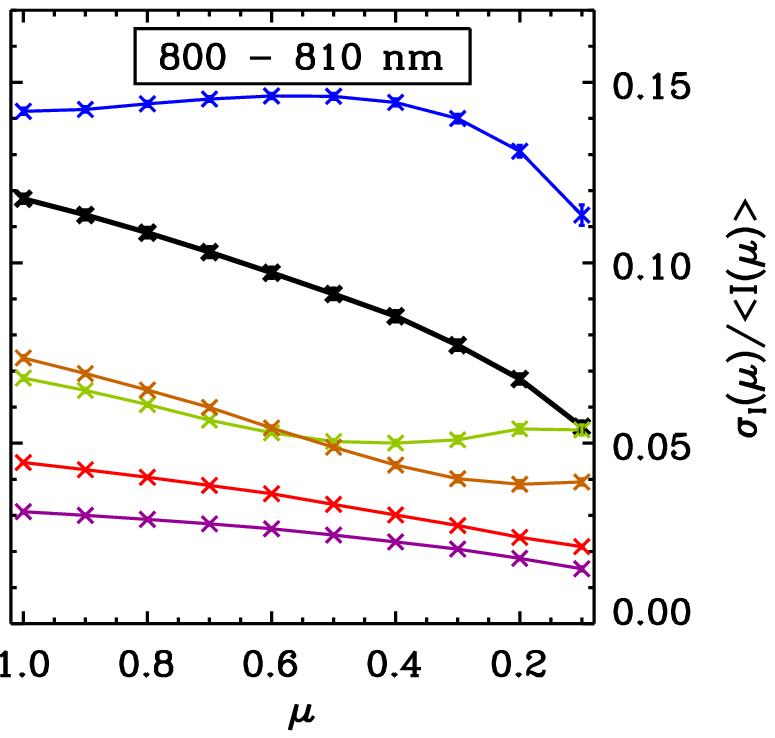}\\
\includegraphics[width=4.4cm]{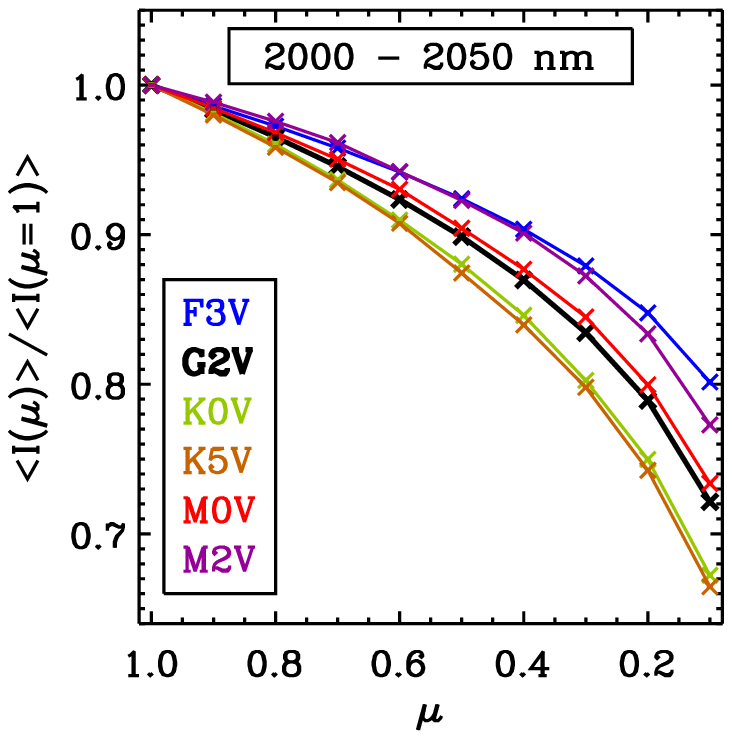}%
\includegraphics[width=4.4cm]{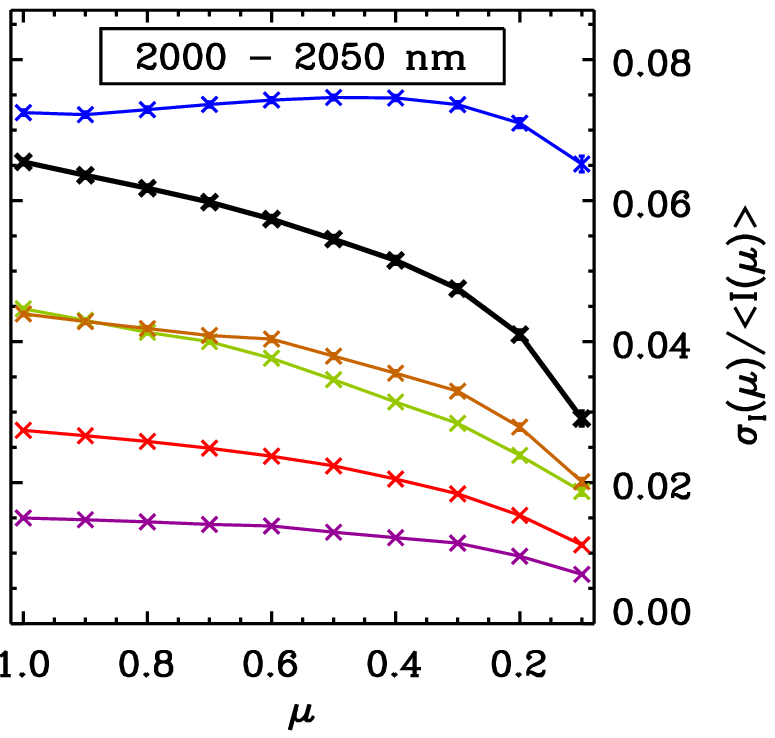}\\
\caption{Centre-to-limb variation of the mean intensity $\langle I \rangle$ ({\it left panels}) and the rms intensity contrast $\sigma_I/\langle I\rangle$ ({\it right panels}) in four different wavelength bands centred at 405, 605, 805, and 2025 nm, respectively. Note the different $y$-scale for each panel.}\label{fig:clv1}
\end{figure}
\begin{figure}
\centering
\includegraphics[width=8.5cm]{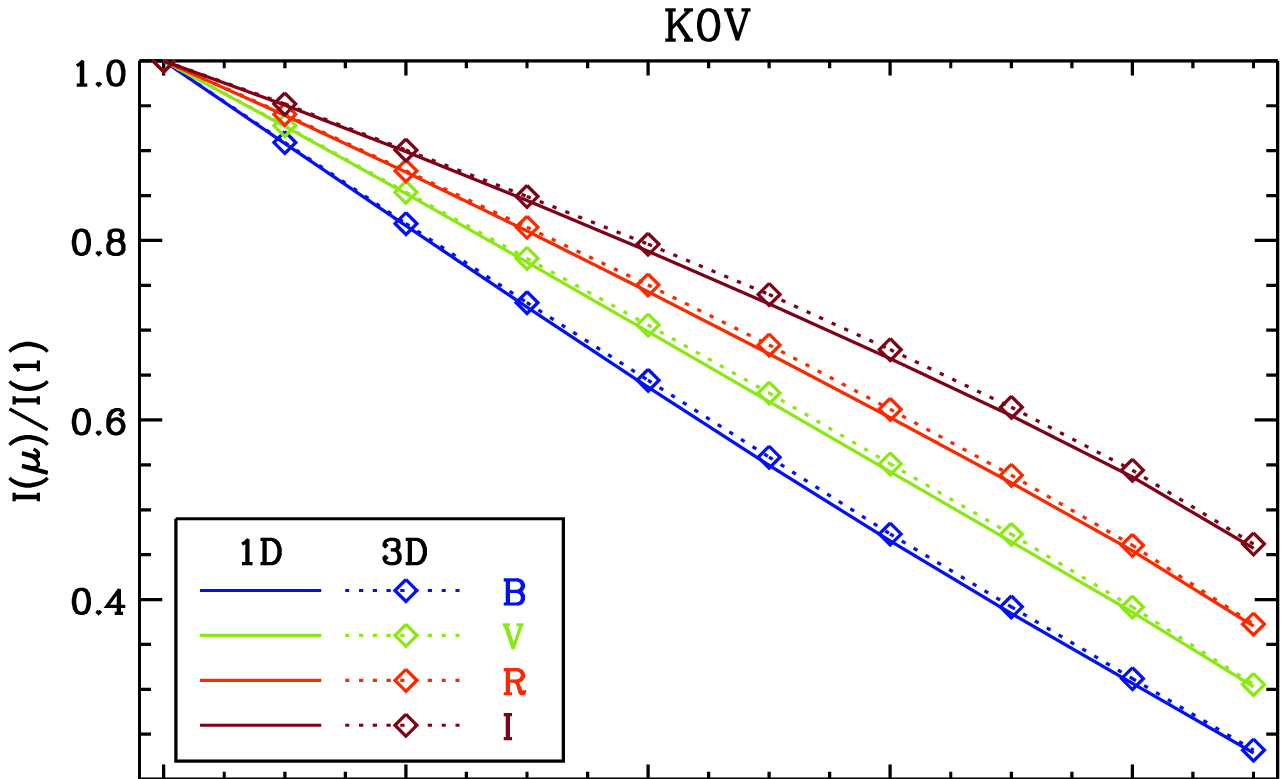}\\\includegraphics[width=8.5cm]{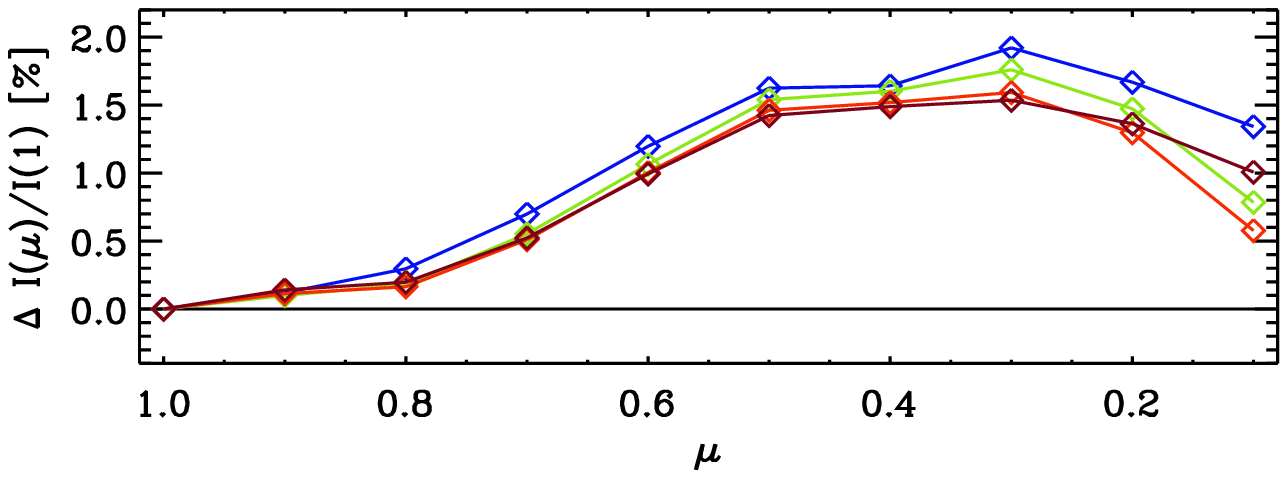}
\caption{Limb darkening. Comparison of the limb darkening between the K0V \texttt{MURaM} simulation (diamonds and dashed curves) and a 1D \texttt{ATLAS} model (solid curves) with similar parameters ($T_{\mathrm{eff}}=4750\,\mathrm{K}$, $\log g[\mathrm{cgs}]= 4.5$, $\vel_{\mathrm{turb}}=1\,\mathrm{km\,s^{-1}}$, solar metallicity) by \citet{Claret2000}. The four passband filters are the Johnson B, V, R, and I passbands \citep{Johnson,Bessell90}. The lower subplot shows the relative differences betwen 1D and 3D results.}\label{fig:clv_comp}
\end{figure}

For the results presented in this section, we solved the radiative transfer using either only the continuum opacity or opacity distribution functions (ODFs) and directly integrated along single rays (with an adaptive increment to resolve the photospheric transition). For the opacities (continuum and ODFs), we used the same \texttt{ATLAS9} data as for the radiative transfer of the hydrodynamical calculations (see Sect.~\ref{sec:sim}). The narrow passbands in which the inclined view (Fig.~\ref{fig:inclined}) and the centre-to-limb variation for all simulations (Fig.~\ref{fig:clv1}) are presented, correspond to single wavelength bins of the \texttt{ATLAS9} opacity data. For the limb darkening calculations in the Johnson filter bands (Fig.~\ref{fig:clv_comp}) several of these wavelength bins (B: 20, V: 25 , R: 35, I: 22) were combined with different weights according to the response functions of the filters \citep{Johnson,Bessell90}.
Figure~\ref{fig:inclined} shows intensity maps (continuum intensity at 400 to 410\,nm) of snapshots of the simulation for a line of sight inclined by $\theta=60^{\circ}$ (i.\,e. $\mu=0.5$) in the $y$-direction. The F- and G-type stars show bright granule edges facing the observer, which are viewed through the more transparent cool intergranular regions in front of them and hence appear brighter. These bright granule edges are less pronounced at this angle in the K-star simulations and absent in the M-star simulations. This can be attributed to the much smaller corrugation of the optical surface.\par
Figure~\ref{fig:clv1} shows the centre-to-limb variation of the mean intensity $\langle I(\mu)\rangle/\langle I(\mu=1)\rangle$ and its normalised standard deviation $\sigma_I(\mu)/\langle I(\mu)\rangle$ (i.\,e. the rms intensity contrast) for four different continuum wavelengths.
In all wavelength bands considered, the limb darkening is strongest in the two K-star simulations. Generally, limb darkening is stronger when the opacity decreases less steeply with height.\footnote{more precisely: when the absorption coefficient $\kappa\varrho$ decreases less steeply with decreasing temperature.} In the lower photospheres of the K stars, the temperature is in the range of 4000 to 5000\,K, for which the opacity becomes almost independent of temperature, while in the photospheres of most other stars the opacity increases rapidly with temperature. Consequently, the opacity drops less steeply with height in the K stars as compared to the other types, resulting in a stronger limb darkening.\par
As the source function (here assumed to be Planckian) depends more strongly on temperature towards shorter wavelengths, the limb darkening is strongest in the blue wavelength band (400\,--\,410\,nm). Our most extreme case is our K5V star, for which the blue band intensity at $\mu=0.1$ is less than 15\% of the disc-centre value.\par
While the limb darkening is qualitatively similar for all stars, the centre-to-limb variation of the intensity contrast changes qualitatively with effective temperature. For most stars, the intensity contrast decreases towards the limb, as the optical surface moves upwards where temperature fluctuations are smaller. However, for the F3V star, the contrast increases with decreasing $\mu$ (except very near to the limb). The reason for this behaviour is the strong corrugation of the optical surface: the ``naked'' granules are separated by deep, optically thin trenches, through which radiation can escape at an inclined angle (see also Fig.~\ref{fig:inclined}). As the granule side walls have less efficient radiative cooling than the granule tops, the temperature contrast between the granule tops and the hot side walls is very high and so, as $\mu$ is decreased, the intensity contrast increases as more bright side walls come into view. At $\mu \lesssim 0.3$ this effect is over-compensated by the normal centre-to-limb decrease of contrast due to the increasing geometrical height of the optical surface. As the corrugation of the optical surface is less pronounced in the cooler stars, the intensity contrast decreases monotonically with decreasing $\mu$ in the G and M stars; the K stars (particularly the K0V star) show, however, an almost constant or even slightly increasing intensity contrast at low $\mu$ at some wavelengths. This is mainly produced by very bright regions behind intergranular lanes. As explained in Sect.~\ref{sec:vort}, the temperature-insensitive regime of the opacity between 4000 and 5000\,K is responsible for a higher sensitivity of $T(\tau=1)$ to density fluctuations. For some extended intergranular lanes with particularly low density this leads to a significant brightening of the granule edges behind them at strongly inclined view. This effect might also be responsible for the fact that the two K-star simulations have similar contrasts, while it is generally decreasing with decreasing effective temperature: in the K5V simulation, the temperature-insensitive opacity regime is reached right at the optical surface in the granules (rather than in the intergranular lanes). Consequently, small density perturbations enhance the rms contrast of the granule intensities and thus the overall intensity contrast of the K5V simulation. \par
Figure~\ref{fig:clv_comp} shows a comparison of the limb darkening of the K0V simulation with that of an \texttt{ATLAS} model with $T_{\mathrm{eff}}=4750\,\mathrm{K}$, $\log g[\mathrm{cgs}]= 4.5$, turbulent velocity $\vel_{\mathrm{turb}}=1\,\mathrm{km\,s^{-1}}$, and solar metallicity by \citet{Claret2000}. The typical relative difference between our 3D calculation and the 1D model is mostly below 2\%. The disc-integrated effect of these deviations on the total stellar radiance is between $0.4$ and $0.5\%$ in all bands. Carrying out the same comparison between our solar model (G2V) and the corresponding 1D results by \citet{Claret2000}, we find relative differences of up to about 8\% (B band) and a disc-integrated effect between 0.2\% (I band) and 1.6\% (B band). We also compared our results to observational data of solar limb darkening by \citet{Neckel} for a limited number of continuum wavelengths between 400 and 900\,nm and found deviations mostly below 5\%. The statistical uncertainties of our limb darkening results are very small: the relative error of $\langle I(\mu)\rangle$ is of the order of $10^{-3}$, estimated from the scatter between the six snapshots. However, we expect somewhat larger systematic errors due to the opacity binning in our simulations with only four bins \citep[][]{Beeck12}.
\begin{table}
\caption{Line parameters.}\label{tab:lines}
\begin{tabular}{lrr}
\hline\hline
Species & $\lambda_{\mathrm{rest}} [\mathrm{nm}]$ & $E_i [eV]\,^{\mathrm{a}}$ \\\hline
Fe\,\textsc{i} & 616.536 & 4.143 \\
Fe\,\textsc{i} & 617.333 & 2.223 \\
Ti\,\textsc{i} & 2223.284 & 1.739 \\\hline
\end{tabular}
\begin{list}{}{}
\item[$^{\mathrm{a}}$] $E_i$ is the excitation potential of the lower level of the transition
\end{list}
\end{table}
%
\section{Spectral line synthesis}\label{sec:lines}
\subsection{Line synthesis and stellar disc integration}\label{sec:met:lines}
\begin{figure}
\centering
\includegraphics[width=8.5cm]{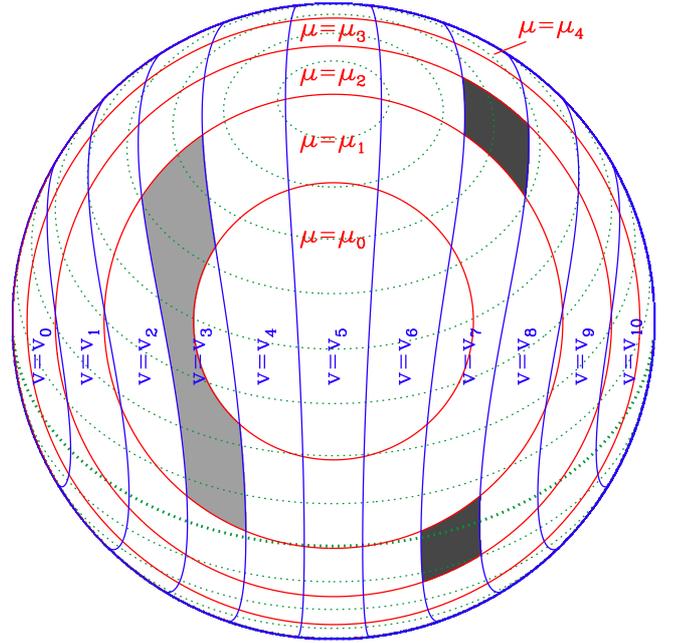}
\caption{Sketch of a stellar disc decomposed into areas of nearly-constant velocity (``$\vel$-stripes'') and surface normal direction (``$\mu$-rings'') for a star with differential rotation in latitude. Red circles indicate the limits between the $\mu$-rings, blue curves indicate the limits between $\vel$-stripes, the dotted green curves show constant latitude circles on the stellar sphere (in steps of $10^{\circ}$). Two bins ($w_{1,3}$ and $w_{2,7}$) are filled in different shades of grey for illustration . The shown star has an inclination of $45^{\circ}$ and rotates differentially with twice the solar equator-to-pole shear ($\alpha=0.4$). In this sketch we used $N_{\mu}=5$ and $N_{\vel}=11$. For the numerical stellar-disc integrations considered in Sect.~\ref{sec:discint} and \ref{sec:comp} we used $N_{\mu}=10$ and $N_{\vel}=51$.}\label{fig:sketch} 
\end{figure}

In order to quantify the effect of the 3D structure on spectral lines,
we chose as three representative lines: an infrared Ti\,\textsc{i} line at 2223\,nm and two optical Fe\,\textsc{i} lines at 616.5 and 617.3\,nm, which are present in all stars of our
effective temperature range, although the titanium line is weak in the two hottest models. These lines are relatively isolated (unblended) and we possess high-resolution archive spectra of F- and G-type main-sequence stars covering spectral ranges containing the two iron lines (see Sect.~\ref{sec:comp}). All lines are magnetically sensitive and are suitable for measuring stellar magnetic fields \citep[][and several follow-up papers]{Richard, JK04}. This will become relevant in the third paper of this series, in which the impact of the magnetic field on convection and spectral lines will be discussed.\par 
The lines were calculated with the line synthesis
code \texttt{SPINOR} \citep{spinor} for six 3D snapshots for each simulation. We considered snapshots 2000 simulation time steps $\delta t$ apart (about 5 -- 7 minutes stellar time).\par
\begin{figure*}
\begin{tabular}{cc}
\includegraphics[width=8.5cm]{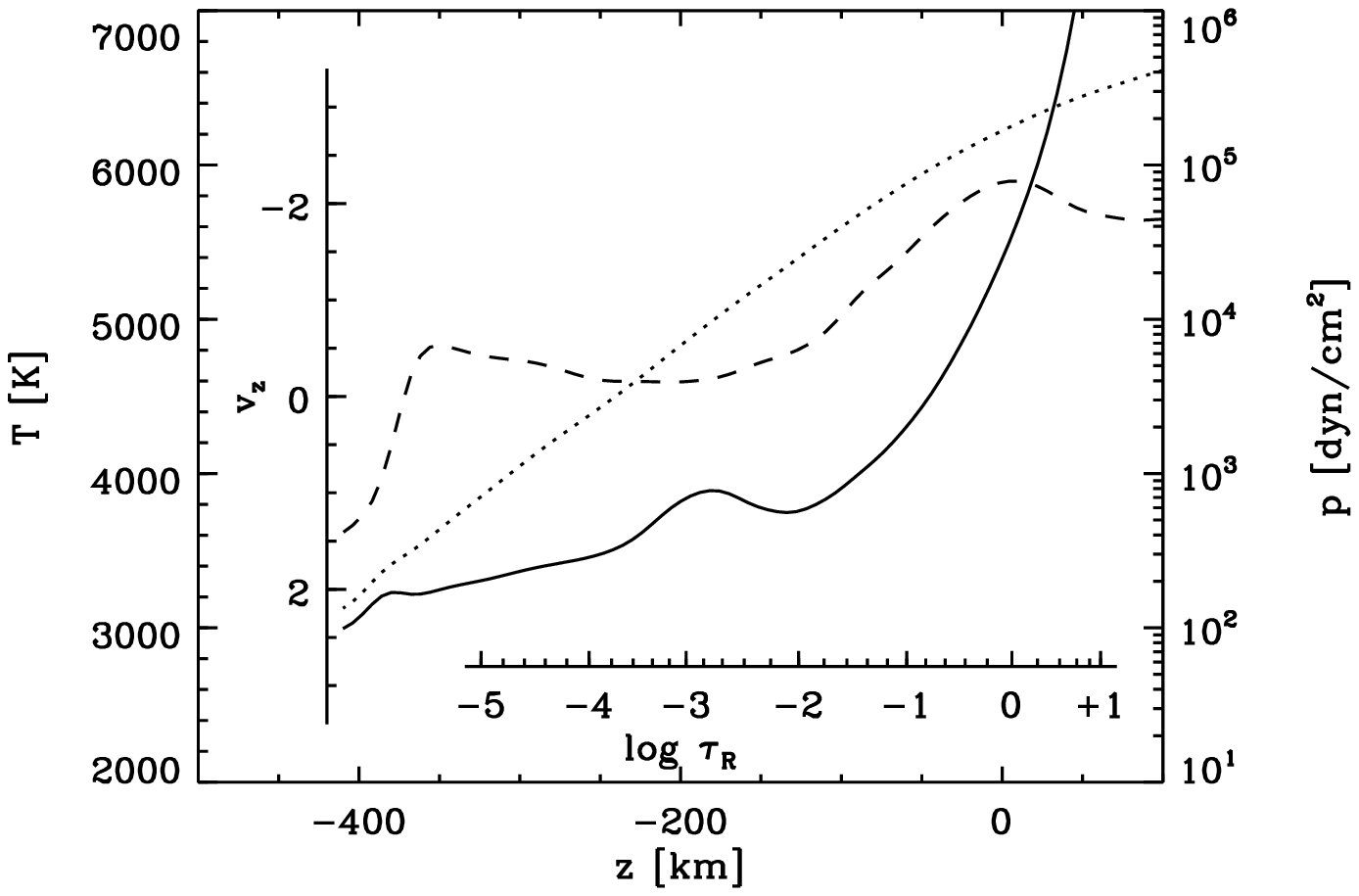} & %
\includegraphics[width=8.5cm]{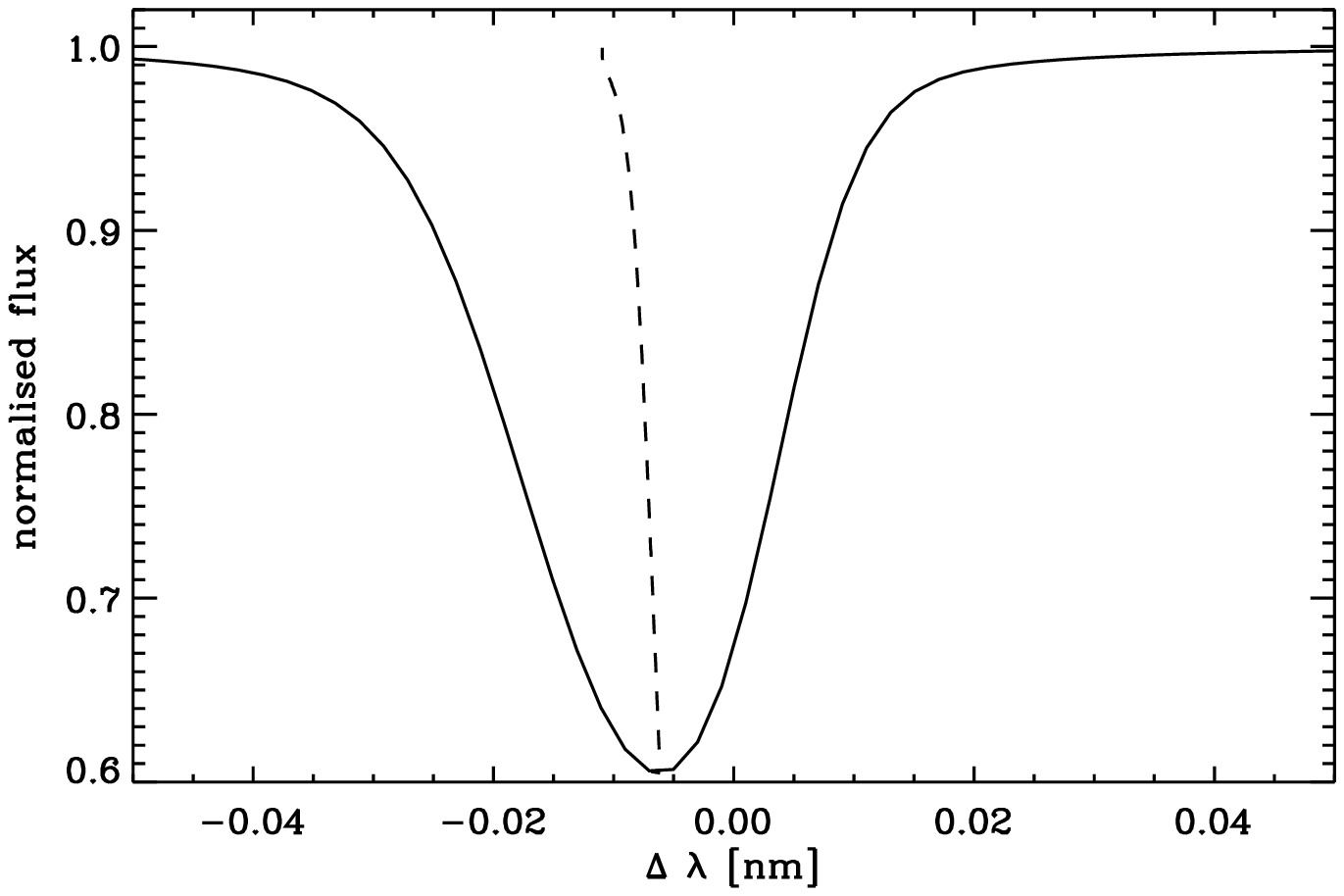}\\
\includegraphics[width=8.5cm]{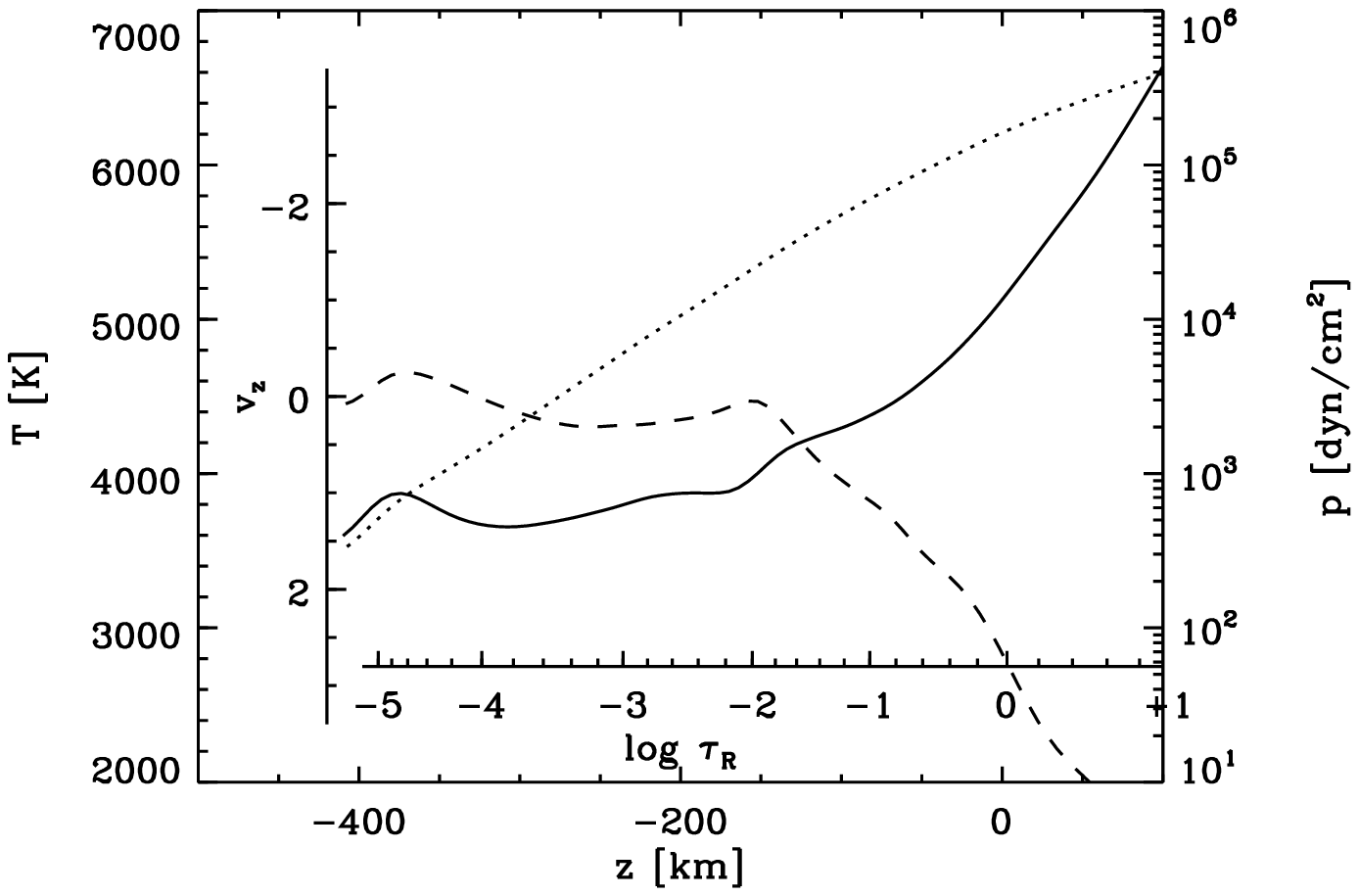} & %
\includegraphics[width=8.5cm]{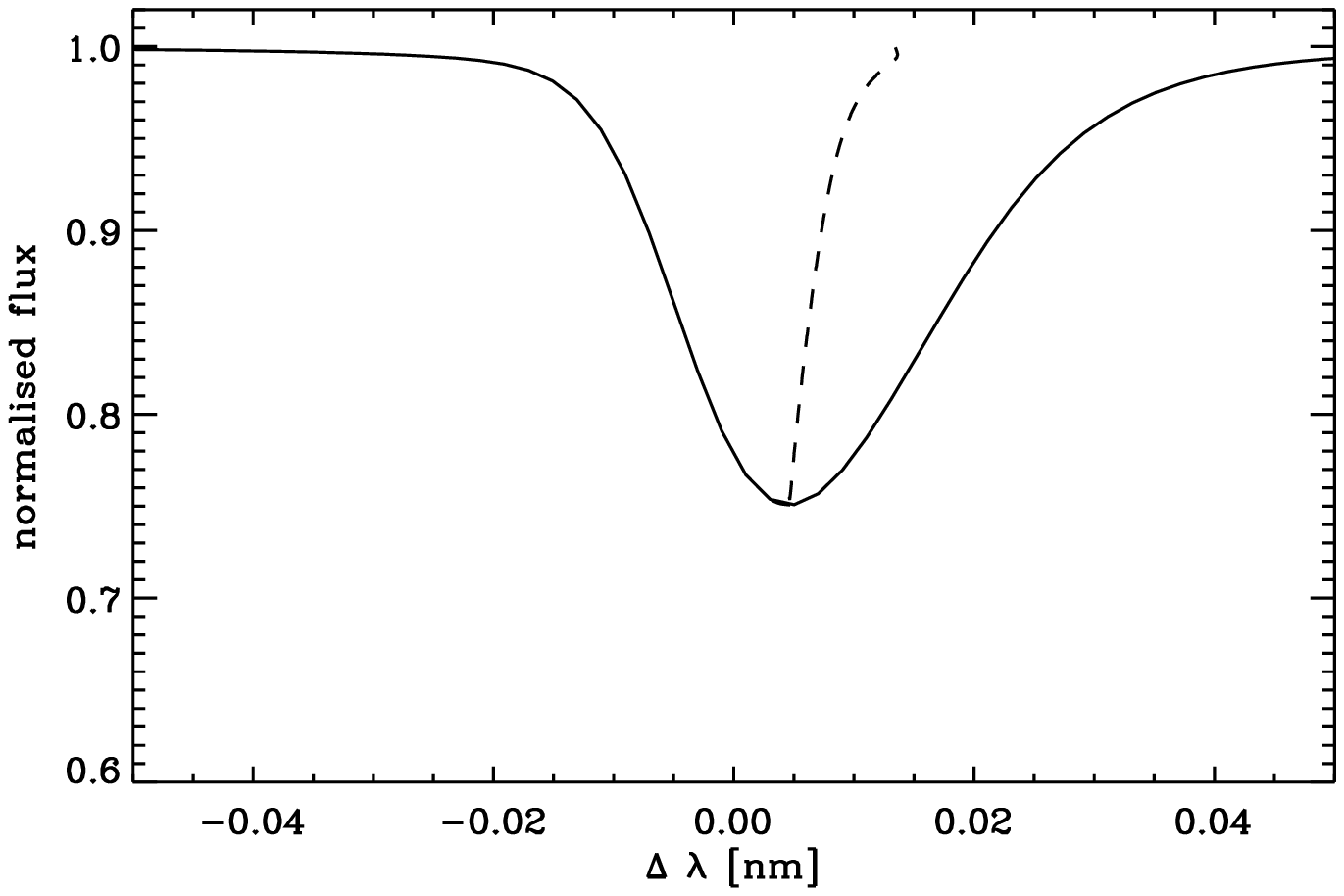}\\
\end{tabular}
\caption{Synthetic line profiles (Ti\,\textsc{i} at 2223\,nm) of single vertical rays of the K0V-star simulation in an upflow region ({\it upper panels}) and in a downflow region ({\it lower panels}). {\it Left panels:} Temperature (solid), pressure (dotted), and vertical component of the fluid velocity (dashed) along the vertical ray. {\it Right panels:} Profiles of the Ti\,\textsc{i} line at 2223\,nm obtained with \texttt{SPINOR} for the corresponding vertical rays ($\Delta\lambda = \lambda-\lambda_{\mathrm{rest}}$; the dashed lines give the bisectors of the profiles).}\label{fig:spatialdis}
\end{figure*}
For comparison with spatially unresolved stellar observations, we integrated the light over a constructed stellar
disc. This integration can be done
semi-analytically by a convolution for rigid-body rotation and homogeneous surface properties \citep{ND90c, Ludwig07}.
Real stars, however, often rotate differentially or have large-scale inhomogeneities such as star spots. In order to cover this general case, we implemented a numerical stellar-disc integration.\par
The locally averaged line profile $I_0(\lambda, \mu)$, which is generally a function of wavelength $\lambda$ and angle $\theta$ (with $\cos\theta=:\mu$) between line of sight and surface normal, is only calculated for a limited number $N_{\mu}$ of reference values $\mu_j$. One considers $I_0(\lambda, \mu)$ to be represented by the profile $I_0(\lambda, \mu_j)$ within a small interval  $\tilde\mu_j\ge \mu> \tilde\mu_{j+1}$ ($j=0,1,\dots,
N_{\mu}-1$), where $\tilde\mu_j>\mu_j>\tilde\mu_{j+1}$. On the stellar disc, these $\mu$ intervals correspond to concentric rings (and a central disc). To include rotation (differential or rigid),
the constructed stellar surface is additionally divided into
``velocity stripes'', which are the projected areas of regions with a
line-of-sight component of the rotation velocity in a given interval $\tilde\vel_k\le\vel<\tilde\vel_{k+1}$ ($k=0,1,...,N_\vel-1$). Figure~\ref{fig:sketch} shows a sketch of a stellar disc with $N_{\mu}=5$ $\mu$-rings and $N_{\vel}=11$ $\vel$-stripes. To obtain the disc-integrated line profile, $F(\lambda)$, we replace the wavelength dependence of $I_0$ by a (mathematically equivalent) line-of-sight-velocity dependence and obtain:
\begin{equation}\label{eqn:discint}
F(\vel)=\sum\limits_{j=0}^{N_{\mu}-1}\sum\limits_{k=0}^{N_{\vel}-1} w_{jk} I_0(\vel+\vel_k, \mu_j)\,,
\end{equation}
where each weight $w_{jk}$ corresponds to the projected area on the stellar disc with $\tilde\mu_j\ge\mu > \tilde\mu_{j+1}$ and $\tilde\vel_k\le \vel < \tilde\vel_{k+1}$ normalised by the total projected disc area. Interval limits were chosen such that the reference values $\mu_j$ and $\vel_k$ are centred in the intervals.\par
For the differential rotation we here use the simple law 
\begin{equation}\label{eqn:alpha}
\Omega=\Omega_{\mathrm{eq}}\cdot\left(1-\alpha\sin^2\theta\right)\,.
\end{equation}
The differential rotation parameter $\alpha$ is positive for solar-like differential rotation \citep[$\alpha_{\odot}\approx 0.2$ at the surface,][]{HH70} and negative for anti-solar differential rotation. With the current implementation of the numerical integration, we are able to cover the range $-0.5 \le \alpha \le +1.0$, which is much wider than the range of today's observed and predicted values of $\alpha$ in stars.\par
It is computationally expensive to use many $\mu$-rings because each value of $\mu_j$ requires a separate 3D line synthesis. In their pioneering work, \citet{ND90c} used $N_{\mu}=3$ and included rigid-body rotation by convolution. Here, we used $N_{\mu}=10$ with $\mu_j=1.0, 0.9,\dots,0.1$.\par 
\begin{figure}
\centering
\includegraphics[height=5.5cm]{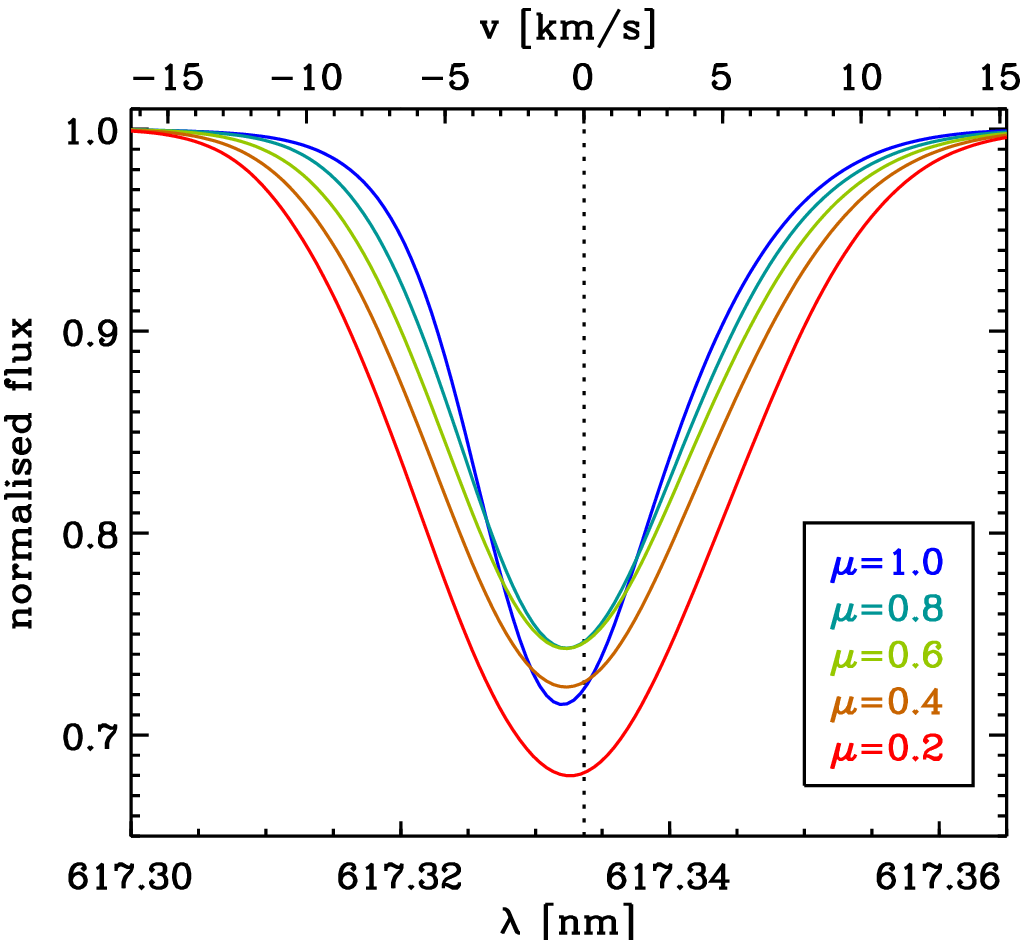}~~\includegraphics[height=5.5cm]{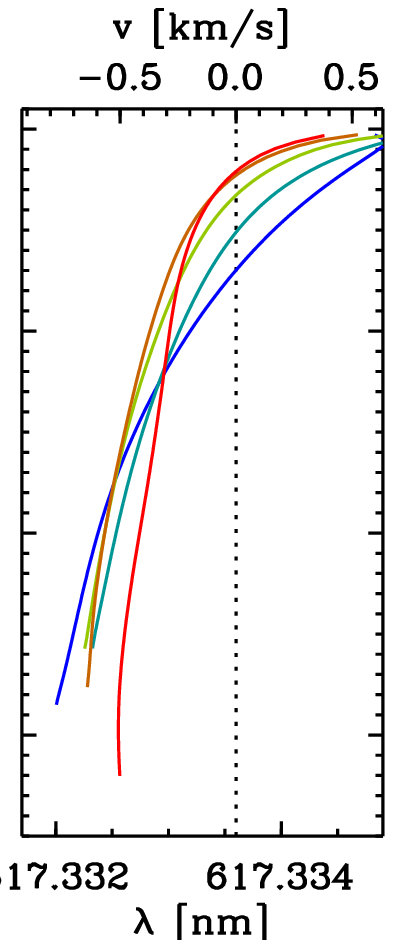}\\
\caption{Centre-to-limb variation of a sample line profile. {\it Left panel:} line profile obtained from the F3V simulation (Fe\,\textsc{i} line at 617.3\,nm; average over six snapshots) for various values of $\mu=\cos \theta$. {\it Right panel:} bisector variation of the profile; same colour code and ordinate labels as {\it left panel}.}\label{fig:clv2}
\end{figure}
\begin{figure*}
\centering
\includegraphics[width=8.5cm]{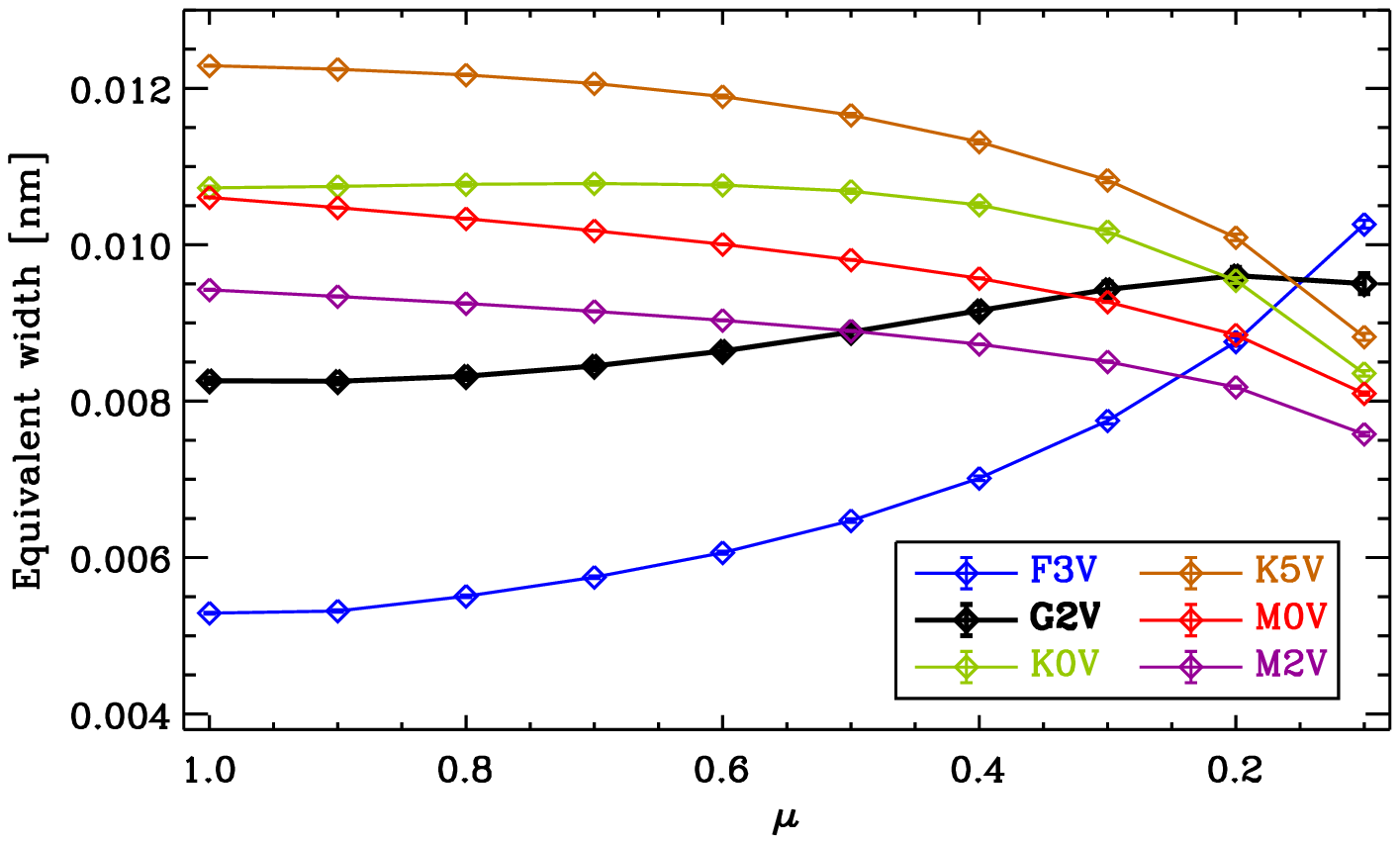}~%
\includegraphics[width=8.5cm]{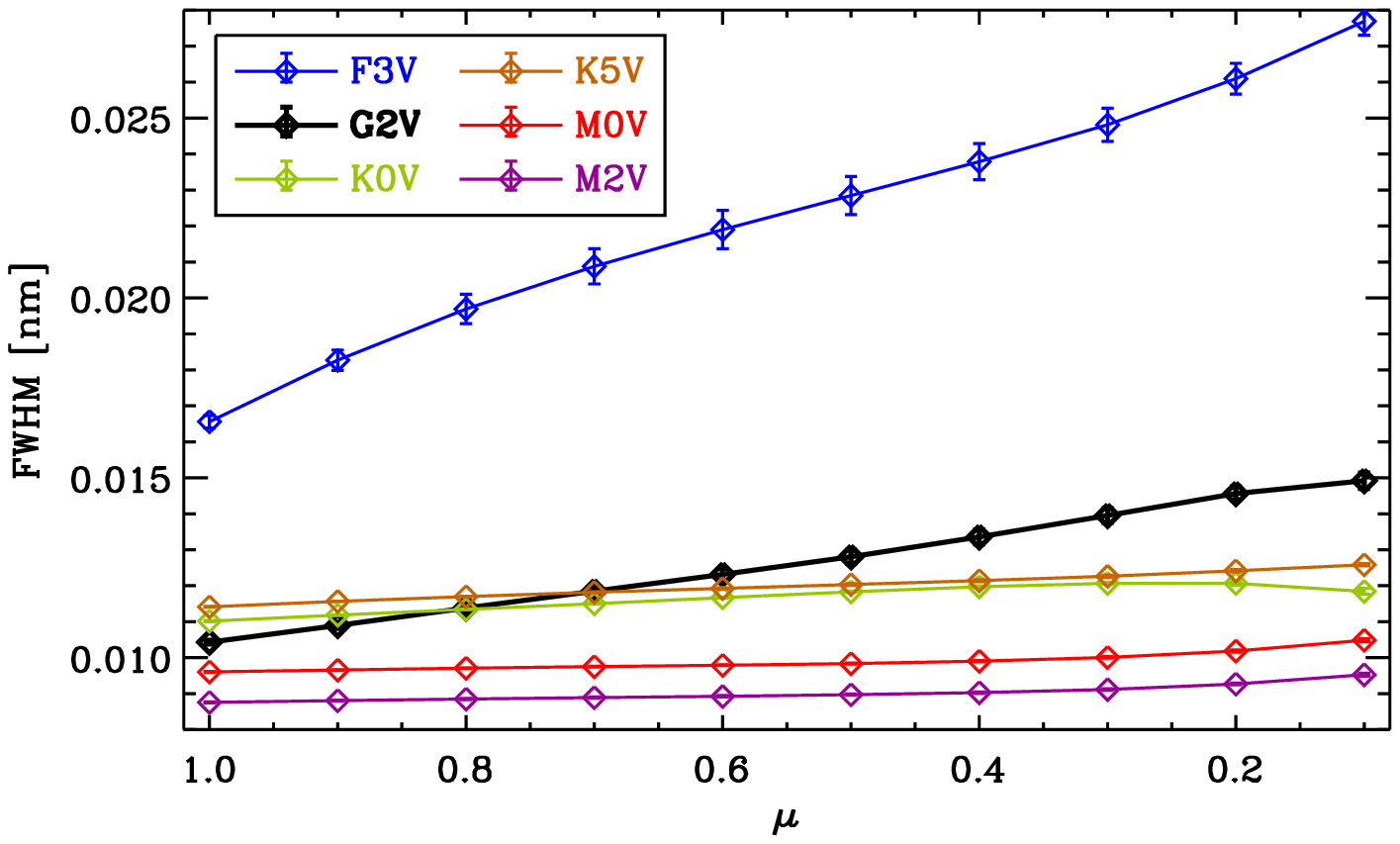}\\[-4mm]
\includegraphics[width=8.5cm]{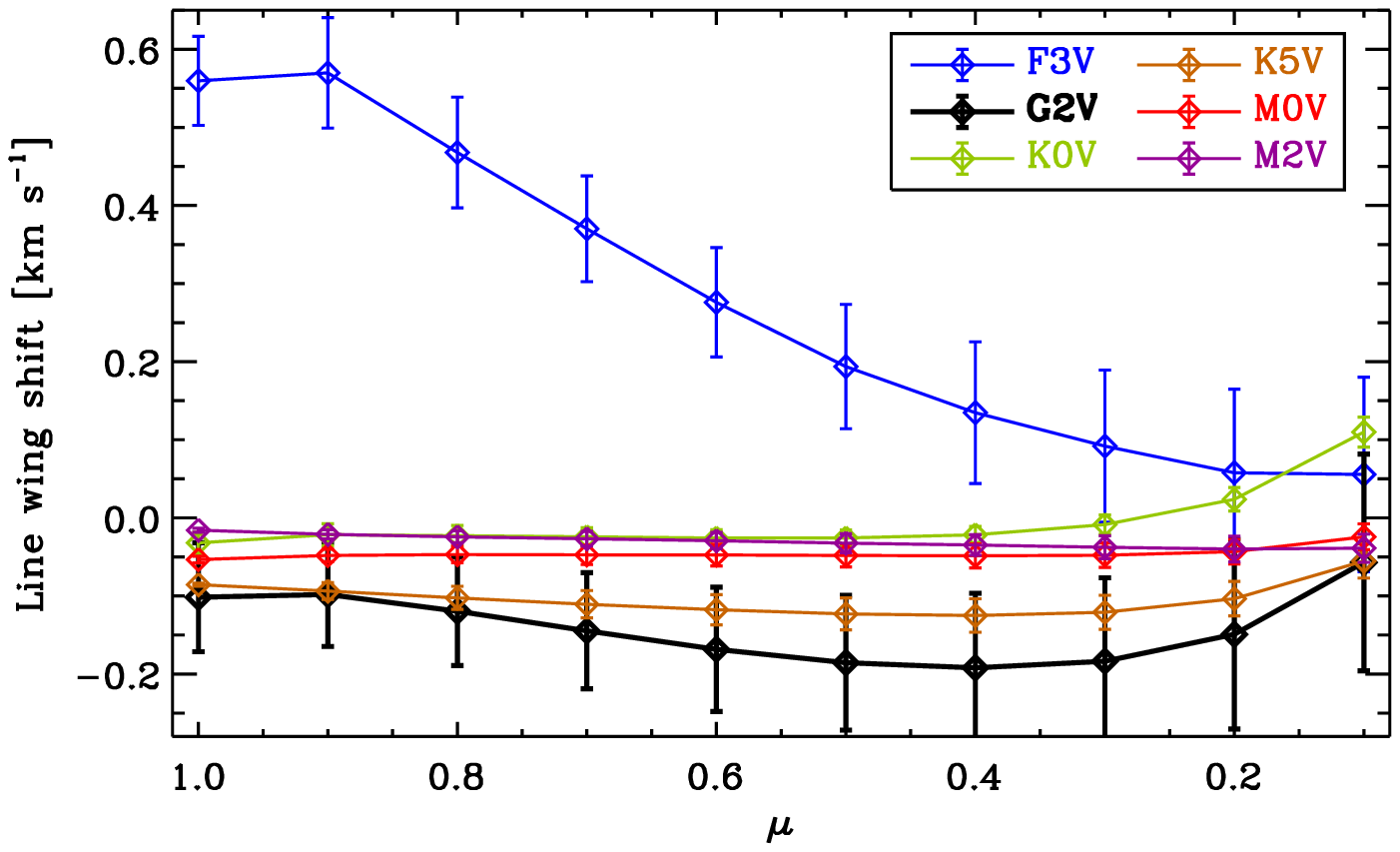}~%
\includegraphics[width=8.5cm]{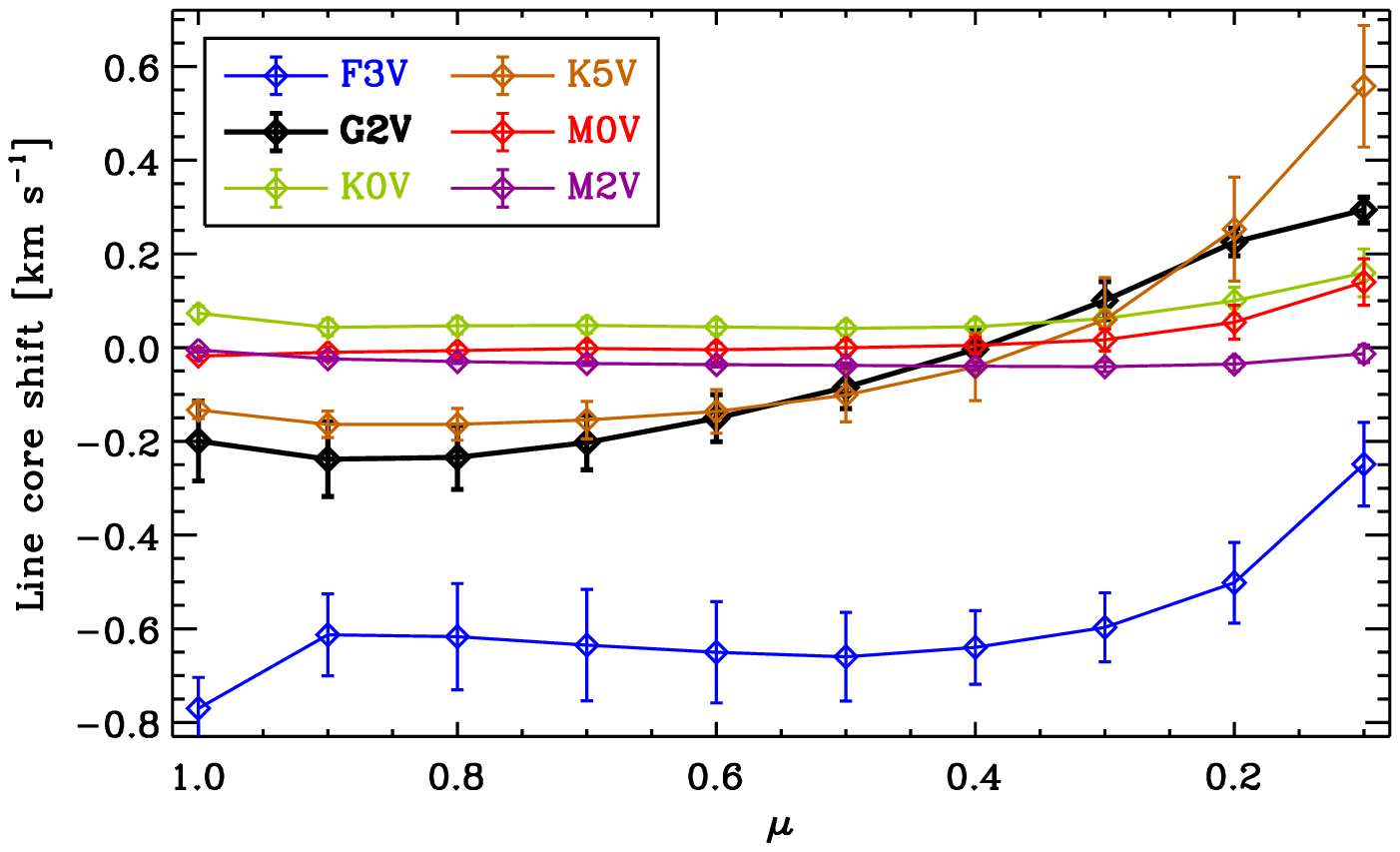}
\caption{Centre-to-limb variation of the Fe\,\textsc{i} line at 617.3\,nm. {\it Top panels:} Variation of equivalent width ({\it right}) and FWHM ({\it left}) of the line as functions of $\mu$. {\it Bottom panels:} Convective Doppler shift (translated into line-of-sight velocity) of the line wings ({\it right}; 5\% level of max. line depth) and the line core ({\it left}) as functions of $\mu$ (positive $=$ redshift). The gravitational redshift is neglected.}\label{fig:clv3}
\end{figure*}
In contrast, the number of $\vel$-stripes has only a minor impact on the computational expense and thus can be chosen almost arbitrarily high. In a few test cases with rigid-body rotation, the relative error between the semi-analytical result and the numerical method was found to be smaller than $10^{-4}$ for $N_{\vel}\ge 50$, which is lower than the typical statistical error due to the limited number of simulation snapshots. For the disc integration presented in this paper, we used $N_{\vel}=51$. The numerical values of $\vel_k$ scale linearly with $\vel_{\mathrm{rot}}\sin i$ (the line-of-sight component of the rotational velocity of the stellar equator; $i$ is the inclination of the rotation axis with respect to the line of sight). Once calculated, the $N_{\mu}\times N_{\vel}$ weights $w_{jk}$ for a given combination of $i$ and $\alpha$ can be used for the integration for any rotational velocity $\vel_{\mathrm{rot}}$ and given local profiles $I_0(\vel,\mu_j)$ .\par
\begin{figure*}
\centering
\includegraphics[height=5.9cm]{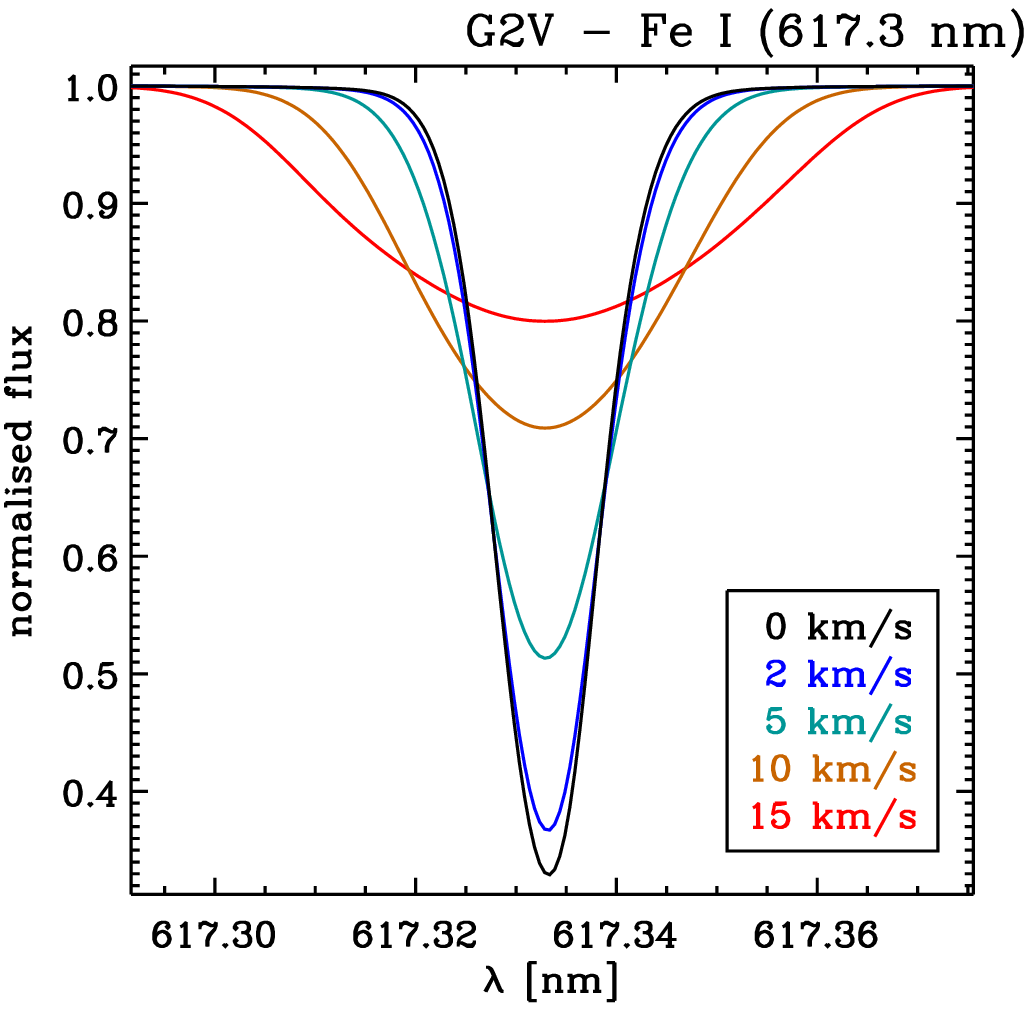}\includegraphics[height=5.9cm]{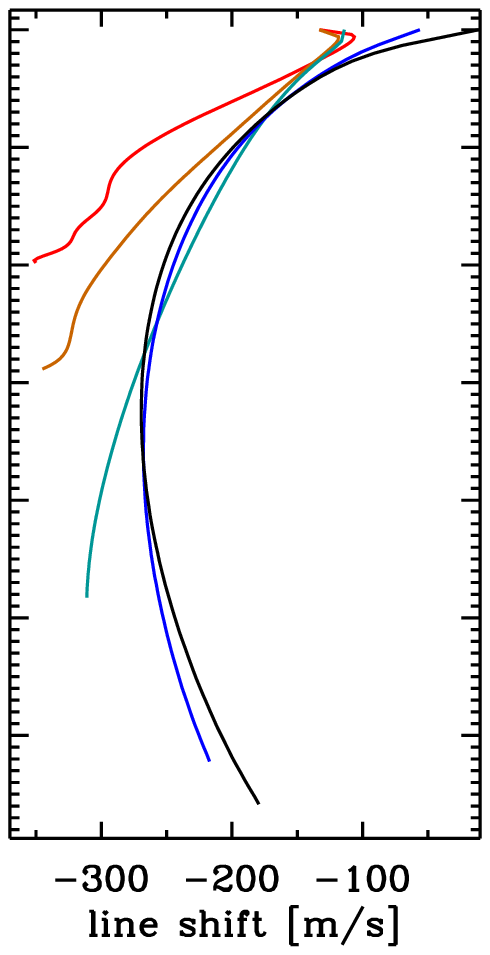}~~\includegraphics[height=5.9cm]{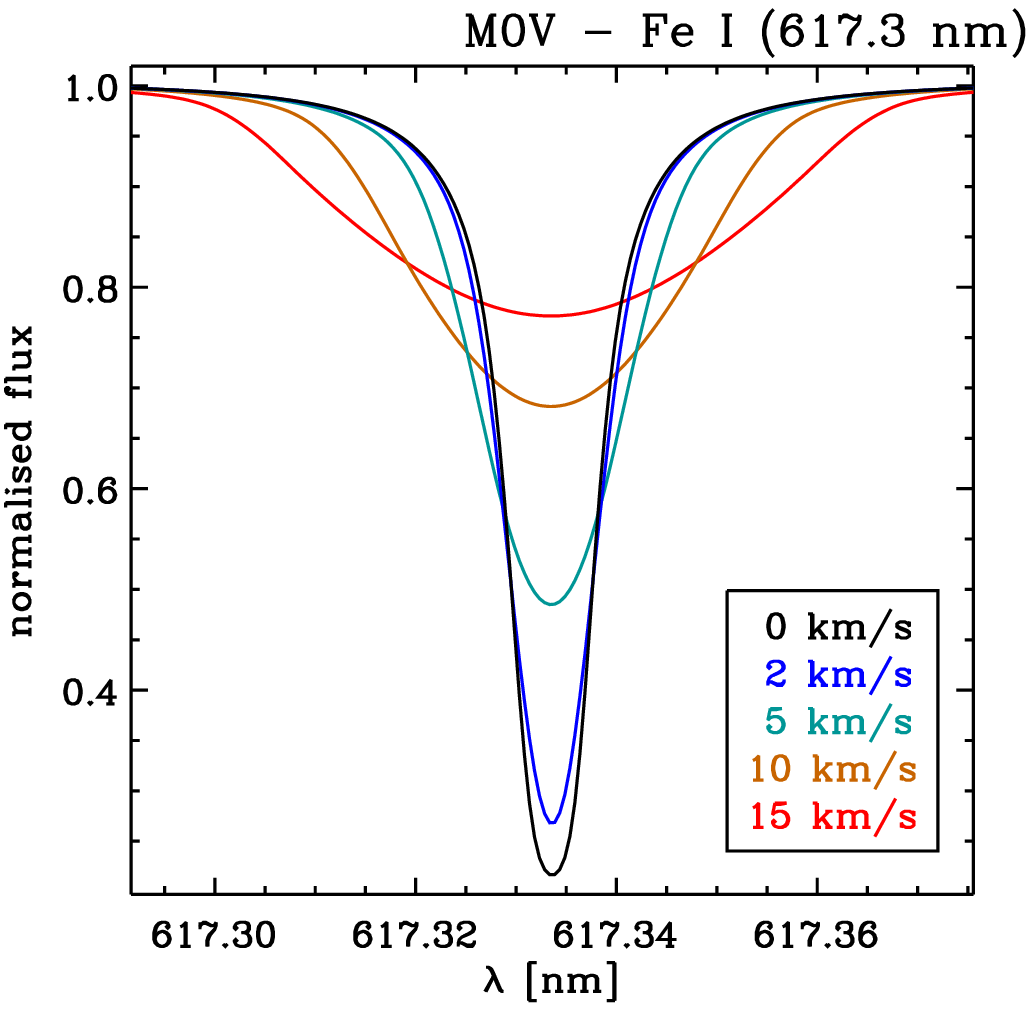}\includegraphics[height=5.9cm]{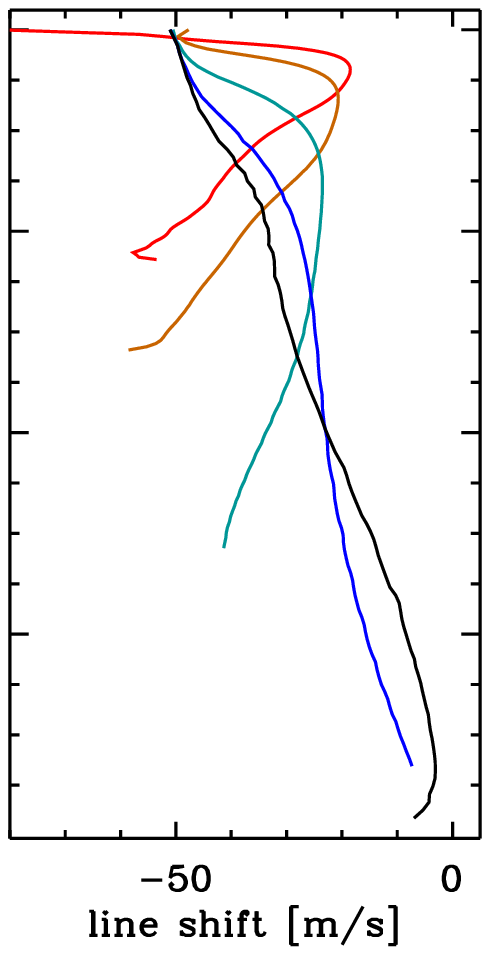}\\[2mm]
\includegraphics[height=5.9cm]{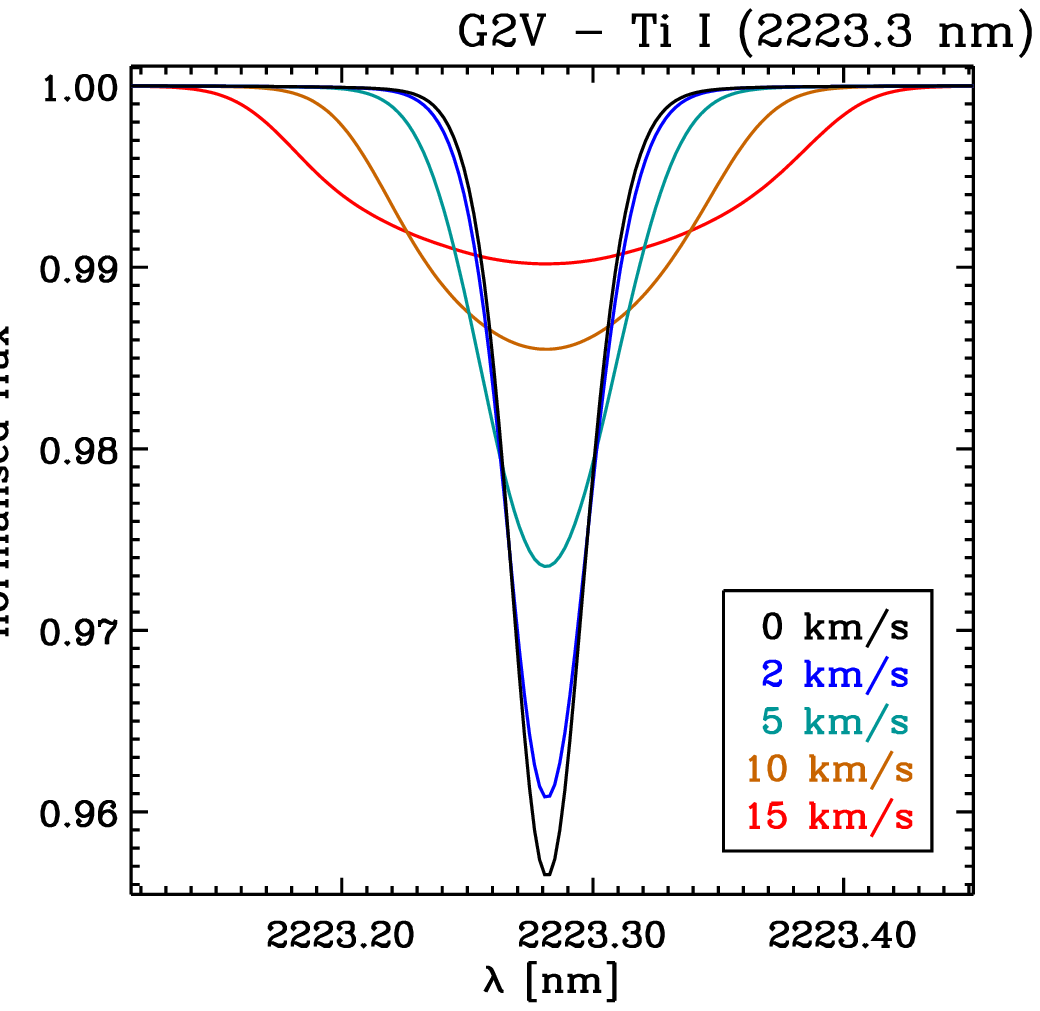}\includegraphics[height=5.9cm]{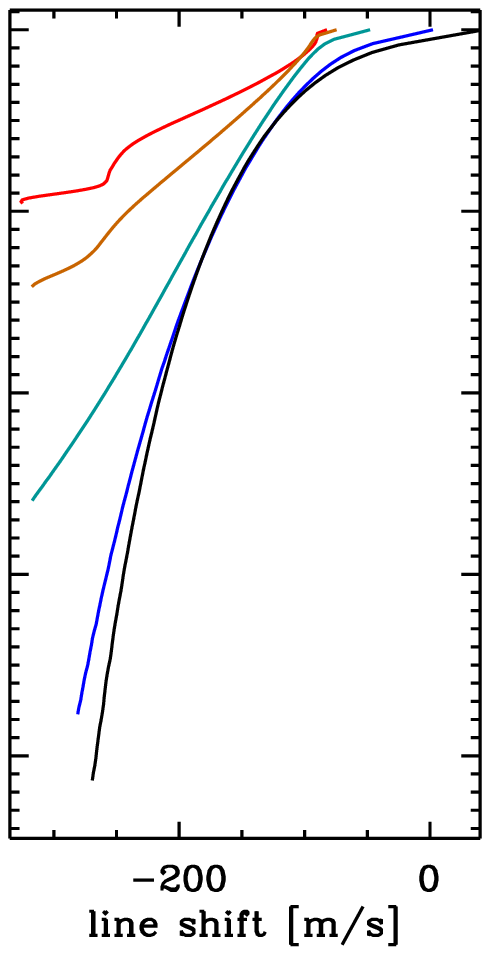}~~\includegraphics[height=5.9cm]{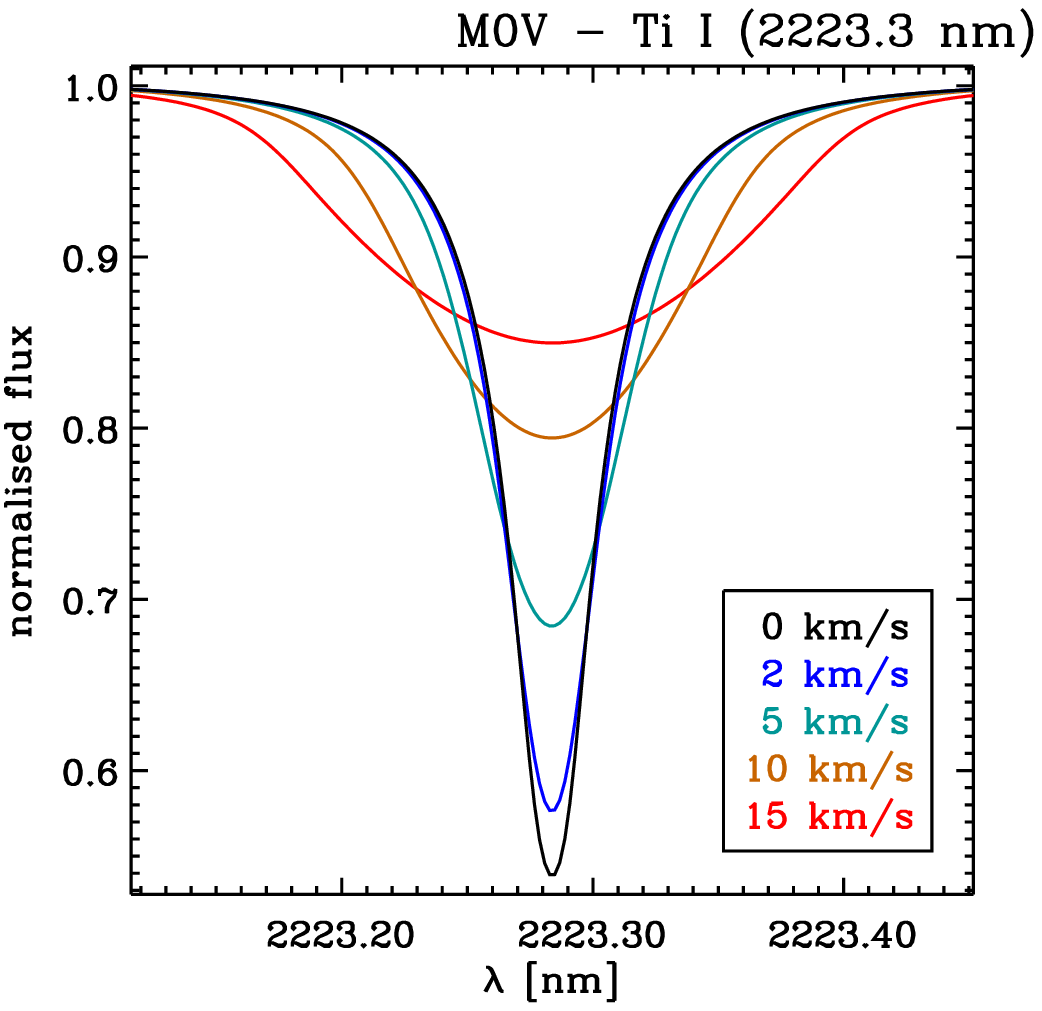}\includegraphics[height=5.9cm]{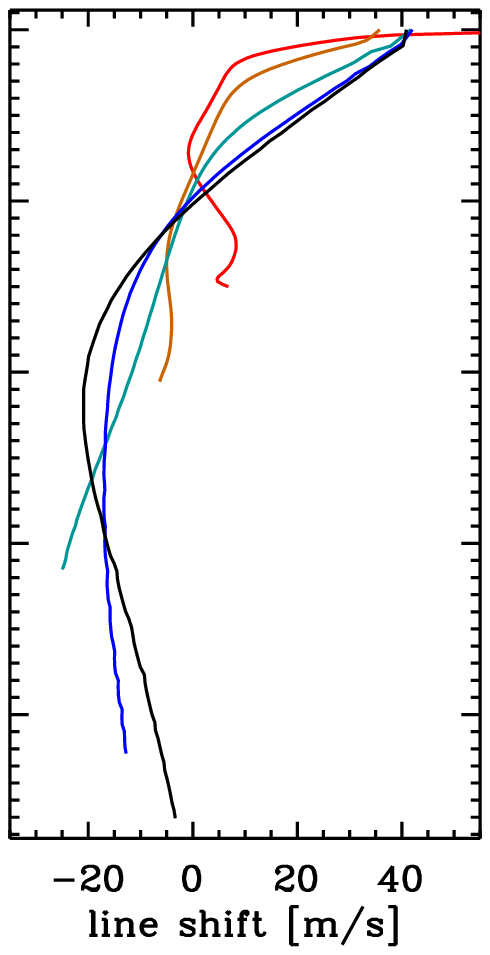}\\
\caption{Rotational broadening and line distortion. {\it Upper panels:} rotational broadening of the Fe\,\textsc{i} line at 617.3\,nm for the G2V ({\it left}) and M0V ({\it right}) simulations. The narrow right subplots of each plot show the bisectors of the lines (same colour code and ordinate labels as other subplot). {\it Lower panels:} same as upper panels, but for the Ti\,\textsc{i} line at 2223\,nm. In all cases, we used solar differential rotation together with five different rotation rates.}\label{fig:rotbroad}
\end{figure*}
\subsection{Spatially resolved line profiles}
Figure~\ref{fig:spatialdis} shows the vertically ($\mu=1$) emerging spectrum of the Ti\,\textsc{i} line at 2223\,nm for two locations (one in an upflow and one in a downflow region) in a snapshot of the K0V simulation. The depth of the line depends on temperature and temperature gradient; it is smaller in the downflow region, where the temperature gradient is lower. While the wings of this line form at $-1<\log\tau_{\mathrm{R}}<0$ in the K0V model, the line core forms at $-2.5<\log\tau_{\mathrm{R}}<-1$. Hence, the depth dependence of the flow velocity entails line asymmetries: in upflow regions, the whole line is blueshifted, but as the flow decelerates with height, the line core is less strongly shifted than the wings. Analogously, the line wings originating in a downflow accelerating with depth are more strongly redshifted than the line core.\par
\subsection{Centre-to-limb variation of line profiles}\label{sec:clv}

Figure~\ref{fig:clv2} illustrates the centre-to-limb variation of the Fe\,\textsc{i} line profile at 617.3 nm averaged over six snapshots of the F3V simulation box. The convective blueshift and line asymmetry are largest at disc centre and decrease towards the limb. For the Sun, this effect has been demonstrated in many observational studies and is referred to as the ``limb effect'' \citep[for a review, see][]{Dravins82}. It is a consequence of the correlation between intensity and velocity: at disc centre, the line-of-sight velocity corresponds to the vertical flow velocity. As upflows (granules) are bright and have a higher area fraction, they contribute more to the average line profile provided that the line depth in up- and downflows is comparable. Therefore, most lines are convectively blue-shifted at disc centre. Near the limb, the line-of-sight velocity is dominated by the horizontal flow component, which has no correlation between its direction and temperature. This results in a more symmetrical line profile with a less shifted line core near the limb. The line wings are, however, somewhat more blue-shifted (near wings) or less red-shifted (extreme wings) near the limb than at the disc centre. As the line wings originate close to the optical surface, this is probably caused by very bright front edges of the granules in this star.\par
The upper panels of Figure~\ref{fig:clv3} show the variation of equivalent width and FWHM with $\mu$ for the same line as in Figure~\ref{fig:clv2}, but for all our simulations. For the cooler simulations, the equivalent width of this line decreases towards the limb as a consequence of the decreasing temperature gradient with respect to optical depth along the inclined optical path. In the two hottest models, however, the line is stronger (higher equivalent width) near the limb (cf.~Fig~\ref{fig:clv2}), because of the high temperature in the lower photospheres: the excitation potential of 2.2\,eV of the lower level of the transition is relatively low and this level becomes depopulated as the temperature rises significantly above 7000\,K. In addition, at this temperature, the first ionisation of iron sets in, so that the abundance of Fe\,\textsc{i} drops rapidly with temperature.\par
The FWHM of the spectral line is growing in all stars towards the limb as a consequence of the increasing contribution of the horizontal flow velocity to the line-of-sight velocity: the rms of the horizontal velocity is up to about three times as high as the rms of the vertical velocity in the photosphere and above (cf. Figure~5 in Paper~I). This leads to a stronger Doppler broadening of the line at low values of $\mu$.\par
The lower panels of Figure~\ref{fig:clv3} show the Doppler shifts of the line wings (5\% of total line depth) and the line core. The line wings are slightly blueshifted for most stars and show a weak dependence on $\mu$. However, for the F3V star, the line wings are strongly redshifted at the disc centre and un-shifted near the limb. This indicates that, at disc centre, the downflow regions contribute considerably to the resulting integrated profile of this line in spite of their lower continuum intensity and area fraction. As discussed above, this line becomes weak due to the high photospheric temperatures and since upflows are much hotter than downflows, the upflows contribute very little to the average line profile. This effect vanishes near the limb, where the line-of-sight velocities correspond rather to the horizontal component of the flows and where the line forms higher in the atmosphere, i.\,e. at lower temperature.\par
The line cores of all stars except M2V show a pronounced increasing redshift (or decreasing blueshift) at decreasing $\mu$. The fact that some line cores are redshifted near the limb is due to a statistical bias owing to reversed granulation: receding flows are more often seen in front of hotter gas above intergranular lanes and approaching flows are more often seen in front of the cooler gas above granules \citep[e.\,g.,][]{Asplund00}. This effect is irrelevant for the line wings, as their formation height is closer to the optical surface.\par
The convective blueshift and its centre-to-limb variation are very important for high-precision radial velocity measurements (e.\,g. for exoplanet detection and characterisation). As this section has shown, the many different mechanisms that determine the convective blueshift for individual spectral lines can only be reproduced by comprehensive 3D calculations \citep[also see][]{Ramirez09, Allende13}.
\subsection{Disc-integrated line profiles}\label{sec:discint}
Using the numerical method outlined in Sect.~\ref{sec:met:lines}, we calculated disc-integrated profiles of the three spectral lines listed in Table~\ref{tab:lines} for six snapshots of each of our six simulations.\par
\begin{figure}
\centering
\includegraphics[width=8.8cm]{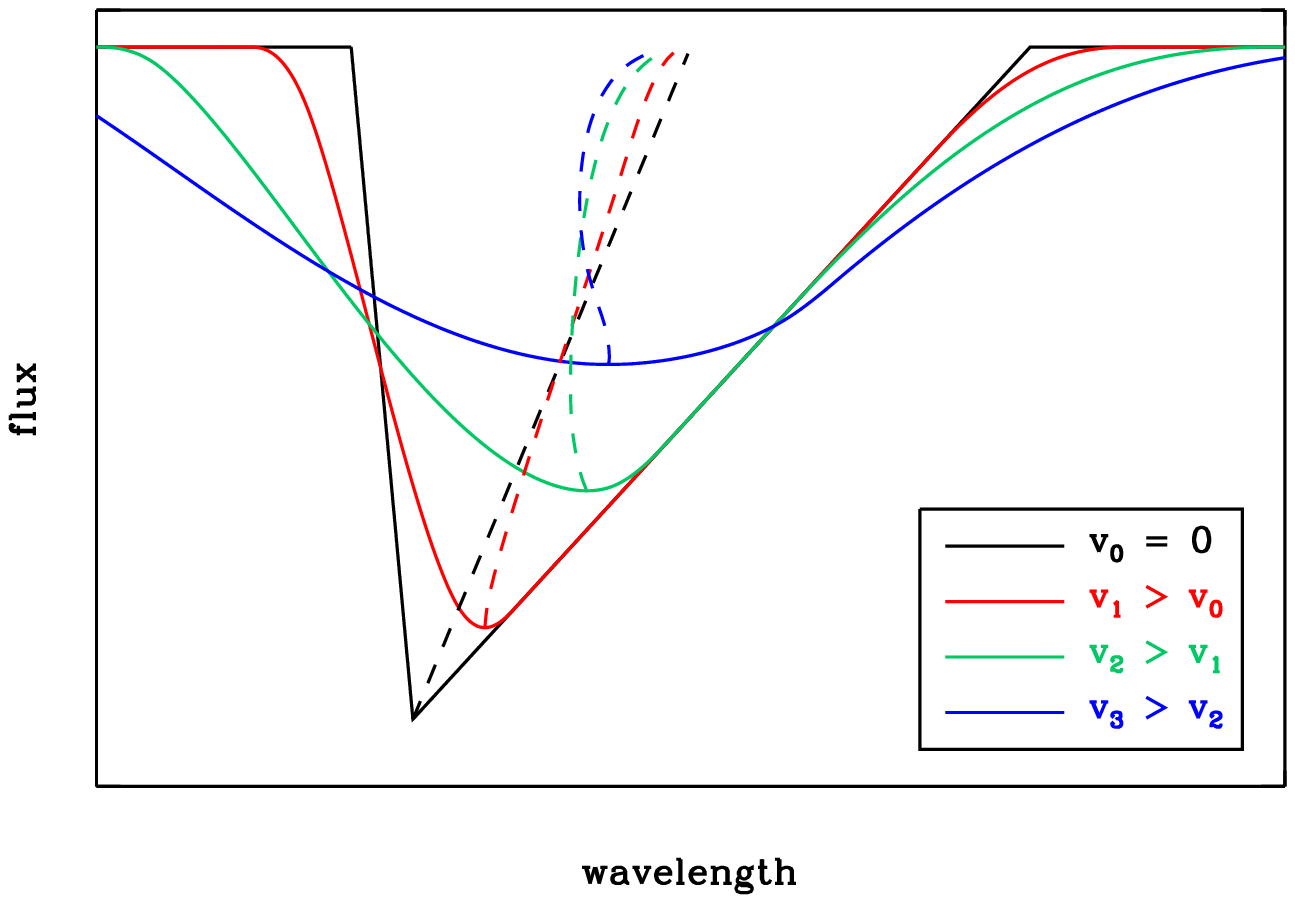}
\caption{Illustration of line broadening and bisector modification by rotation. The black solid line shows an artificial line profile with a straight (but inclined) bisector (black dashed curve). The coloured curves were obtained by convolving the black curve with a (rigid-body) rotation profile of different width, simulating different velocities. Note that the bisectors (dashed curves) are modified by the convolution although the rotation profile is perfectly axis-symmetric.}\label{fig:bisec_illu}
\end{figure}

 
Figure~\ref{fig:rotbroad} shows the profiles of two lines (Fe\,\textsc{i} at 617.3\,nm and Ti\,\textsc{i} at 2223.3\,nm) for the solar (G2V) and the M0V simulations for solar-like differential rotation ($\alpha=0.2$, Eq.~(\ref{eqn:alpha})) and various values of $\vel_{\mathrm{rot}} \sin i$ (i.\,e., the projection of the equatorial rotation velocity onto the line of sight). As we have discussed in the previous section, the convective flows and their correlation with temperature generally cause line asymmetries and shifts (with respect to the rest wavelength of the line). These effects are visible in the disc-integrated profiles as well and best illustrated by line bisectors, which are generally curved and shifted. Stellar rotation broadens the spectral line. The line shift due to rotation is symmetric with respect to the projected rotation axis on the stellar disc (see Fig.~\ref{fig:sketch}) and does therefore not produce an additional line asymmetry. However, asymmetries of the local profiles (due to convection) can be modified by rotation in a non-trivial fashion. This is illustrated in Figure~\ref{fig:bisec_illu}: here, a $\mu$-independent arteficial line profile is convolved by a rotation profile to mimic stellar rigid-body rotation ($\alpha=0$). The bisector depends on rotation velocity and evolves from a straight line into curved C- and S-shapes when the rotation velocity is increased.\par
\begin{figure}
\centering
\includegraphics[width=4.3cm]{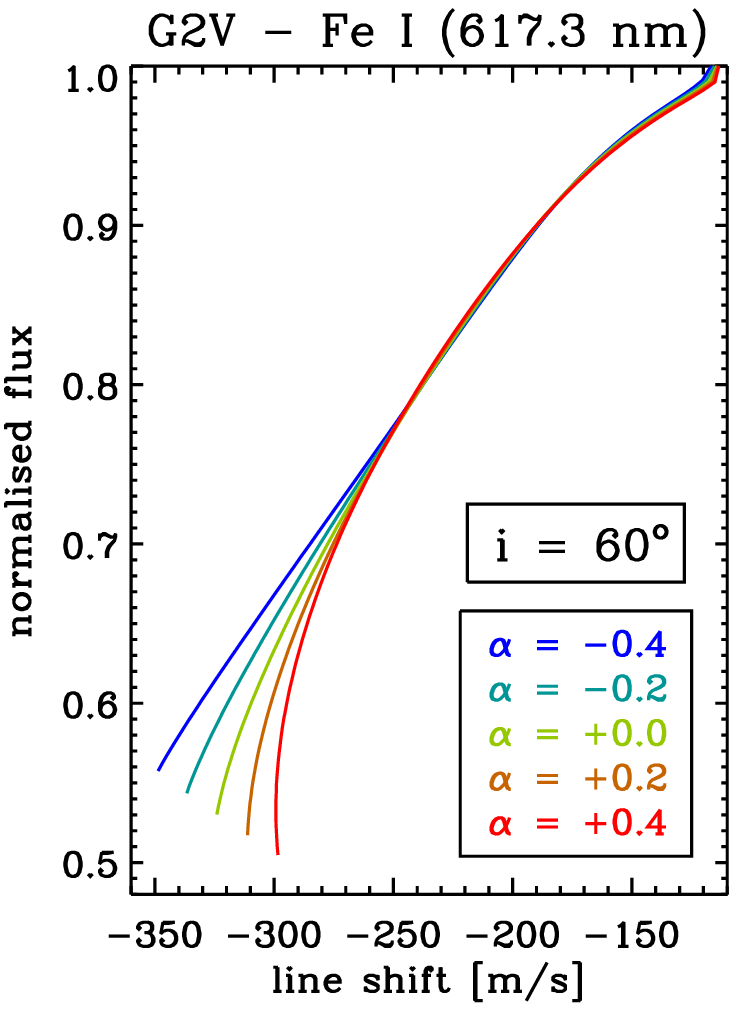}~~\includegraphics[width=4.3cm]{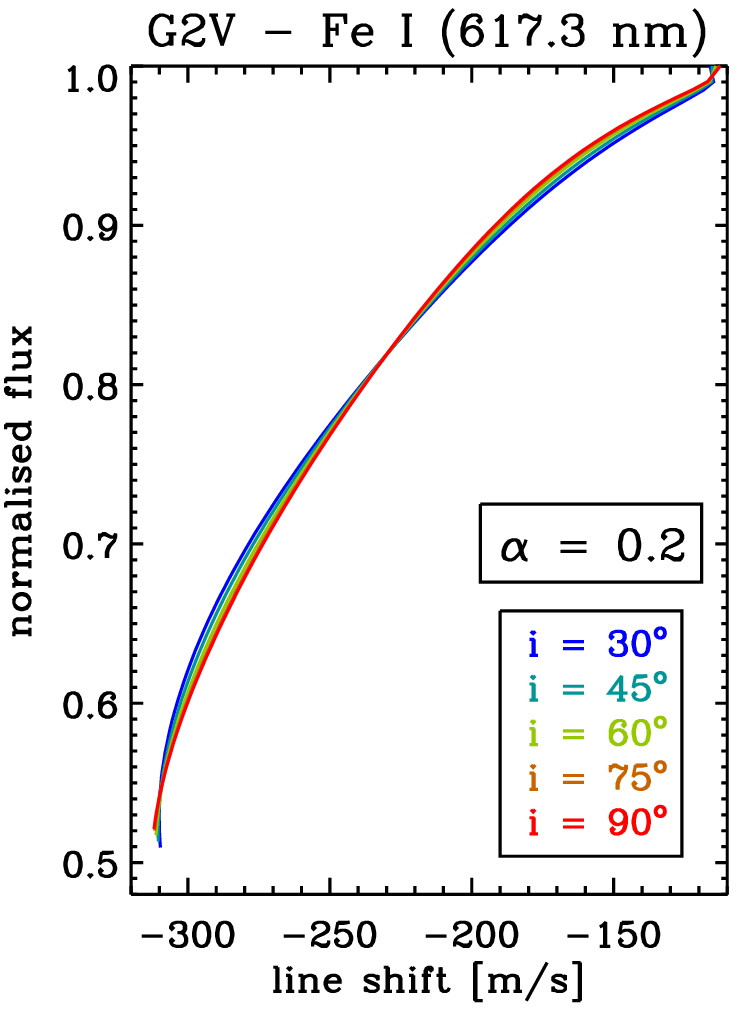}\\[3mm]
\includegraphics[width=4.3cm]{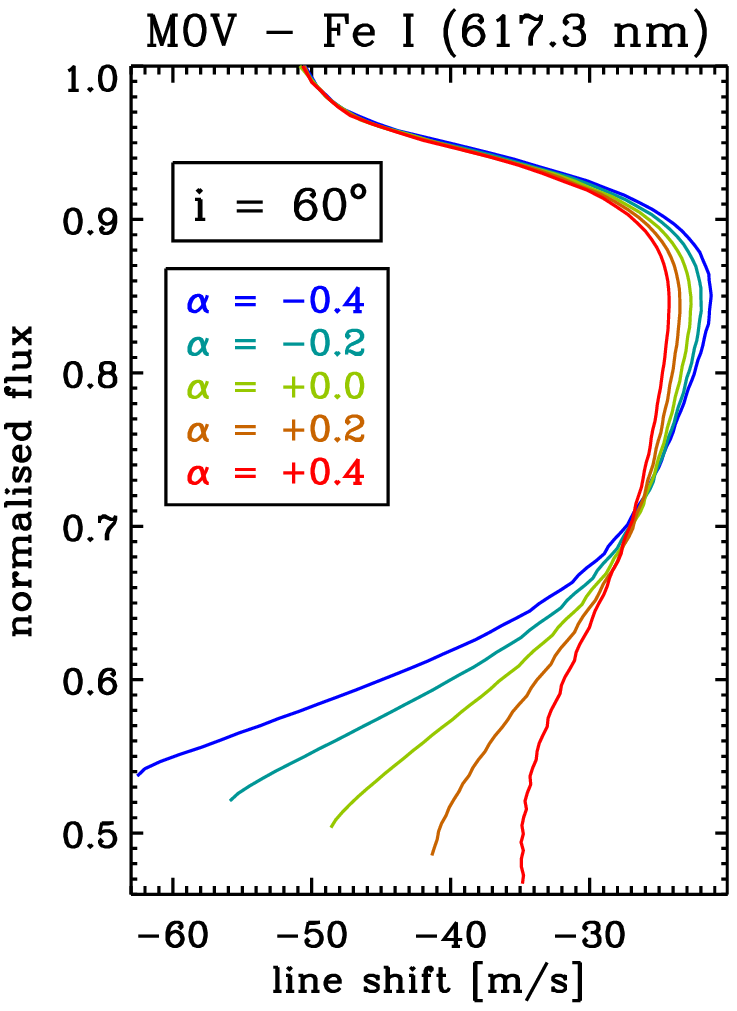}~~\includegraphics[width=4.3cm]{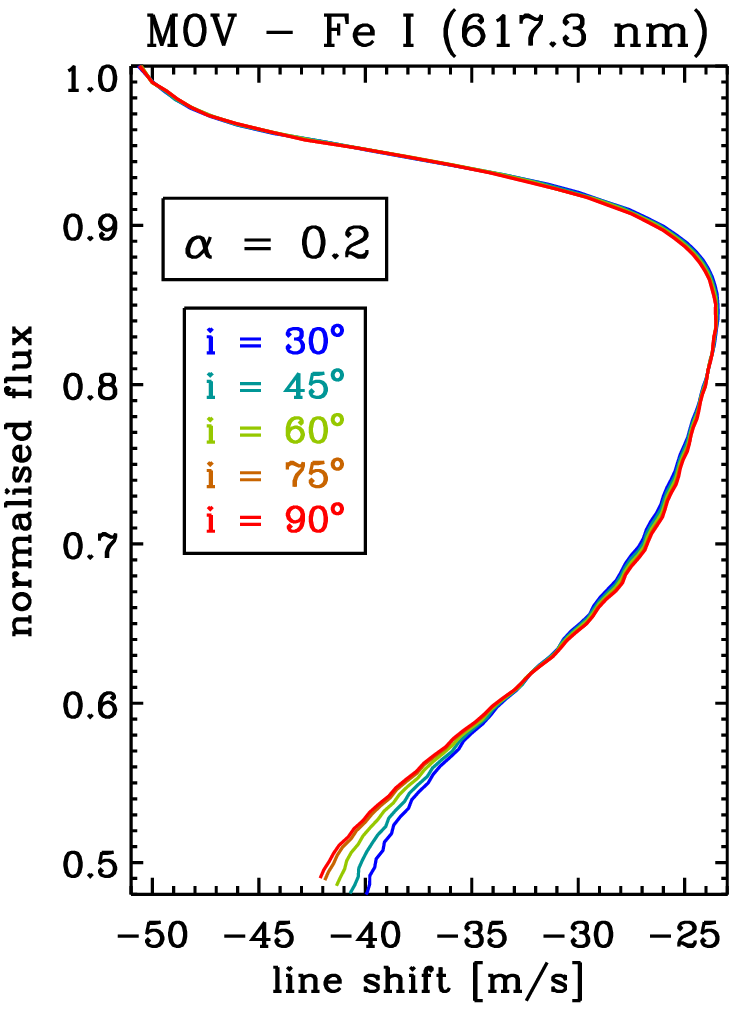}\\
\caption{Bisector variation with inclination $i$ and differential rotation parameter $\alpha$ of the Fe\,\textsc{i} line at 617.3\,nm for the G2V and M0V simulations. For the variation of $\alpha$ ({\it left panels}) the inclination was fixed at $i=60^{\circ}$; for the variation of $i$ ({\it right panels}) the differential rotation parameter was fixed at $\alpha=+0.2$. For all calculations $\vel\sin i= 5\,\mathrm{km\,s^{-1}}$ was used.}\label{fig:bisvar}
\end{figure}

Figure~\ref{fig:bisvar} shows the dependence of the bisector of the Fe\,\textsc{i} line at 617.3\,nm on the differential rotation parameter $\alpha$ as defined in Eq.~(\ref{eqn:alpha}) and on the inclination $i$ of the rotation axis of the star at a rotation speed of $\vel_{\mathrm{rot}}\sin i=5\,\mathrm{km\,s^{-1}}$. When $\alpha$ is varied, $i=60^{\circ}$ is fixed, when $i$ is varied, $\alpha=0.2$ is fixed. While the impact of $i$ on the bisector shape is very subtle, the effect of $\alpha$ is more pronounced. The effect is similar in magnitude for all inclinations between 15 and 90$^{\circ}$ at constant $\vel_{\mathrm{rot}}\sin i$. In contrast, we find that the effect of inclination is proportional to $|\alpha|$ and thus vanishes for rigid-body rotation ($\alpha=0$) as expected. The dependence of the bisector shape on $\alpha$ and $i$ can be explained by the different contribution of differently shifted and inclined projected surface elements of the star. With our numerical method (see Sect.~\ref{sec:met:lines} and Fig.~\ref{fig:sketch}), the weights $w_{jk}$ change with $\alpha$ and (if $\alpha \ne 0$) with $i$. Similarly a continuous contribution function $w(\mu,\vel)$ would depend on both parameters. This implies that, at least for inactive stars with homogeneous convective surface structure, the investigation of the bisector shape of individual lines can yield constraints on the rotational velocity and the differential rotation parameter.
\subsection{Comparison to observational data}\label{sec:comp}
For a preliminary comparison of our disc-integrated line profiles, we considered three F stars from a set of observations of main-sequence stars obtained with the CES spectrograph mounted on the 3.5-m telescope at Calar Alto observatory \citep{Reiners06,ammrei12}.\par
\begin{figure*}
\centering
\includegraphics[width=5.9cm]{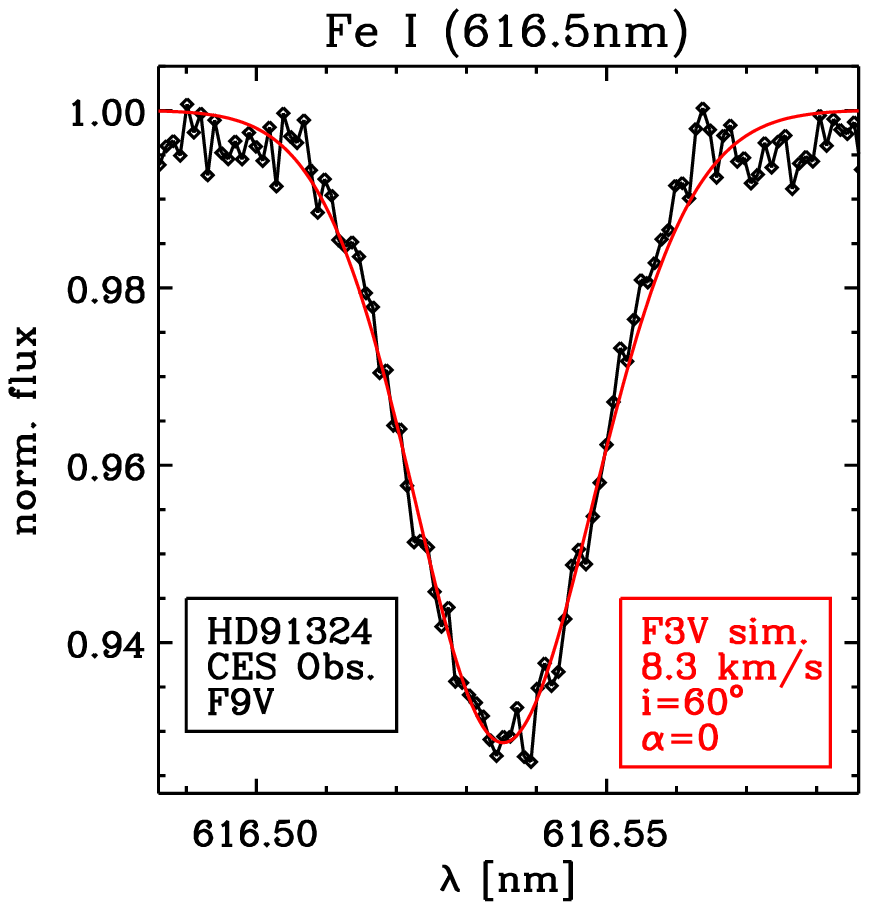}~~\includegraphics[width=5.9cm]{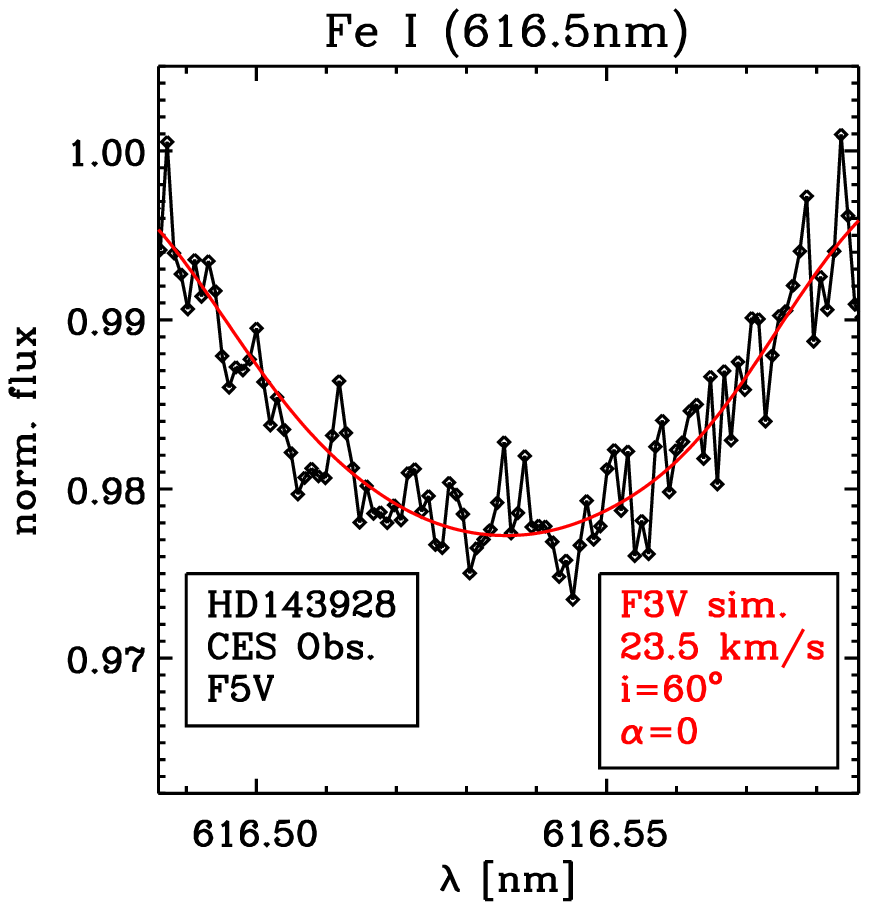}~~\includegraphics[width=5.9cm]{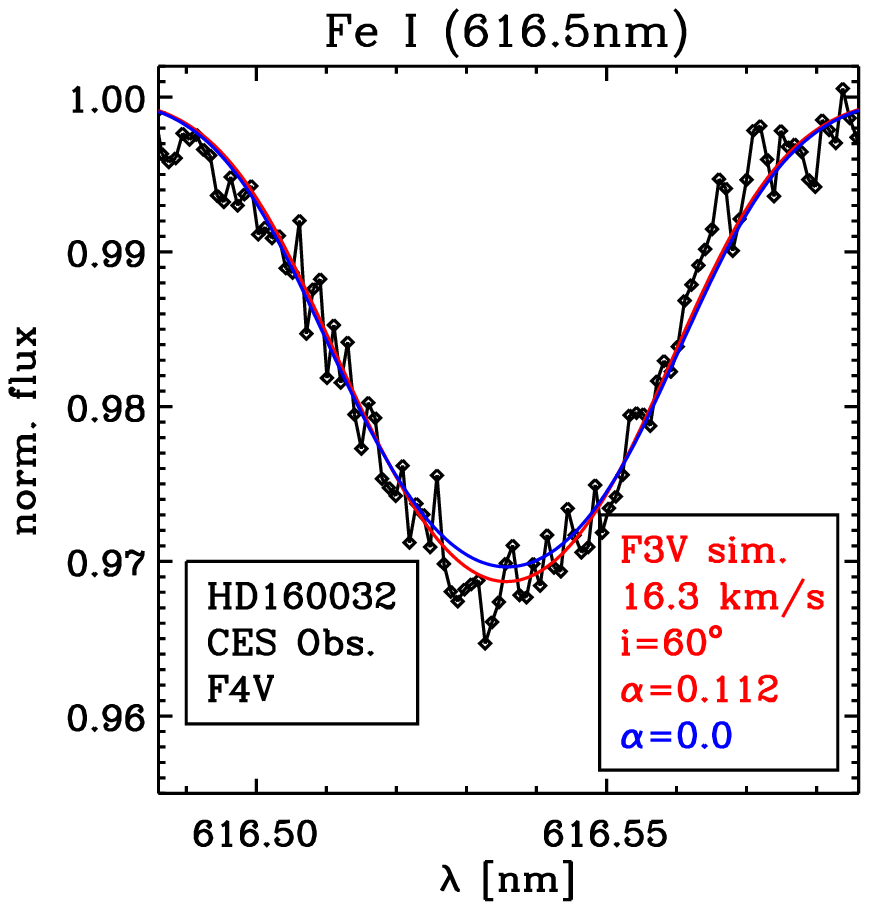}\\
\includegraphics[width=5.9cm]{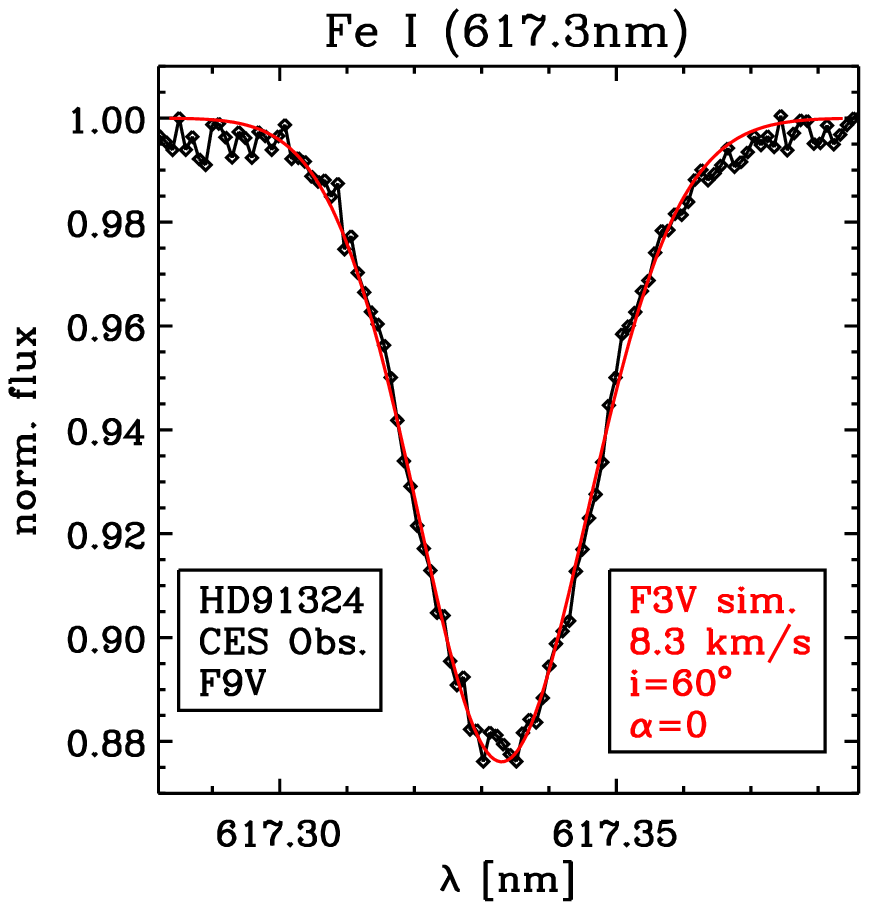}~~\includegraphics[width=5.9cm]{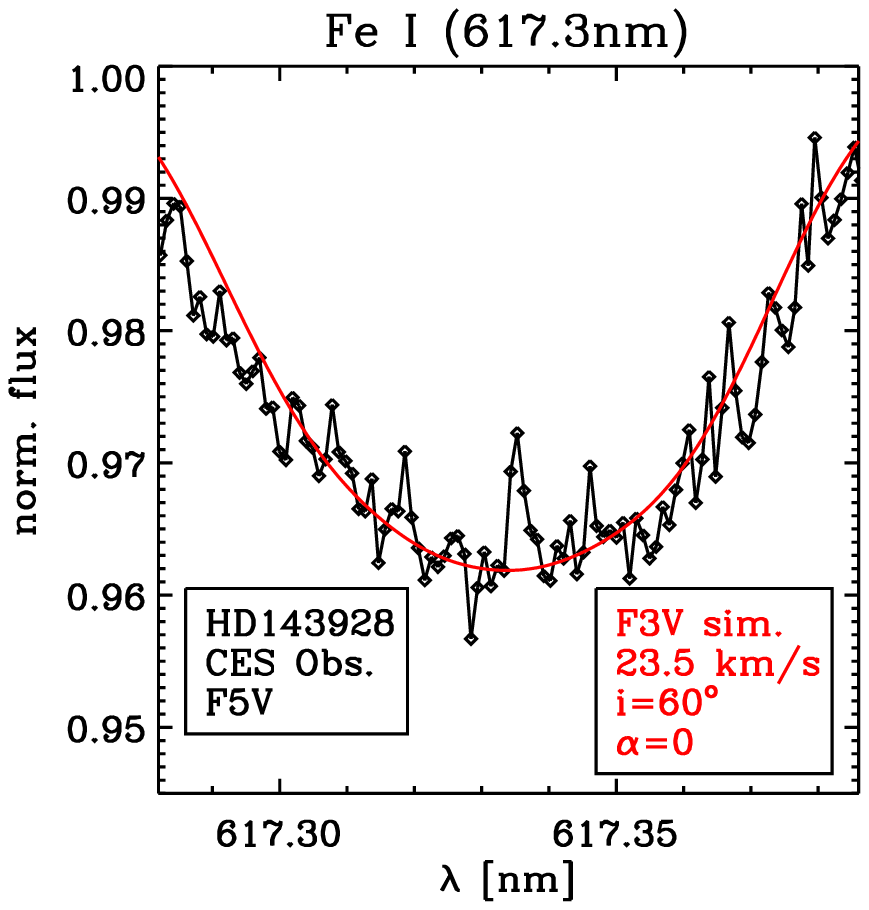}~~\includegraphics[width=5.9cm]{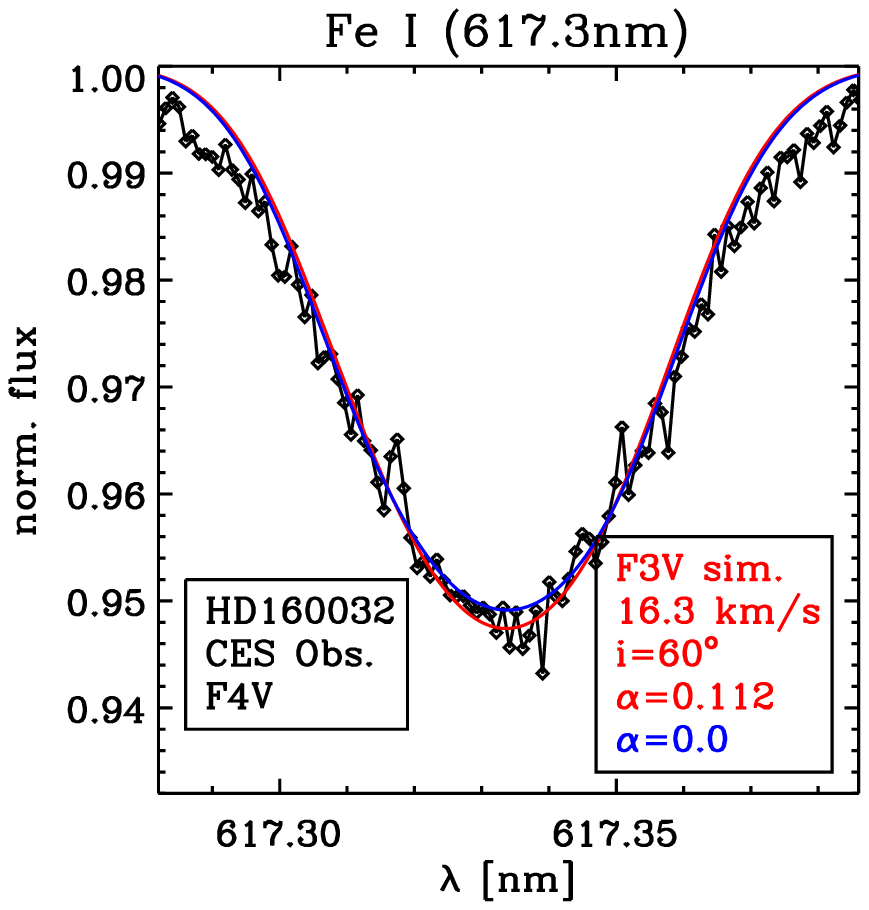}\\
\caption{Profiles of the two Fe\,\textsc{i} lines of three F stars (black lines and diamonds) in comparison to disc-integrated profiles from the F3V simulation (red curves). For the rotational broadening of the synthesised lines, an inclination of the rotation axis of $i=60^{\circ}$. For a better comparison, the line profiles were multiplied by a factor such that the depth of the line core come into agreement. For the projected rotational velocities $\vel_{\mathrm{rot}}\sin i$ and differential rotation parameter $\alpha$ the values by \citet{ammrei12} were used; for comparison, the blue curve in the {\it right panels} shows a disc-integrated profile with the same $\vel_{\mathrm{rot}}\sin i$ but $\alpha=0$.}\label{fig:comp_prof}
\end{figure*}

Figure~\ref{fig:comp_prof} shows a comparison of disc-integrated synthetic spectra with their observational counterpart. As we have only a limited number of spectral types and only one set of elemental abundances, the depth of the synthetic spectral line did not match the observed ones and was multiplied by a factor in order to facilitate a better comparison between model and observation. For the unknown inclinations, we assumed $i=60^{\circ}$ as the ``average case''. The projected rotational velocity $\vel_{\mathrm{rot}}\sin i$ and the differential rotation parameter $\alpha$ was set to the values published by \citet{ammrei12}. The observed data have typical signal-to-noise ratios of several hundred and a spectral resolving power of $R\approx 230,000$. The instrumental broadening was mimicked by a convolution of the synthetic spectra with a gaussian. In general, the synthetic spectra provide a good representation of the observations. Some deviations are likely due to the mismatch in effective temperature, surface gravity, and chemical composition between stars and simulations as well as to magnetic activity and surface inhomogeneities in the observed stars, which are not taken into account so far.\par
The preliminary comparison to observed data in this section merely serves as an illustration and an outlook. A more detailed analysis of observational data is beyond the scope of this paper and will follow in a subsequent paper.

\section{Summary}

The surface convection of six main-sequence stars was simulated with the 3D magneto-hydrodynamics code \texttt{MURaM}. The first paper of this series, Paper~I, dealt with the overall structure of the convective (sub-)surface layers with an emphasis on the optically thick part of the simulation domain and on average depth profiles of thermodynamic quantities and energy fluxes. Here, we analysed the properties of the granules in terms of their sizes, intensities, vertical velocities, and lifetimes in analogy to the analysis of solar granules in high spatial resolution observations. Furthermore, we calculated the profiles of three spectral lines and computed disc-integrated line profiles for stars with various rotation parameters (rotation velocity, inclination, differential rotation parameter).\par
In agreement with earlier studies, we find that the granules of main-sequence stars become smaller as effective temperature decreases and gravitational acceleration increases. The mean size is several thousand kilometers in our hottest model, an F3V star, and only a few hundred kilometers in our coolest simulation, an M2V star. Correlations between granule size, brightness, and vertical velocity were found. For instance, large granules have, on average, a lower vertical velocity than small ones and bright granules have, on average, a higher vertical velocity than darker ones. The mean lifetime of the granules decreases for decreasing effective temperature. As in various previous studies on solar granulation, we find an exponentially decreasing lifetime distribution for long-lived granules (lifetime of more than about 3 minutes) in all stars. The mean lifetime is decreasing along the model sequence from  ${\sim}8\,\min$ for F3V to ${\sim}2\,\min$ for M2V. However, the short-lived granules are more frequent than this exponential law predicts.  For each individual star, we find the lifetime of granules rather uncorrelated to other (time averaged) granule properties.\par
In all simulations, we find vertical vortex features, which sometimes show up as non-magnetic bright points due to the reduction of gas pressure caused by the centrifugal force. Owing to the relatively weak dependence of opacity on temperature between 4000 and 5000\,K, they become particularly bright in the K-star simulations.\par
The spectral line profiles vary substantially over the stellar disc, mainly due to the 3D character of the photospheric flows. The local line profiles are asymmetric owing to the variation of the line-of-sight velocity along the optical path. In disc-integrated spectra, the line asymmetries are strongly modulated by stellar rotation and we find a substantial modulation of the line bisectors not only by the rotation velocity but also by the differential rotation and, to a lesser extend, by the inclination. A first comparison to observed stellar line profiles gives a good qualitative match.\par
In a subsequent paper, we will investigate how photospheric magnetic fields of various field strengths affect the surface convection and spectral line profiles.
\begin{acknowledgements}
The authors acknowledge research funding by the {\it Deutsche Forschungsgemeinschaft (DFG)} under the grant {\it SFB 963/1, project A16}. BB acknowledges financial support by the {\it International Max Planck Research School }(IMPRS) {\it on Physical Processes in the Solar System and Beyond at the Universities of Braunschweig and G\"ottingen}. AR acknowledges research funding from {\it DFG} grant {\it RE 1664/9-1}. The authors thank M. van Noort for help with the \texttt{SPINOR} code.  
\end{acknowledgements}
\bibliography{beecketal}
\end{document}